\documentstyle[12pt,doublespace,epsf,here,rotating]{report}

\newcommand{\dspexp}[1]{\mbox{$e$\raisebox{1.5ex}{$\displaystyle{#1}$}}}
\def\approxle{\,\raisebox{-0.625ex}{$\stackrel{<}{\sim}$}\,}
\def\approxge{\,\raisebox{-0.625ex}{$\stackrel{>}{\sim}$}\,}

\begin{document}

\pagestyle{empty}

\vspace*{0.5in}
\begin{center}
{\Large \bf  Meson Properties in the Quark Model:  \\
A Look at Some Outstanding Problems} \\[0.4in]
by \\[0.3in]
{\bf Harry G. Blundell, \hspace{0.025in} B.Sc.} \\[0.6in]
A thesis submitted to \\
the Faculty of Graduate Studies and Research \\
in partial fulfilment of \\
the requirements for the degree of \\
Doctor of Philosophy  \\[0.4in]
Ottawa-Carleton Institute for Physics \\ 
Department of Physics \\
{\bf Carleton University} \\
Ottawa, Ontario, Canada \\
July 19, 1996 \\[0.3in]
\copyright Harry G. Blundell, 1996
\end{center}

\clearpage

\pagestyle{plain}
\pagenumbering{roman}
\setcounter{page}{2}

\vspace*{-0.5in}
\begin{center}
The undersigned recommend to \\
the Faculty of Graduate Studies and Research\\
acceptance of the thesis\\[.5in]
{\large \bf  Meson Properties in the Quark Model:  \\
A Look at Some Outstanding Problems}\\[.5in]
submitted by {\bf Harry G. Blundell, B.Sc.} \\
in partial fulfilment of the requirements for \\
the degree of Doctor of Philosophy \\[.5in]
\rule{3in}{.5pt}\\
{\bf Chair, Department of Physics}\\[.5in]
\rule{3in}{.5pt}\\
{\bf Thesis Supervisor}\\[.5in]
\rule{3in}{.5pt}\\
{\bf External Examiner}\\[.5in]
Carleton University\\[.2in]
Date \rule{1.5in}{.5pt}
\end{center}

\newpage

\begin{center}
{\huge \bf  Abstract}\\[0.5in]
\end{center}
\addcontentsline{toc}{section}{Abstract}

This thesis examines three problems of the quark model, split between two
general areas.

The first area involves models of meson decay.  The calculations of decay
widths in two such models, the $^3P_0$ model and the flux-tube breaking model,
were automated so that they could be easily done for any decay.  The models
were then used to investigate two particular problems.  The first is the nature
of the $f_4(2220)$ -- although tentatively identified as the $^3F_4$ $s\bar{s}$
meson by the Particle Data Group, its identity has been uncertain since its
discovery in 1983.  An exhaustive calculation of the strong decay modes of the
$^3F_2$ and $^3F_4$ $s\bar{s}$ mesons was performed in order to examine the
possibility that the state is a meson.  It was found that the $f_4(2220)$
cannot be the $^3F_2$ $s \bar{s}$ meson, and is unlikely to be the $^3F_4$,
although given the uncertainties of the models the latter possibility cannot be
ruled out.  Instead the following explanation is proposed: that the broad state
seen in hadron beam experiments is the $^3F_4$ $s\bar{s}$ meson, and the narrow
state seen in $J/\psi$ radiative decay is a glueball.  Further experimental
data is needed to finally identify the $f_4(2220)$.

The second problem investigated using meson decay models is the
determination of the mixing angle between the $K_1(1270)$ and $K_1(1400)$
mesons.  This was done by comparing predictions of the meson decay models for
five partial decay widths and two ratios of D to S amplitudes to experimental
data.  A mixing angle of approximately $44^\circ$ was extracted, and the
implications to the quark model Hamiltonian are discussed.

The third problem falls in the second general area: final state interactions
(FSI's).  The effect of the strong interactions between the outgoing pions was
examined for the reaction $\gamma \gamma \rightarrow \pi \pi$ near
threshold.  For charged pions, the experimental data agrees with the results
calculated both with and without the effects of FSI's -- better data is need to
distinguish between the two.  However, agreement was not obtained for neutral
pions.  It is believed that the discrepancy may be due to the effects of
resonance production.  If this proves to be the case, then the techniques used
in this work could be extended to try to understand a long-standing puzzle: the
nature of the unexplained structures seen in the cross sections of $\gamma
\gamma \rightarrow$ two vector mesons.

\newpage

\begin{center}
{\huge \bf Acknowledgments}\\[1.2cm]
\end{center} 
\addcontentsline{toc}{section}{Acknowledgments}

This work was made possible by the support of a large number of people.  I
would especially like to thank my supervisor Stephen Godfrey, for doing those
things that supervisors do.  Also our collaborator (on the $\gamma \gamma \to
\pi \pi$ work) Eric Swanson.  A large number of other people answered physics
questions (or tried to) -- in particular Peter Watson and Mike Doncheski (Mike
also read this thesis for me).  A number of people answered computer questions
-- in particular Mike Boyce, Matthew Jones, Alan Barney and Wade Xiong.  The
theory post-docs and my fellow grad students in physics kept life interesting
-- in particular Mike Boyce, Alan Dekok, Mike Doncheski, Larry Gates, Matthew
Jones, Ted Lawrence, Greg Stuart, Steve Taylor, and Sherry Towers.  And
finally, thanks to Deborah Schneider, for inspiring me to actually finish.

\newpage

\tableofcontents
\newpage
\addcontentsline{toc}{section}{List of Tables}
\listoftables
\newpage
\addcontentsline{toc}{section}{List of Figures}
\listoffigures

\clearpage

\pagenumbering{arabic}
\pagestyle{myheadings}
\setcounter{page}{1}

\chapter{Introduction}
\markright{Chapter 1.  Introduction}

Particle physics is the study of the particles that make up the universe, and
the interactions that take place between them.  It is also referred to as
elementary particle physics, by which we mean that we would ideally like to
understand the universe in terms of its ``elementary'' constituents (should
such things exist), that are not composed of other particles.

Why study particle physics?  The questions ``What is the universe made of?''
and ``How does it work?'' are of fundamental interest for their own sake.
In addition, understanding the universe as it is now might help us answer the 
question ``Where did it come from?''

Aside from these weighty questions, what about immediate material benefits?  
Like much of fundamental research, it is impossible to know now what benefits 
might be realized from particle physics.  However, it is worth
noting that in 1897 the quest to understand the universe led to the discovery
of the electron by J.J. Thompson.  It is safe to say that this early particle
physicist could not have predicted the huge effect his work would have on 
humanity.

\section{The Standard Model of Particle Physics}

We assume that the reader is familiar with modern physics.  Familiarity with
the field of particle physics would also be useful, but hopefully not
necessary; in addition to the introduction to the field that follows,
Appendix~\ref{a:tools} contains descriptions of some of the terms and tools of
particle physics that may be useful to specialists in other fields.

\subsection{The Particles that Constitute Matter}

At present, the elementary particles that make up matter are thought to be the
quarks (six of them, plus their antiparticles) and the leptons (again, six of
them, plus antiparticles).  We think of them as elementary because, so far, we
do not have any indication that they are not.  The quarks are the up ($u$),
charm ($c$) and top ($t$), all with electric charge $+\frac{2}{3}$ (on a scale
where the electron has charge $-1$), and the down ($d$), strange ($s$) and
bottom ($b$), all with charge $-\frac{1}{3}$.  The leptons are the electron
(e), mu ($\mu$) and tau ($\tau$), all with charge $-1$, and the three
corresponding neutrinos, $\nu_{\rm e}$, $\nu_\mu$ and $\nu_\tau$, all with
charge 0.  We say corresponding because all these particles are usually thought
of as being arranged in doublets,
\[
{\rm Quarks}:\;
\begin{array}{r} {\rm charge}\; +\frac{2}{3}:\;\;\;\; \\ 
{\rm charge}\; -\frac{1}{3}: \;\;\;\;\end{array}
\left(\begin{array}{c} u \\ d \end{array} \right),
\left(\begin{array}{c} c \\ s \end{array} \right),
\left(\begin{array}{c} t \\ b \end{array} \right),
\]
\begin{equation}
{\rm Leptons}:\;
\begin{array}{r} {\rm charge}\;\;\;\;\; 0:\;\;\;\; \\ 
{\rm charge}\; -1: \;\;\;\;\end{array}
\left(\begin{array}{c} \nu_{\rm e} \\ {\rm e} \end{array} \right),
\left(\begin{array}{c} \nu_\mu \\ \mu \end{array} \right),
\left(\begin{array}{c} \nu_\tau \\ \tau \end{array} \right).\label{particles}
\end{equation} 
As one moves to the right in the first, second and fourth rows\footnote{Each
column is called a generation; there appear to be only three generations.}, the
particles become more massive.  The particles in the third row, the neutrinos,
have masses smaller than the experiments can currently measure, and may be
massless.  The present experimental values of the masses\footnote{Note that the
masses in Table~\ref{t:masses} are given in units of energy (eV), when
the units of mass would normally be eV/$c^2$.  We are following
the particle physics convention of setting $c \equiv h\!\!\!^- \equiv 1$, which
expresses mass and momentum in units of energy, and length and time in units of
energy$^{-1}$.  When converting from these `natural units' to more standard
units at the end of a calculation, factors of $h\!\!\!^-$ and $c$ are inserted
as needed to give the required units.} are listed in Table~\ref{t:masses}
\cite{montanet94:review}.  Two different masses are listed for the quarks.  The
current mass is the mass that appears in the Lagrangian describing the strong
interaction (see below).  The constituent mass is the effective mass the quark
has when it is bound inside a hadron (the hadrons are the strongly interacting
particles -- see below) -- the numbers given are approximate because they
depend on the hadron model used.

\begin{table}[t]
\vspace*{-0.3cm}
\begin{center}
\begin{tabular}{|c|r|r||c|r|} \hline
Quark & \multicolumn{2}{c||}{Quark Mass} & Lepton & Lepton Mass \hspace{0.7cm}
\\ \cline{2-3} 
& Current \hspace{0.5cm} & Constituent & & \\ \hline \hline
$u$ & 2 to 8 MeV & $\sim 300$ MeV & $\nu_{\rm e}$ & $<5.1$ eV (95\% CL)\\
$c$ & 1.0 to 1.6 GeV & $\sim 1.5$ GeV & $\nu_\mu$ & $< 0.16$ MeV (90\% CL)\\
$t$ & $180 \pm 12$ GeV & $\sim 180$ GeV & $\nu_\tau$ & $< 31$ MeV (95\% CL)\\
$d$ & 5 to 15 MeV & $\sim 300$ MeV & e & $0.51099906(15)$ MeV\\
$s$ & 100 to 300 MeV & $\sim 500$ MeV & $\mu$ & $105.658389(34)$ MeV\\
$b$ & 4.1 to 4.5 GeV & $\sim 5$ GeV & $\tau$ & $1777.1^{+0.4}_{-0.5}$ MeV\\\hline
\end{tabular}
\\
\end{center}
\vspace{-0.4cm}
\caption[Masses of the quarks and leptons]{Masses of the quarks and leptons.
The numbers in parentheses denote the 1 standard deviation uncertainty in the 
end digits of the value.  The $\pm$ errors are also 1 standard deviation 
uncertainties.  CL means confidence limit (e.g.\ we are 95\% sure that 
$m_{\nu_{\rm e}} < 5.1$ eV).}
\label{t:masses}
\end{table}

Associated with each type of quark and lepton is its antiparticle, which has
the same mass, but opposite quantum numbers (such as charge).  Antiparticles
are denoted by an overbar (e.g.\ $\bar{u}$).  Each type of quark or lepton is
referred to as a flavour; a particle and its antiparticle have opposite
flavour.  Each quark and lepton has an intrinsic angular momentum, called spin,
of $\frac{1}{2}$, making them fermions (they obey Fermi-Dirac statistics, which
hold for particles of half-integral spin).

In addition to their electric charge, each quark has an additional ``charge''
referred to as colour (but having absolutely nothing to do with the colours of
the everyday world).  There are three possible values of a colour charge, plus
the three anti-colours of the antiquarks.  It appears to be a property of
nature that coloured objects cannot exist freely by themselves, so quarks are
confined inside hadrons in configurations that
produce an object with no net colour.

\subsection{The Interactions}

There are four different types of interactions that take place between the
particles; to each of them there corresponds one or more gauge bosons (they
obey Bose-Einstein statistics, which hold for particles of integral spin), a
type of particle that carries the effects of the interaction when it is
exchanged between the two particles involved.

The electromagnetic interaction occurs between objects with electric charge,
including neutral particles composed of charged constituents (e.g. the neutron
interacts via its magnetic moment).  It is carried by the photon $\gamma$,
which is massless and has a spin of 1 (making it a vector particle).  As a
quantum of electromagnetic radiation, it will be most familiar to the reader as
visible light.  The theory governing the electromagnetic interaction is called
Quantum Electrodynamics (QED).

The weak interaction occurs between leptons and quarks, and involves the
interacting particles being transformed into their partners in the doublets of
particles shown in (\ref{particles}).\footnote{For the quarks at least this is
a simplification: while the quarks prefer to be transformed into their
partners, they may also be transformed into one of the other two quarks in the
same row as their partners.  It is not clear if the leptons also have this type
of cross-generational mixing.}  It is carried by the $Z^0$ and $W^\pm$ bosons
(spin 1) with the charges shown and masses of $91.187\pm 0.007$ GeV and
$80.33\pm 0.15$ GeV, respectively.  Because the weak interaction deals with the
flavour of the involved particles, the theory that describes it is sometimes
called Quantum Flavourdynamics.  The theories of the electromagnetic and weak
interactions have been combined into a single theory, called the Electroweak
Theory.  The weak interaction is the cause of certain decays, such as the decay
of the neutron ${\rm n} \to {\rm p}\: {\rm e}^-\: \bar{\nu}_e$.

The strong interaction occurs between coloured objects, including colourless
particles composed of coloured constituents.  It is carried by eight gluons
$g$, which have spin 1, no mass and no charge.  However, they themselves have
colour, which complicates the theory of the strong interaction, Quantum
Chromodynamics (QCD).  The strong interaction binds the quarks together to form
hadrons, including the nucleons.  The residual strong forces between the
colourless nucleons bind the nucleus together.

The particles and interactions discussed above, and a spontaneous symmetry
breaking mechanism (called the Higgs mechanism) that gives the fermions their
masses make up what is called the Standard Model of Particle Physics.  The
Higgs mechanism gives rise to the Higgs boson, a massive ($>58.4$ GeV, 95\%
CL), neutral scalar (spin 0) particle, which has not yet been seen.

The remaining interaction, gravity, is well described by a classical theory, 
General Relativity, but so far no way has been found to incorporate it into
a quantum field theory (such as the electroweak theory or QCD).  That has not 
prevented physicists from naming its hypothetical carrier the graviton --
it would be massless, neutral, and have spin 2.

One drawback of the standard model is that it has 19 (23 if the neutrinos have
mass) parameters that must be fitted to the data (e.g.\ the 12 quark and lepton
masses).  This doesn't mean that the theory is wrong, but suggests that 
there is probably a more complete theory out there. 
The electroweak theory has been extremely successful; despite extensive 
experimental programs designed to test it, no statistically 
significant deviations from it have been found.

Calculations in the electroweak theory are carried out perturbatively, in
powers of the fine structure constant $\alpha = 1/(137.0359895\pm 0.0000061)$
for a momentum transfer of zero (at higher momentum transfers $\alpha$
increases somewhat; at the $Z$ mass it is approximately 1/128).  Because
$\alpha$ is small, the perturbation theory converges reasonably well.  However,
for QCD, the situation is rather different.  Because the gluons carry a colour
charge, they can interact among themselves, leading to a coupling constant
$\alpha_s$ that {\em decreases} for larger momentum transfer, and increases as
the momentum transfer goes towards zero.  This behaviour at large momentum
transfer is known as asymptotic freedom, and means that perturbative QCD is
valid for high momentum transfers, but breaks down at the lower momentum
transfers where the coupling becomes strong.

Unfortunately, the strong interactions found in the everyday world (the binding
of quarks into nucleons, and of these into atomic nuclei) occur at low energies
(and hence low momentum transfers).  At present, lattice methods are the only
rigorous way to do non-perturbative calculations with QCD.  However, while
great progress has been made using these techniques, it is not yet clear
how long it will be until reliable results can be obtained, especially for
complicated problems such as the calculation of decay widths of excited states.

\section{The Hadrons}

The six quarks and six leptons (plus their antiparticles) may make up matter,
but only three of them make up the everyday matter around us.  All of the
matter in our everyday world is made up of atoms.  An atom consists of a dense
core, the nucleus, surrounded by a cloud of electrons.  The structure of the
electron cloud is responsible for the chemical properties of the atom.  

The nucleus is made up of a number of nucleons (protons (charge +1) and,
usually, neutrons (charge 0)) bound together by the strong interaction.  The
atomic number of the atom is given by the number of protons; a neutral atom
also has the same number of electrons.  The number of neutrons is very
approximately equal to the number of protons.  Both protons and neutrons are
made up of three light quarks: the proton from two $u$'s and a $d$, and the
neutron from a $u$ and two $d$'s.  (Note how the charges of the quarks add to
give the correct nucleon charges.)  Everyday matter is made up of these three
constituents (e$^-$, $u$ and $d$) because being the lightest members of their
rows in (\ref{particles}), they are stable against decay\footnote{Actually, a
$u$ quark can decay into a $d$ via the weak interaction, and vice versa.  However, a $u$ or $d$ always remains, so taken together they are stable.};
neutrinos are also common, but because they only interact via the weak
interaction they interact with everyday matter very rarely, and we are not
aware of them.

Protons and neutrons are examples of baryons, one of the two known types of
structures which quarks can form.  Baryons are made up of three quarks ($qqq$);
an anti-baryon would be made up of three antiquarks.  The other known type of
structure is the meson, which is made up of a quark and an antiquark,
$q\bar{q}$ (so an anti-meson is just a meson).  The complicated structure of
QCD means that groups of quarks can be bound together only for certain
configurations which can have no net colour (so they are in a colour singlet
state).  It also means that the attractive force between coloured objects is
huge, so they are always confined together into colourless objects.  This
property is called confinement.

In a baryon each quark is a different colour, and the three colours mixed
together produce a colourless object.  In a meson the quark has a colour and
the antiquark has the corresponding anti-colour to cancel it, again producing a
colourless object.  It may also be possible to have bound $qqqqqq$ or
$q\bar{q}q\bar{q}$ states, etc., but no observed states have been firmly
identified as such.  Even more interesting would be glueballs (states
consisting only of gluons) or hybrids (states consisting of both quarks and
gluons -- see below).  Once again, no observed states have been firmly
identified with these structures.  These strongly-interacting states built up
of quarks and gluons are collectively called hadrons.

In order to explain glueballs and hybrids, it helps to introduce the flux-tube
picture of QCD.  At higher energies, where perturbative QCD is valid, the
picture of a gluon as a discrete particle exchanged by quarks is appropriate.
However, at lower energies where the coupling becomes stronger there are
indications that a more appropriate picture is that of tubes of chromoelectric
flux connecting the quarks \cite{kogut75:hamiltonian}.  These could be thought
of as being similar to the lines of an electric field connecting electrically
charged objects, but because of the self-interacting nature of the gluons the
lines collapse down to a tube.  This picture incorporates confinement -- if the
flux tubes have a certain energy per unit length, then as the quarks get
further apart the energy of the tube would go up linearly, making it impossible
to separate the quarks.

A meson or baryon is specified by the flavours of the quarks, the space
wavefunction of the quarks, and the wavefunction describing how the quark spins
combine.  Much like a hydrogen atom, the space wavefunction can have angular
and/or radial excitations which are identified as different meson or baryon
states.  It has been theorized (see for example Reference
\cite{isgur85:flux-tube}) that the flux-tube can also be excited, either
vibrationally (like a vibrating string) or topologically (more complicated
topologies of the flux-tube than just a single line).  The resulting particles
would be meson and baryon hybrids.  The glueball can also be thought of in this
picture as a flux-tube connected into a loop, with no quarks present.

The experimental identification of hadrons other than baryons and mesons
would be an important confirmation of our understanding of QCD.

\subsection{The Mesons}

As mentioned above, a meson is specified by its flavour, and its spin and space
wavefunctions.  The flavour wavefunctions that we will deal with in this work
are given in Appendix~\ref{a:flavours}.  The spin wavefunction is particularly
simple: the two spins of $\frac{1}{2}$ can add to give $S=0$ (a singlet) or
$S=1$ (a triplet); for more details see Appendix~\ref{a:spinoverlap}.  Like the
space wavefunction of the hydrogen atom, meson space wavefunctions also have a
radial quantum number $n$ and a quantum number for orbital angular momentum,
$L$ (but unlike the hydrogen atom, $n$ and $L$ are independent).  The total
angular momentum is given by $\vec{J}=\vec{L}+\vec{S}$.  The space and spin
state of a meson is often specified by spectroscopic notation, $n^{2S+1}L_J$,
where the $2S+1$ is 1 or 3 for the singlet or triplet, respectively, and L is
given by a letter (S, P, D, F, G, H,... being 0, 1, 2, 3, 4, 5,...).  If $n=1$
it will often be neglected in the notation.  It is also useful to give the
parity and charge conjugation\footnote{Parity, charge conjugation, and other
conservation laws and invariance principles are discussed briefly in
Appendix~\ref{a:conslaws}.} (where applicable) quantum numbers of the state --
these are usually expressed with $J$ as $J^{\cal{PC}}$.

In Table~\ref{t:mesonspectrum} we list the known mesons constructed of 
the light $u$, $d$ and $s$ quarks (which are all that we will be
concerned with in this thesis) according to the Particle Data Group
\cite{montanet94:review}.  The rows have different space and spin
states and the columns represent different flavours.  For the $I=1$ and
$I=\frac{1}{2}$ column headings the quark states listed are considered to be
different charge states of the same meson (and the antiparticles), because of
the isospin symmetry between the $u$ and $d$ quarks.  For the $I=0$ column the
two mesons in each row are listed together because the proportion of
$\frac{1}{\sqrt{2}}(u\bar{u}+d\bar{d})$ and $s\bar{s}$ states in each is not
the same for all of the rows.  The ${\cal C}$ assignment in $J^{\cal PC}$ only
refers to the neutral members of the row.  The names of the mesons are related
to their flavours, the optional subscript after the name gives $J$, and the
optional number in parentheses after the name is the mass of the meson in MeV.
\begin{table}[t]
\newcommand{\BA}{@{\vline}}
\newcommand{\SP}{\hspace*{0.15cm}}
\vspace{-0.3cm}
\begin{center}
\begin{tabular}{|c\BA c\BA c\BA r@{,}@{\SP} l\BA c|} \hline
$n^{2S+1}L_J$\SP & \SP$J^{\cal{PC}}$\SP & $I=1$ & \multicolumn{2}{c|}{$I=0$} & 
$I=\frac{1}{2}$ \\
& & \SP(-$u\bar{d}$, $\frac{1}{\sqrt{2}}(u\bar{u}$-$d\bar{d})$, $d\bar{u})$\SP 
& \multicolumn{2}{c|}{$(\frac{1}{\sqrt{2}}(u\bar{u}+d\bar{d}), s\bar{s})$} &
(-$u\bar{s}$, -$d\bar{s}$, -$s\bar{d}$, $s\bar{u})$ \\ 
\hline\hline
$1^1S_0$ & $0^{-+}$ & $\pi$ & $\eta$ & $\eta'$ & $K$ \\ 
$1^3S_1$ & $1^{--}$ & $\rho$ & $\omega$ & $\phi$ & $K^*(892)$ \\ 
$1^1P_1$ & $1^{+-}$ & $b_1(1235)$ & $h_1(1170)$ & $h_1(1380)^\heartsuit$\SP & 
\SP$K_1(1270),K_1(1400)^\dagger$ \\ 
$1^3P_0$ & $0^{++}$ & $a_0(980)$ & $f_0(1300)$ & $f_0(980)$ & $K_0^*(1430)$ \\ 
$1^3P_1$ & $1^{++}$ & $a_1(1260)$ & $f_1(1285)$ & $f_1(1510)$ & 
$K_1(1270),K_1(1400)^\dagger$ \\ 
$1^3P_2$ & $2^{++}$ & $a_2(1320)$ & $f_2(1270)$ & $f_2'(1525)$ & 
$K_2^*(1430)$ \\ 
$1^1D_2$ & $2^{-+}$ & $\pi_2(1670)$ & & & $K_2(1770)$ \\ 
$1^3D_1$ & $1^{--}$ & $\rho(1700)$ & $\omega(1600)$ & & $K^*(1680)^\ddagger$\\ 
$1^3D_2$ & $2^{--}$ & & & & $K_2(1820)$ \\ 
$1^3D_3$ & $3^{--}$ & $\rho_3(1690)$ & $\omega_3(1670)$ & $\phi_3(1850)$ & 
$K_3^*(1780)$ \\ 
$1^3F_4$ & $4^{++}$ & $a_4(2040)^\heartsuit$ & $f_4(2050)$ & 
$f_4(2220)^\heartsuit$ & $K_4^*(2045)$ \\ 
$2^1S_0$ & $0^{-+}$ & $\pi(1300)$ & $\eta(1295)$ & & $K(1460)^\heartsuit$ \\ 
$2^3S_1$ & $1^{--}$ & $\rho(1450)$ & $\omega(1420)$ & $\phi(1680)$ & 
$K^*(1410)^\ddagger$ \\ 
$2^3P_2$ & $2^{++}$ & & \SP$f_2(1810)^\heartsuit$ & $f_2(2010)$ & 
$K_2^*(1980)^\heartsuit$\\ 
$3^1S_0$ & $0^{-+}$ & $\pi(1770)^\heartsuit$ & $\eta(1760)^\heartsuit$ & & 
$K(1830)^\heartsuit$ \\ \hline
\end{tabular}
\end{center}
\vspace{-0.4cm}
\caption[The known meson spectrum for the light quarks]{The meson spectrum for
the light quarks.  This table is mostly taken from the Review of Particle
Properties \cite{montanet94:review} by the Particle Data Group. \\
$^\heartsuit$ These states are not included in the Meson Summary Table of the
Review of Particle Properties. \\ 
$^\dagger$ The Particle Data Group believes that the $1^1P_1$ and $1^3P_1$ 
strange mesons are nearly equal ($45^\circ$) mixtures of the $K_1(1270)$ and 
$K_1(1400)$.  See Section~\ref{3:k1} and Appendix~\ref{a:mixings}. \\ 
$^\ddagger$ The Particle Data Group notes that the $K^*(1410)$ could be
replaced by the $K^*(1680)$ as the $2^3S_1$ strange meson.}
\label{t:mesonspectrum}
\end{table}

\subsection{The Quark Model}
\label{1:quarkmodel}

Because perturbative QCD does not work at low energies, and non-perturbative
calculations have yet to produce detailed results, we try to calculate the
properties of hadrons using models inspired by QCD rather than the full theory
itself.  The quark model\footnote{In this thesis, when we refer to the quark
model, we mean the constituent quark model in particular.} is one such attempt,
and a very successful one.

In a quark model of a meson, the wavefunction describing the relative motion of
the quark and antiquark is obtained by solving the Schr\"{o}dinger equation
with a Hamiltonian inspired by QCD.  As an example of such a model, consider
that of Godfrey and Isgur
\cite{godfrey85:mesons}.  Their effective potential, 
$V_{q\bar{q}} (\vec{p},\vec{r})$, contains the effects of a Lorentz-vector
one-gluon-exchange interaction at short distances and a Lorentz-scalar linear
interaction that models confinement.  $V_{q\bar{q}} (\vec{p},\vec{r})$
was found by equating the scattering amplitude of free quarks, using a
scattering kernel with the desired Dirac structure, with the effects of the
effective potential.  Because of relativistic effects the potential is momentum
dependent in addition to being coordinate dependent.  To first order in
$(v/c)^2$, $V_{q\bar{q}} (\vec{p},\vec{r})$ reduces to the standard
nonrelativistic result:
\begin{equation}
V_{q\bar{q}} (\vec{p},\vec{r}) \to V(\vec{r}) =
H^{\rm conf}_{q\bar{q}}  +H^{\rm cont}_{q\bar{q}} + H^{\rm ten}_{q\bar{q}}
+ H^{\rm SO}_{q\bar{q}}, 
\end{equation}
where
\begin{equation}
H^{\rm conf}_{q\bar{q}} = C + br + {{\alpha_s(r)} \over r}
\vec{F}_q \cdot \vec{F}_{\bar{q}}  
\end{equation}
includes the spin-independent linear confinement term and Coulomb-like
(from one-gluon-exchange) interaction,
\begin{equation}
H^{\rm cont}_{q\bar{q}} = - {{ 8\pi} \over 3} 
{{ \alpha_s(r)} \over{m_{q}m_{\bar{q}} } } \vec{S}_q \cdot \vec{S}_{\bar{q}}
\;\delta^3 (\vec{r}) \; \vec{F}_q \cdot \vec{F}_{\bar{q}} 
\end{equation}
is the contact part of the colour-hyperfine interaction,
\begin{equation}
H^{\rm ten}_{q\bar{q}} = -{ {\alpha_s(r)}\over {m_q m_{\bar{q}} } } 
{1\over{r^3}} \left[{ 
{ { 3\vec{S}_q\cdot \vec{r} \; \vec{S}_{\bar{q}} \cdot \vec{r} } \over {r^2} } 
- \vec{S}_q \cdot \vec{S}_{\bar{q}} 
}\right] \; 
\vec{F}_q \cdot \vec{F}_{\bar{q}}
\end{equation}
is the tensor part of the colour-hyperfine interaction, and 
\begin{equation}
H^{\rm SO}_{q\bar{q}} = H^{\rm SO(CM)}_{q\bar{q}} + 
H^{\rm SO(TP)}_{q\bar{q}} \label{spinorbit}
\end{equation}
is the spin-orbit interaction with
\begin{equation}
H^{\rm SO(CM)}_{ q\bar{q}} =-{{\alpha_s(r)}\over {r^3}}
\left({ 
{ {\vec{S}_q} \over {m_q m_{\bar{q}} }} 
+ { {\vec{S}_{\bar{q}} }\over {m_q m_{\bar{q}}} } 
+ { {\vec{S}_q} \over {m_q^2 } } 
+ { {\vec{S}_{\bar{q}} }\over {m_{\bar{q}}^2 } } 
}\right) 
\cdot \vec {L} \; \vec{F}_q \cdot \vec{F}_{\bar{q}} 
\end{equation}
its colour magnetic piece arising from one-gluon-exchange and
\begin{equation}
H^{\rm SO(TP)}_{q\bar{q}} =- {1\over{2r}}
{ { \partial H_{q\bar{q}}^{conf} } \over{\partial r} }
\left({ { {\vec{S}_q} \over {m_q^2 } } 
+ { {\vec{S}_{\bar{q}} }\over {m_{\bar{q}}^2 } } }\right) \cdot \vec {L}
\end{equation}
the Thomas precession term.  In these formulae, $m_q$ ($m_{\bar{q}}$) and
$\vec{S}_q$ ($\vec{S}_{\bar{q}}$) are the mass and spin of the quark
(antiquark), $L$ is the orbital angular momentum between the quark and
antiquark, $\alpha_s(r)$ is the running coupling constant of QCD, and the $\vec{F}$'s
are related to the Gell-Mann $\lambda$-matrices and have expectation value
$\langle \vec{F}_q \cdot \vec{F}_{\bar{q}} \rangle = -4/3$ for a meson.

Godfrey and Isgur's model is actually a {\em relativized} quark model, as
opposed to a nonrelativistic quark model.  They include relativistic effects by
replacing the kinetic energy term
$m_q+m_{\bar{q}}+\frac{p^2}{2m_q}+\frac{p^2}{2m_{\bar{q}}}$ in the Hamiltonian
by the relativistic term $\sqrt{p^2+m_q^2}+\sqrt{p^2+m_{\bar{q}}^2}$ ($p$ is
the relative momentum\footnote{See Appendix~\ref{a:relcoord} for the
definitions of relative and CM coordinates.} in the CM frame), by keeping $m/E$
factors, and by applying relativistic smearing to the potentials.  They also
include the effects of annihilation interactions via gluons for the isoscalar
mesons.

The quark model has been highly successful at mapping the meson spectrum.
Associated models of meson decay have also been quite successful at predicting
the decays \cite{godfrey85:mesons,kokoski87:meson}.

\section{The Work of this Thesis}

We mentioned above that perturbative QCD does not work at low energies, and
that non-perturbative calculations have yet to produce detailed results.  (In
fact, the problem arises with any strongly-interacting field theory.)  Until we
can make quantitative predictions of hadronic properties using QCD, we cannot
say that we understand the theory, nor can we make significant headway in this
important regime.

In order to provide some quantitative predictions for the low-energy regime, an
industry has sprung up that calculates hadron properties by using models
inspired by QCD, rather than the full theory itself.  There are three purposes
of such a program.  The first is to shed light on the dynamics of the model,
and hence (hopefully) on QCD, by determining which facets of the model are
important in predicting results that agree with experiment.  The second is to
understand the properties of the hadrons themselves.  The third is to use these
calculated properties to map the hadron spectrum.  If we understand the meson
and baryon spectra well, and there are extra states that do not fit into them,
then these states must be something new (hybrids, glueballs, etc.)!
Confirmation of these extra states would then feed back to the first purpose,
understanding the dynamics of QCD.

This thesis deals with properties of mesons
calculated in the quark model, which is one of these attempts to approximate
QCD.  
The meson spectrum, with its large number of states to be studied, provides an
excellent testing ground for our understanding of QCD.  We examine two general
areas in the quark model: models of meson decay, and final state interactions.

In the first area the properties we calculate are the decay widths\footnote{See
Appendix~\ref{a:crosssecwid} for an introduction to decay widths, resonances
and cross-sections.} for the strong decays of mesons.  Because meson resonances
are studied experimentally by examining the particles to which they decay,
decay widths are vital in their identification.  As an example of this we study
the identity of the $f_4(2220)$ state -- although tentatively identified as the
$^3F_4$ $s\bar{s}$ meson by the Particle Data Group (see
Table~\ref{t:mesonspectrum}), its identity has been uncertain since its
discovery in 1983.  We also investigate the mixing\footnote{See
Appendix~\ref{a:mixings} for in introduction to mixing.} between the
$K_1(1270)$ and $K_1(1400)$ mesons, in order to see if the dynamics of the
model are responsible for the mixing, or whether it occurs via a separate
mechanism that has not yet been included.  This is done by comparing the
models' predictions for the decay widths of these mesons to experimental data.
These works have already been published
\cite{blundell96:xi,blundell96:properties}.

In the second area we investigate the effects of the final state
interactions in the reaction $\gamma \gamma \to \pi \pi$.  This is again an
attempt to understand the quark model dynamics responsible for the
interaction.  In addition, if we understand the situation in $\gamma \gamma
\rightarrow$ two pseudoscalar\footnote{A pseudoscalar particle has 
$J^{\cal P}=0^-$, as opposed to a scalar which has $J^{\cal P}=0^+$.}  mesons,
we can then approach the situation in $\gamma
\gamma \rightarrow$ two vector mesons with more confidence; the structures 
seen in the cross-sections of these processes are not well understood and are a
long-standing puzzle.

In Chapter~\ref{2:models} we describe in detail the models of meson decay we
are interested in, and the means by which we applied them.  In
Chapter~\ref{3:mesonapp} we fit the free parameter of the models, and
investigate a number of choices that must be made in their application.  We
then apply the models to the first two chosen problems.  In
Chapter~\ref{4:finalstate} we describe in detail the techniques we use for
calculating the effects of final state interactions.  In Chapter~\ref{5:ggpp}
we apply the techniques to our third chosen problem.  In
Chapter~\ref{6:conclusions} we conclude.

\chapter{Two Models of Meson Decay}
\markright{Chapter 2.  Two Models of Meson Decay}
\label{2:models}

\section{The $^3P_0$ Model of Meson Decay}
\label{2:3P0}

The $^3P_0$ model \cite{leyaouanc73:naive,roberts92:general}, also known as the
Quark-Pair Creation (QPC) model, is applicable to OZI-allowed\footnote{See
Appendix~\ref{a:ozirule} for an explanation of the OZI rule.} strong decays of
a meson into two other mesons, which are expected to be the dominant decay
modes of a meson if they are allowed.  It can also be used to describe two-body
OZI-allowed strong decays of other hadrons.

\begin{figure}[t]
\vspace{-0.3cm}
\begin{center}
\makebox{\epsfxsize=6.0in\epsffile{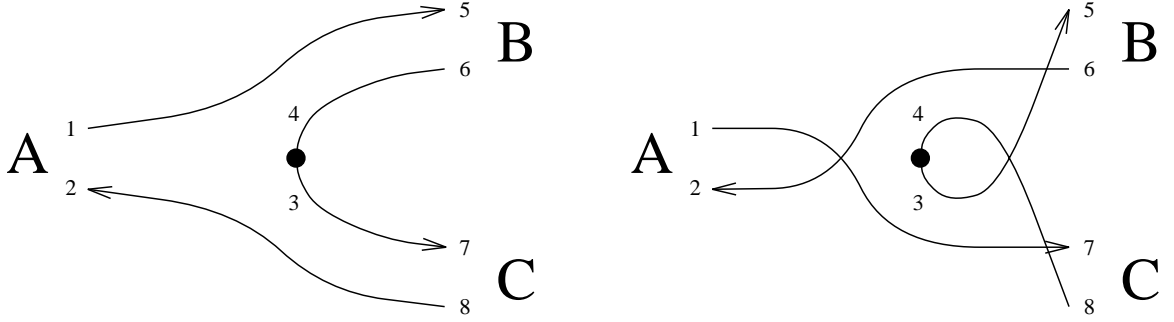}}
\end{center}
\vspace{-0.5cm}
\caption[The two possible diagrams contributing to the meson decay $A
\rightarrow B C$ in the $^3P_0$ model]{The two possible diagrams contributing 
to the meson decay $A
\rightarrow B C$ in the $^3P_0$ model.  In many cases only one of these
diagrams will contribute.}
\label{f:twodiag}
\end{figure}

In this model, meson decay occurs when a quark-antiquark pair
(labelled 3 and 4) is produced from the vacuum in a state suitable for
quark rearrangement to occur, as in Figure~\ref{f:twodiag}.
The created pair will have the quantum numbers of the vacuum, $J^{\cal
PC}=0^{++}$, suggesting that they are in a $^3P_0$ state.  There is one
undetermined parameter $\gamma$ in the model -- it represents the probability
that a quark-antiquark pair will be created from the vacuum.  The pair is
assumed to be created in an SU(N) flavour singlet, where the number of
flavours N is arbitrary, since changing it will just rescale the value of
$\gamma$ needed (see Appendix~\ref{a:flavouroverlap}).  The rest of the model
is just the description of the overlap of the initial meson ($A$) and the
created pair (sometimes referred to by 0) with the two final mesons ($B$,$C$),
to calculate the probability that rearrangement (and hence decay) will occur.

For the meson wavefunction we use a mock meson defined by \cite{hayne82:beyond}
\begin{eqnarray}
|A(n_A \mbox{}^{2S_A+1}L_A \,\mbox{}_{J_A M_{J_A}}) (\vec{P}_A)
\rangle &\equiv& \sqrt{2 E_A}\: \sum_{M_{L_A},M_{S_A}}\! \langle L_A M_{L_A}
S_A M_{S_A} | J_A M_{J_A} \rangle \nonumber\\
&&\times \int\! {\rm d}^3\vec{p}_A\; \psi_{n_A L_A M_{L_A}}\!(\vec{p}_A)\;
\chi^{1 2}_{S_A M_{S_A}}\, \phi^{1 2}_A\;\, \omega^{1 2}_A \nonumber\\
&&\times |q_1({\scriptstyle \frac{m_1}{m_1+m_2}}\vec{P}_A+\vec{p}_A)\:
\bar{q}_2({\scriptstyle \frac{m_2}{m_1+m_2}}\vec{P}_A-\vec{p}_A)\rangle. 
\label{mockmeson}
\end{eqnarray}
Note that given our field-theory conventions of
Appendix~\ref{a:fieldtheorycon}, the mock meson is normalized relativistically
to
\begin{equation}
\langle A(n_A \mbox{}^{2S_A+1}L_A \,\mbox{}_{J_A M_{J_A}}) (\vec{P})\:|\:
A(n_A \mbox{}^{2S_A+1}L_A \,\mbox{}_{J_A M_{J_A}}) (\vec{P'})\rangle 
= 2E_A \:\delta^3(\vec{P}-\vec{P'}), 
\label{mockmesonnorm}
\end{equation}
but uses nonrelativistic spinors and CM coordinates.  The meson has CM momentum
$\vec{P}_A$, and the relative momentum of the q\={q} pair $\vec{p}_A$ is
integrated over all values.  The other quantities are as follows: $n_A$ is the
radial quantum number; $|L_A,M_{L_A}\rangle$, $|S_A,M_{S_A}\rangle$ and
$|J_A,M_{J_A}\rangle$ are the quantum numbers of the orbital angular momentum
between the two quarks, the total spin of the two quarks, and the total angular
momentum of the meson, respectively; $\langle L_A M_{L_A} S_A M_{S_A} | J_A
M_{J_A} \rangle$ is a Clebsch-Gordan coefficient\footnote{See
Appendix~\ref{a:cgcoeffsnj} for an explanation of Clebsch-Gordan coefficients
and Wigner $nj$ symbols.}; $E_A$ is the total energy of the meson; $\chi^{1
2}_{S_A M_{S_A}}$, $\phi^{1 2}_A$ and $\omega^{1 2}_A$ are the appropriate
factors for combining the quark spins to obtain $|S_A,M_{S_A}\rangle$, the
flavours to obtain the correct meson flavour, and the colours to obtain a
colour singlet, respectively; $\psi_{n_A L_A M_{L_A}}\!(\vec{p}_A)$ is the
relative wavefunction of the quarks in momentum-space; $m_1$ and $m_2$ are the
masses of the quark and antiquark respectively; and finally, $|q_1(\vec{p_1})
\: \bar{q}_2(\vec{p_2})\rangle$ is the basic state of a free quark and
antiquark.

Now we can consider the decay $A \rightarrow B C$.  Define the S matrix
\begin{equation}
S \equiv I -2\pi i \:\delta(E_f-E_i)\:T
\end{equation}
and then
\begin{equation}
\langle f|T|i\rangle \equiv \delta^3(\vec{P}_f-\vec{P}_i)\: M^{M_{J_A} M_{J_B}
M_{J_C}} 
\label{texpectation}
\end{equation}
which gives, using relativistic phase space\footnote{Note that we have used
relativistic phase space (leading to Eq.~\ref{width}) and a relativistic
normalization (Eq.~\ref{mockmesonnorm}).  We will also consider other choices
of phase space/normalization in Section~\ref{3:parameters}.} and a meson
wavefunction normalized as in Eq.~\ref{mockmesonnorm}, the decay width in the
CM frame
\begin{equation}
\Gamma = \pi^2 \frac{P}{M_A^2}\frac{{\cal S}}{(2J_A+1)} \sum_{M_{J_A},M_{J_B},
M_{J_C}} |M^{M_{J_A} M_{J_B} M_{J_C}}|^2.  \label{width}
\end{equation}
Here $P$ is the magnitude of the momentum of either outgoing meson, $M_A$ is
the mass of meson $A$, ${\cal S} \equiv 1/(1+\delta_{BC})$ is a statistical
factor that is needed if $B$ and $C$ are identical particles, and 
$M^{M_{J_A} M_{J_B} M_{J_C}}$ is the decay amplitude.

For the transition operator we use
\begin{equation}
T = - 3 \gamma \sum_m\: \langle 1m\,1\!-\!m|00
\rangle \int\!{\rm d}^3\vec{p}_3\; {\rm d}^3\vec{p}_4\:
\delta^3(\vec{p}_3+\vec{p}_4)\:
{\cal Y}^m_1({\scriptstyle \frac{\vec{p}_3-\vec{p}_4}{2}})\;
\chi^{3 4}_{1 -\!m}\; \phi^{3 4}_0\;\, \omega^{3 4}_0\;
b^\dagger_3(p_3)\; d^\dagger_4(p_4). 
\label{tmatrix}
\end{equation}
Here the momenta of both the created quark (3) and antiquark (4) are integrated
over all values, with the constraint that their total momentum is zero.  Other
quantities are as follows: $\gamma$ is the one undetermined parameter in the
model\footnote{Our value of $\gamma$ is higher than that used by Kokoski and
Isgur \cite{kokoski87:meson} by a factor of $\sqrt{96 \pi}$ due to different
field theory conventions, constant factors in $T$, etc.  The calculated values
of the widths are, of course, unaffected.}; $b^\dagger_3(p_3)$ and
$d^\dagger_4(p_4)$ are the creation operators of the created quark and
antiquark, respectively; ${\cal Y}^m_l(\vec{p}) \equiv p^l\,
Y^m_l(\theta_p,\phi_p)$ is a solid harmonic that gives the momentum-space
distribution of the created pair\footnote{The presence of the solid harmonic
can be demonstrated by examining the matrix element for the creation of a
quark-antiquark pair from the vacuum with the scalar operator constructed from
the field: $\langle q(p_q) \: \bar{q}(p_{\bar{q}}) \:|\: \Psi(x')
\bar{\Psi}(x'') \:|\: 0 \rangle$.  The solid harmonic and the Clebsch-Gordan
coefficient can both be extracted from the spinor portion of the result.}; and
the other quantities are as in Eq.~\ref{mockmeson}.  Here the spins and
relative orbital angular momentum of the created quark and antiquark are
combined in a $^3P_0$ state to give the pair the overall $J^{\cal PC}=0^{++}$
quantum numbers.

Combining Eqs.~\ref{mockmeson}, \ref{texpectation} and \ref{tmatrix} gives for 
the amplitude in the CM frame (after doing the colour wavefunction 
overlap, and transforming one of the spin wavefunction overlaps -- 
see Appendix~\ref{a:overlaps}):
\begin{eqnarray}
M^{M_{J_A} M_{J_B} M_{J_C}}(\vec{P}) \;&=&\; \gamma\;\sqrt{8 E_A E_B E_C}
\sum_{\renewcommand{\arraystretch}{.5}\begin{array}[t]{l}
\scriptstyle M_{L_A},M_{S_A},M_{L_B},M_{S_B},\\
\scriptstyle M_{L_C},M_{S_C},m
\end{array}}\renewcommand{\arraystretch}{1}\!\!
\langle L_A M_{L_A} S_A M_{S_A} | J_A M_{J_A} \rangle \nonumber\\
&&\times \langle L_B M_{L_B} S_B M_{S_B} | J_B M_{J_B} \rangle \:
\langle L_C M_{L_C} S_C M_{S_C} | J_C M_{J_C} \rangle \nonumber\\
&&\times \langle 1m\,1\!-\!\!m|00 \rangle \:
\langle \chi^{1 4}_{S_B M_{S_B}} \chi^{3 2}_{S_C M_{S_C}} |
\chi^{1 2}_{S_A M_{S_A}} \chi^{3 4}_{1 -\!m} \rangle \nonumber \\
&&\times \left[ \langle \phi^{1 4}_B \phi^{3 2}_C | \phi^{1 2}_A 
\phi^{3 4}_0 \rangle\:
I(\vec{P},m_1,m_2,m_3) \right. \label{amplitude}\\
&&\left. +(-1)^{1+S_A+S_B+S_C} \:
\langle \phi^{3 2}_B \phi^{1 4}_C | \phi^{1 2}_A \phi^{3 4}_0 \rangle\:
I(-\vec{P},m_2,m_1,m_3) \right].  \nonumber
\end{eqnarray}
The two terms in the last factor correspond to the two possible diagrams in
Figure~\ref{f:twodiag} -- in the first diagram the quark in $A$ ends up $B$; in
the second it ends up in $C$.  The superscripts within the spin and flavour
wavefunction overlaps serve to identify the (anti)quarks on the left side with
the corresponding (anti)quarks on the right side.  The momentum space integral
$I(\vec{P},m_1,m_2,m_3)$ is given by
\begin{eqnarray}
I(\vec{P},m_1,m_2,m_3) &=& \int\!{\rm d}^3\vec{p}\;
\psi^*_{n_B L_B M_{L_B}}\!
({\scriptstyle \frac{m_3}{m_1+m_3}}\vec{P}+\vec{p})\;
\psi^*_{n_C L_C M_{L_C}}\!
({\scriptstyle \frac{m_3}{m_2+m_3}}\vec{P}+\vec{p}) \nonumber \\
&&\times \psi_{n_A L_A M_{L_A}}\!
(\vec{P}+\vec{p})\;
{\cal Y}^m_1(\vec{p}) \label{3P0integral}
\end{eqnarray}
where we have taken $\vec{P} \equiv \vec{P_B} = - \vec{P_C}$.

\section{Calculating Meson Decay Widths with the $^3P_0$ Model}
\label{2:3P0calc}

In order to calculate the decay amplitudes of the $^3P_0$ model, we
use the techniques of Roberts and Silvestre-Brac
\cite{roberts92:general}.  These techniques require that the radial
portions of the meson space wavefunctions be expressible in certain functional
forms, which encompass simple harmonic oscillator (SHO) wavefunctions.  In what
follows, we assume that the radial portions of the wavefunctions are expressed
as linear combinations of the first $N+1$ SHO radial wavefunctions (see
Appendix~\ref{a:space}),
\begin{equation}
R_{L_A}(p_A) = \sum_{a=0}^N \,d^A_a \,R_{n_A=a,L_A}^{\rm SHO}(p_A).
\label{shoexpansion}
\end{equation}
The $\beta_A$, etc. found below are the oscillator parameters of the 
wavefunctions.

The decay amplitudes of Eq.~\ref{amplitude} are converted to partial wave
amplitudes by means of a recoupling calculation (see 
Appendix~\ref{a:partialwaves}).  Then the whole expression for the
amplitudes, including the integrals of Eqs.~\ref{3P0integral} and
\ref{recoupling}, is converted into a sum over angular momentum quantum
numbers.  The graphical methods for manipulating angular momentum expressions
found in Zare \cite{zare88:angular} were particularly useful in this regard.
Our result, very similar to that of Roberts and Silvestre-Brac, is
\begin{eqnarray}
\lefteqn{M^{S L}(P) = 
\gamma\:\sqrt{\frac{8 E_A E_B E_C}{3}} \:(-i)^{L_A+L_B+L_C} \:
(-1)^{1+S_B+S+J_B+J_C} } \nonumber \\
&& \times \sum_{S'} (-1)^{S'} \left[\begin{array}{ccc}
\frac{1}{2} & \frac{1}{2} & S_A \\
\frac{1}{2} & \frac{1}{2} & 1 \\
S_B & S_C & S'
\end{array} \right]
\sum_{L'} \left[\begin{array}{ccc}
S_B & L_B & J_B \\
S_C & L_C & J_C \\
S' & L' & S
\end{array} \right]
\sum_{L''} \widehat{L''} \left\{\begin{array}{ccc}
S_A & L_A & J_A \\
L'' & S' & 1
\end{array} \right\} \nonumber \\
&& \times \left\{\begin{array}{ccc}
S' & L' & S \\
L & J_A & L''
\end{array} \right\}
\left[\langle \phi^{1 4}_B \phi^{3 2}_C | \phi^{1 2}_A 
\phi^{3 4}_0 \rangle\:
\varepsilon(L_A,L_B,L_C,L,L',L'',P,m_1,m_2,m_3) \right. + \nonumber \\
&& \left. (-1)^{1+S_A+S_B+S_C+L} \:
\langle \phi^{3 2}_B \phi^{1 4}_C | \phi^{1 2}_A \phi^{3 4}_0 \rangle\:
\varepsilon(L_A,L_B,L_C,L,L',L'',P,m_2,m_1,m_3) \right] \nonumber \\
&&
\end{eqnarray}
where the modified $9j$ symbol is defined in terms of the $9j$ symbol 
\begin{equation}
\left[\begin{array}{ccc}
j_1 & j_2 & J_{12} \\
j_3 & j_4 & J_{34} \\
J_{13} & J_{24} & J
\end{array} \right] \equiv 
\widehat{J_{12}} \,\widehat{J_{34}} \,\widehat{J_{13}} \,\widehat{J_{24}}
\,\widehat{J} 
\left\{\begin{array}{ccc}
j_1 & j_2 & J_{12} \\
j_3 & j_4 & J_{34} \\
J_{13} & J_{24} & J
\end{array} \right\}, 
\end{equation}
and where $\widehat{J} \equiv \sqrt{2J+1}$,
\begin{eqnarray}
\lefteqn{\varepsilon(L_A,L_B,L_C,L,L',L'',P,m_1,m_2,m_3) \equiv
\frac{1}{2 \, \beta_A^{L_A+\frac{3}{2}} \,\beta_B^{L_B+\frac{3}{2}}\, 
\beta_C^{L_C+\frac{3}{2}}}\: 
\frac{\dspexp{-F^2P^2}}{G^{L_A+L_B+L_C+4}}} \nonumber \\
&& \times \!\sum_{l_1,l_2,l_3,l_4}\! C_{l_1}^{L_B} \,C_{l_2}^{L_C} \,
C_{l_3}^1 \,
C_{l_4}^{L_A} \,\left(x-{\scriptstyle \frac{m_1}{m_1+m_3}}\right)^{l_1} 
\left(x-{\scriptstyle \frac{m_2}{m_2+m_3}}\right)^{l_2}
(x-1)^{l_3} x^{l_4} \nonumber \\
&& \times \!\sum_{l_{12},l_5,l_6,l_7,l_8} \!(-1)^{l_{12}+l_6} \,\widehat{l_5} 
\left[\begin{array}{ccc}
l_1 & l_1' & L_B \\
l_2 & l_2' & L_C \\
l_{12} & l_6 & L'
\end{array} \right] 
\left[\begin{array}{ccc}
l_3 & l_3' & 1 \\
l_4 & l_4' & L_A \\
l_7 & l_8 & L''
\end{array} \right] 
\left\{\begin{array}{ccc}
L & l_{12} & l_5 \\
l_6 & L'' & L'
\end{array} \right\} \nonumber \\
&& \times B_{l_1l_2}^{l_{12}} \,B_{Ll_{12}}^{l_5} \,B_{l_1'l_2'}^{l_6} 
\,B_{l_3l_4}^{l_7} \,B_{l_3'l_4'}^{l_8} \:\sum_{\lambda,\mu,\nu} 
D_{\lambda\mu\nu} \: I_\nu(l_5,l_6,l_7,l_8,L'') \:
\frac{P^{l_1+l_2+l_3+l_4+2\lambda+\nu}}{G^{2\mu+\nu-l_1-l_2-l_3-l_4}}
\nonumber \\
&& \times \left(\frac{l_1'+l_2'+l_3'+l_4'+2\mu+\nu+1}{2}\right)!, \\
\lefteqn{F^2 \equiv \frac{1}{2} \left[ \frac{x^2}{\beta_A^2}+ 
\frac{\left(x-{\scriptstyle \frac{m_1}{m_1+m_3}}\right)^2}{\beta_B^2}+
\frac{\left(x-{\scriptstyle \frac{m_2}{m_2+m_3}}\right)^2}{\beta_C^2} 
\right], } \\
\lefteqn{G^2 \equiv \frac{1}{2} \left[\frac{1}{\beta_A^2}+\frac{1}{\beta_B^2} +
\frac{1}{\beta_C^2} \right], } \\
\lefteqn{x \equiv \frac{\left({\scriptstyle \frac{m_1}{m_1+m_3}}\beta_C^2+
{\scriptstyle \frac{m_2}{m_2+m_3}} \beta_B^2\right) \beta_A^2}
{\beta_A^2\beta_B^2+\beta_A^2\beta_C^2+\beta_B^2\beta_C^2}, } \\
\lefteqn{C_{l_1}^l \equiv \sqrt{\frac{4\pi(2l+1)!}{(2l_1+1)!\left[2(l-l_1)+1
\right]!}}, } \label{cfunc} \\
\lefteqn{B_{l_1l_2}^l \equiv \frac{(-1)^l}{\sqrt{4\pi}}\,\widehat{l_1}\,
\widehat{l_2}
\left(\begin{array}{ccc}
l_1 & l_2 & l \\
0 & 0 & 0
\end{array} \right), }\\
\lefteqn{I_{2p}(l_5,l_6,l_7,l_8,L'')\equiv(-1)^{L''} (2p)! \;\widehat{l_5}\,
\widehat{l_6}\,\widehat{l_7}\,\widehat{l_8} \sum_{\sigma=0}^p \frac{4^\sigma
(4\sigma+1)(p+\sigma)!}{(2p+2\sigma+1)!\,(p-\sigma)!} } \nonumber \\
&& \times \left(\begin{array}{ccc}
2\sigma & l_5 & l_7 \\
0 & 0 & 0
\end{array} \right) 
\left(\begin{array}{ccc}
2\sigma & l_6 & l_8 \\
0 & 0 & 0
\end{array} \right) 
\left\{\begin{array}{ccc}
l_5 & l_6 & L'' \\
l_8 & l_7 & 2\sigma
\end{array} \right\}, \\
\lefteqn{I_{2p+1}(l_5,l_6,l_7,l_8,L'')\equiv2\,(-1)^{L''+1} (2p+1)! \;
\widehat{l_5}\,
\widehat{l_6}\,\widehat{l_7}\,\widehat{l_8} \sum_{\sigma=0}^p \frac{4^\sigma
(4\sigma+3)(p+\sigma+1)!}{(2p+2\sigma+3)!\,(p-\sigma)!} } \nonumber \\
&& \times \left(\begin{array}{ccc}
2\sigma+1 & l_5 & l_7 \\
0 & 0 & 0
\end{array} \right) 
\left(\begin{array}{ccc}
2\sigma+1 & l_6 & l_8 \\
0 & 0 & 0
\end{array} \right) 
\left\{\begin{array}{ccc}
l_5 & l_6 & L'' \\
l_8 & l_7 & 2\sigma+1
\end{array} \right\}, 
\end{eqnarray}
where $l_1'\equiv L_B-l_1$, $l_2'\equiv L_C-l_2$, $l_3'\equiv 1-l_3$, 
$l_4'\equiv L_A-l_4$ and 
$D_{\lambda\mu\nu}$ can be extracted (by equating coefficients of 
$k^{2\lambda} \,q^{2\mu}\,(\vec{k}\cdot \vec{q})^\nu$) from
\begin{eqnarray}
\lefteqn{\sum_{\lambda,\mu,\nu} D_{\lambda\mu\nu} \:k^{2\lambda} \,q^{2\mu} 
\,(\vec{k}\cdot \vec{q})^\nu \equiv \sum_{a,b,c=0}^N (-1)^{a+b+c} \,d_a^A \,d_b^B 
\,d_c^C } \nonumber \\
&&\times \sqrt{8 \,a! \,b! \,c! \,
\Gamma(a+L_A+{\scriptstyle \frac{3}{2}}) \,
\Gamma(b+L_B+{\scriptstyle \frac{3}{2}})\,
\Gamma(c+L_C+{\scriptstyle \frac{3}{2}})} \nonumber \\
&&\times \sum_{m_a=0}^a \sum_{m_b=0}^b \sum_{m_c=0}^c (-1)^{m_a+m_b+m_c}
\frac{1}{\beta_A^{2m_a} \beta_B^{2m_b} \beta_C^{2m_c}}
\frac{1}{(a-m_a)! (b-m_b)! (c-m_c)!} \nonumber \\
&&\times \frac{1}{\Gamma(m_a+L_A+\frac{3}{2}) \Gamma(m_b+L_B+\frac{3}{2})
\Gamma(m_c+L_C+\frac{3}{2})} \nonumber \\
&&\times \sum_{a_1=0}^{m_a} \sum_{a_2=0}^{m_a-a_1} \sum_{b_1=0}^{m_b} 
\sum_{b_2=0}^{m_b-b_1} \sum_{c_1=0}^{m_c} \sum_{c_2=0}^{m_c-c_1} 
\frac{2^{a_3+b_3+c_3}}
{a_1! \,a_2! \,a_3! \,b_1! \,b_2! \,b_3! \,c_1! \,c_2! \,c_3!}\: 
x^{a_3+2a_1} \nonumber  \\
&&\times \left(x-{\scriptstyle \frac{m_1}{m_1+m_3}}\right)^{b_3+2b_1} 
\left(x-{\scriptstyle \frac{m_2}{m_2+m_3}}\right)^{c_3+2c_1}
(k^2)^{a_1+b_1+c_1} \,(q^2)^{a_2+b_2+c_2} \,(\vec{k}\cdot 
\vec{q})^{a_3+b_3+c_3}, \nonumber \\
&&\end{eqnarray}
where $a_3\equiv m_a-a_1-a_2$, $b_3\equiv m_b-b_1-b_2$ and 
$c_3\equiv m_c-c_1-c_2$.

These expressions are general: they can be used for any meson decay where the
radial portion of the wavefunctions can be expanded in terms of SHO radial
wavefunctions.  In order to obtain symbolic results in terms of the various
parameters, we coded these expressions into routines for the symbolic
computation package Mathematica \cite{wolfram91:mathematica}.  In principle,
these routines are limited only by the size of the symbolic problem that
results, and the available computer resources -- in practice, decays involving
mesons with more complicated wavefunctions (i.e.\ having high $N$ in
Eq.~\ref{shoexpansion}, or high $L$) would have to be done numerically.  The
overlaps of the colour, flavour and spin wavefunctions of the mesons and the
created pair are calculated using the techniques of Appendix~\ref{a:overlaps}.

\section{The Flux-tube Breaking Model of Meson Decay}
\label{2:fluxtube}

In the flux-tube picture a meson consists of a quark and antiquark connected by
a tube of chromoelectric flux, which can be treated as a vibrating string.  For
mesons the string is in its vibrational ground state.  (Vibrational excitations
of the string would correspond to a type of meson hybrid, particles whose
existence have not yet been confirmed.)  Meson decay occurs when the flux-tube
breaks at a point, and a quark-antiquark pair is created from the vacuum to
connect to the free ends of the flux-tubes, leaving a final state consisting of
two mesons.

\begin{figure}[t]
\vspace{-0.5cm}
\begin{center}
\makebox{\epsfysize=1.5in\epsffile{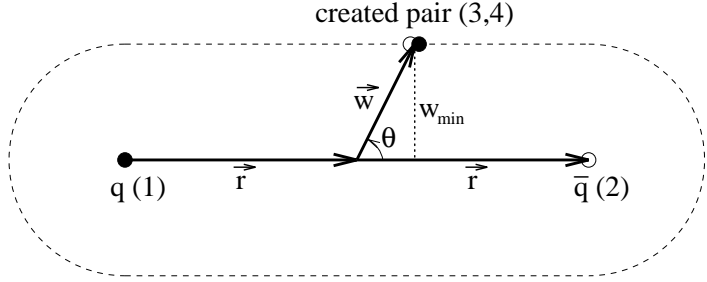}}
\end{center}
\vspace{-0.5cm}
\caption[The position-space coordinates used in the flux-tube model]
{The position-space coordinates used in the flux-tube model.  The 
cigar-shaped dashed line shows a possible surface of constant 
$w_{\rm min}$.}
\label{f:fluxgeom}
\end{figure}

The flux-tube breaking model of meson decay is similar to the
$^3P_0$ model, but extends it by considering the actual dynamics of the
flux-tubes.  This is done by including a factor representing the overlap of 
the flux-tube of the initial meson with those of the two outgoing mesons.  
Kokoski and Isgur \cite{kokoski87:meson} have
calculated this factor by treating the flux-tubes as vibrating strings.  They
approximate the rather complicated result by replacing the undetermined
parameter $\gamma$ in the $^3P_0$ model with a function of the location of the
created quark-antiquark pair, and a new undetermined parameter $\gamma_0$:
\begin{equation}
\gamma(\vec{r},\vec{w}) = \gamma_0 \:\dspexp{-{\scriptstyle\frac{1}{2}}b 
w_{\rm min}^2}.
\end{equation}
Here $b$ is the string tension, where a value of $0.18$~${\rm GeV}^2$ is
typically used, and $w_{\rm min}$ is the shortest distance from the line
segment connecting the original quark and antiquark to the location at which
the new quark-antiquark pair is created from the vacuum (see
Figure~\ref{f:fluxgeom}):
\begin{equation}
w_{\rm min}^2=\left\{ \begin{array}{ll}
w^2 \sin^2{\theta}, & \mbox{if $r \geq w\left|\cos{\theta}\right|$}\\
r^2+w^2-2rw \left|\cos{\theta}\right|, &
\mbox{if $r < w\left|\cos{\theta}\right|$}
\end{array}
\right.. \label{wmindef}
\end{equation}

To incorporate this into the $^3P_0$ model, we first Fourier transform
Eq.~\ref{3P0integral} so that the integral is over position-space.  We then 
pull
the parameter $\gamma$ (from Eq.~\ref{amplitude}) inside the integral, and replace it 
by the function of
position $\gamma(\vec{r},\vec{w})$.  The expression for the amplitude in the
flux-tube model is then the same as that of Eq.~\ref{amplitude} except that
$\gamma$ is replaced by $\gamma_0$, and $I(\vec{P},m_1,m_2,m_3)$ is replaced by
\begin{eqnarray}
\lefteqn{I^{\rm ft}(\vec{P},m_1,m_2,m_3)=-\frac{8}{(2\pi)^{\frac{3}{2}}}
\int\!{\rm d}^3\vec{r}\int\!{\rm d}^3\vec{w}\;\:
\psi^*_{n_B L_B M_{L_B}}\!
(-\vec{w}-\vec{r})\;
\psi^*_{n_C L_C M_{L_C}}\!
(\vec{w}-\vec{r}) } \nonumber\\
&&\times{\cal Y}^m_1\!\!\left(\left[\left(\vec{P}+i\vec{\nabla}_{\vec{r}_A}
\right)
\psi_{n_A L_A M_{L_A}}\!(\vec{r}_A)\right]_{\vec{r}_A=-2\vec{r}}\right)
\dspexp{-{\scriptstyle\frac{1}{2}}b w_{\rm min}^2}\; 
\dspexp{i\vec{P}\cdot\left(m_+\vec{r} + m_-\vec{w}\right)} 
\;\;\;\;\;\;\;\;\;\;\;
\label{fluxintegral}
\end{eqnarray}
where $m_+ = \frac{m_1}{m_1+m_3}+\frac{m_2}{m_2+m_3}$, $m_- =
\frac{m_1}{m_1+m_3}-\frac{m_2}{m_2+m_3}$ and the $\psi$'s are now the relative
wavefunctions in position-space.  The last exponential function in
Eq.~\ref{fluxintegral} is the combination of the plane waves representing the
CM motion of mesons $B$ and $C$ (the plane wave for $A$ is unity, since it is
at rest in the CM frame).  It comes from Fourier-transforming the
delta-functions implicit in Eq.~\ref{mockmeson} that specify that the CM
momenta of the mesons are held fixed to constant values.  Note also that we
chose to apply the gradient in the argument of the solid harmonic to $\psi_{n_A
L_A M_{L_A}}$ only.  We could have manipulated the Fourier-transform\footnote{A
word of explanation: $\frac{\vec{p}_3-\vec{p}_4}{2}$ in Eq.~\ref{tmatrix} is
rewritten as $\vec{p}_0$, the relative momentum of the created pair.  Depending
on which delta functions from the anticommutators of the annihilation and
creation operators we use to replace $\vec{p}_0$ before doing the
Fourier-transform, we end up with different combinations of $\vec{p}_A$,
$\vec{p}_B$ and $\vec{p}_C$ in the argument of the solid harmonic.  Each
$\vec{p}_i$ becomes a gradient applied to $\psi_{n_i L_i M_{L_i}}$ during
the Fourier-transform.} to apply it to just $\psi^*_{n_B L_B M_{L_B}}$ or
$\psi^*_{n_C L_C M_{L_C}}$, or some combination of all three.  In fact since
$B$ and $C$, being lighter than $A$, tend to be less excited, choosing the
wavefunctions of $B$ or $C$ would in general be simpler.

\section{Calculating Meson Decay Widths with the Flux-tube Breaking Model}
\label{2:fluxcalc}

In order to calculate the decay amplitudes of the flux-tube breaking model,
there are a number of integrals that must be evaluated: two 3-dimensional
integrals from Eq.~\ref{fluxintegral} and, if
we choose to use the recoupling calculation to convert to partial wave
amplitudes (see 
Appendix~\ref{a:partialwaves}), another 2-dimensional integral
from Eq.~\ref{recoupling}.

We wish to develop general routines to calculate any decay amplitude.  This
prevents us from making the common assumptions that can simplify
Eq.~\ref{fluxintegral}, such as neglecting mass differences between the quarks,
restricting the orbital angular momenta of each meson to be 0 or 1 so that
their spherical harmonics can be expressed as dot products of their argument
and a constant vector, or simplifying the definition of $w_{\rm min}$ to remove
the complication that arises for $r < w\left|\cos{\theta}\right|$ (i.e.\ the
definition currently used for $r \geq w\left|\cos{\theta}\right|$ would be used
for all values).  The main goal of simplifying Eq.~\ref{fluxintegral} is to
allow more of the integrals to be done analytically, to reduce the computer
time required for a calculation.

In order to save ourselves two integrations, we choose to use the Jacob-Wick
formula to convert to partial wave amplitudes (see
Appendix~\ref{a:partialwaves}).  (Remember that this involves choosing
$\vec{P}$ to lie along $\hat{z}$, the z axis.)  This leaves us with integrals
over the variables $r$, $\theta_r$, $\phi_r$, $w$, $\theta_w$, and $\phi_w$.
The divided form of Eq.~\ref{wmindef} suggests that we will not be able to do
the integrals over any of the variables on which it depends.  This immediately
rules out $r$ and $w$.  If $\vec{r}$ and $\vec{w}$ are expressed in the same
coordinate system, the angle between them ($\theta$) is given by
\begin{equation}
\cos{\theta} = \sin{\theta_r} \sin{\theta_w} \cos{(\phi_r-\phi_w)}+
\cos{\theta_r} \cos{\theta_w}
\end{equation}
which presents a problem for the other variables as well.  The solution is to
use the freedom to choose the coordinate system of $w$ to set $\hat{z}_w$ to
lie along $\vec{r}$.  We also choose to orient $\hat{x}_w$ and $\hat{y}_w$ such
that the original $\hat{z}=\hat{z}_r$ axis has $\phi_w=0$ in the $w$ coordinate
system.  Then $\theta=\theta_w$, and
\begin{eqnarray}
\vec{r}\cdot\vec{P}&=& rP \cos{\theta_r} \nonumber \\
\vec{w}\cdot\vec{P}&=& wP \left[\sin{\theta_r} \sin{\theta_w} \cos{\phi_w}+
\cos{\theta_r} \cos{\theta_w}\right].
\end{eqnarray}

A wrinkle with this approach is that in Eq.~\ref{fluxintegral} we also have
expressions like $\vec{w}-\vec{r}$, where the vectors must be expressed in the
same coordinate system.  The expression will have to be replaced by
$R^{-1}(\phi_r,\theta_r,\pi)\:\vec{w}-\vec{r}$, where $R(\phi_r,\theta_r,\pi)$
is the active rotation of a vector in the $\vec{r}$ coordinate system to one in
the $\vec{w}$ coordinate system.  To evaluate the angular part of an expression
such as $\psi^*_{n_B L_B M_{L_B}}\!
(-R^{-1}(\phi_r,\theta_r,\pi)\:\vec{w}-\vec{r})$, we first use the following
identity for solid harmonics
\begin{equation}
{\cal Y}^m_l(\alpha_1 \vec{r}_1+\alpha_2 \vec{r}_2) = \sum_{l_1=0}^l 
C_{l_1}^l  \alpha_1^{l_1} \alpha_2^{l-l_1} \!\!\sum_{m_1=-l_1}^{l_1} 
\langle l_1 m_1 (l-l_1) (m-m_1)| l m\rangle \,{\cal Y}^{m_1}_{l_1}(\vec{r}_1)
\,{\cal Y}^{m-m_1}_{l-l_1}(\vec{r}_2),  \label{rot1}
\end{equation}
where $C_{l_1}^l$ is defined in Eq.~\ref{cfunc}.  Then to handle the rotation
we make use of the following result
\begin{equation}
{\cal Y}^{*m}_l(R^{-1}(\phi_r,\theta_r,\pi)\:\vec{w}) = 
[{\cal R}(\phi_r,\theta_r,\pi) {\cal Y}^m_l(\vec{w})]^* =
\sum_{m'=-l}^l  D^l_{mm'}(\phi_r,\theta_r,\pi) 
\,{\cal Y}^{*m'}_l(\vec{w}), \label{rot2}
\end{equation}
where $D^l_{mm'}(\phi_r,\theta_r,\pi)$ is a rotation matrix.

The final wrinkle is that the substitution for $\vec{P}$ required in the
evaluation of $I^{\rm ft}(-\vec{P},m_2,m_1,m_3)$ requires more than just taking
$P\to-P$ when two different coordinate systems are used for $\vec{r}$ and
$\vec{w}$.  The solution is to first use the symmetry properties of the angular
parts of the wavefunctions to obtain
\begin{equation}
I^{\rm ft}(-\vec{P},m_2,m_1,m_3)=(-1)^{1+L_A+L_B+L_C} 
I^{\rm ft}(\vec{P},m_2,m_1,m_3), \label{lsymmetry}
\end{equation}
after which our particular choice for the $\vec{w}$ coordinate system can be 
used to evaluate the dot products.  At this stage, the expression for
the partial wave amplitudes is (before applying Eqs.~\ref{rot1} and \ref{rot2})
\begin{eqnarray}
\lefteqn{M^{S L}(P) = \gamma_0\:\frac{\sqrt{32 \pi (2 L+1)E_A E_B E_C}}
{2 J_A +1} \!\!
\sum_{\renewcommand{\arraystretch}{.5}\begin{array}[t]{l}
\scriptstyle M_{L_A},M_{S_A},M_{L_B},M_{S_B},M_{J_B},\\
\scriptstyle M_{L_C},M_{S_C},M_{J_C},m
\end{array}}\renewcommand{\arraystretch}{1} } \nonumber \\
&& \times \langle L 0 S (M_{J_B}\!+\!M_{J_C})|J_A (M_{J_B}\!+\!M_{J_C})\rangle 
\:
\langle J_B M_{J_B} J_C M_{J_C} | S (M_{J_B}\!+\!M_{J_C}) \rangle \nonumber \\
&& \times \langle L_A M_{L_A} S_A M_{S_A} | J_A  (M_{J_B}\!+\!M_{J_C})\rangle 
\:
\langle L_B M_{L_B} S_B M_{S_B} | J_B M_{J_B} \rangle \nonumber \\
&& \times \langle L_C M_{L_C} S_C M_{S_C} | J_C M_{J_C} \rangle \:
\langle 1m\,1\!-\!\!m|00 \rangle \:
\langle \chi^{1 4}_{S_B M_{S_B}} \chi^{3 2}_{S_C M_{S_C}} |
\chi^{1 2}_{S_A M_{S_A}} \chi^{3 4}_{1 -\!m} \rangle \nonumber \\
&& \times \left[ \langle \phi^{1 4}_B \phi^{3 2}_C | \phi^{1 2}_A 
\phi^{3 4}_0 \rangle\:
I^{\rm ft}(P\hat{z},m_1,m_2,m_3) \right. \nonumber \\
&& +(-1)^{L_A+L_B+L_C+S_A+S_B+S_C} \:
\langle \left. \phi^{3 2}_B \phi^{1 4}_C | \phi^{1 2}_A \phi^{3 4}_0 \rangle\:
I^{\rm ft}(P\hat{z},m_2,m_1,m_3) \right]
\end{eqnarray}
where
\begin{eqnarray}
\lefteqn{I^{\rm ft}(P\hat{z},m_1,m_2,m_3) = -\frac{8}{(2\pi)^{\frac{3}{2}}}
\int_0^\infty {\rm d}r \int_0^\infty {\rm d}w \int_{-1}^1 
{\rm d}(\cos{\theta_w}) \:r^2 \,w^2 \:\dspexp{-{\scriptstyle\frac{1}{2}}
b w_{\rm min}^2} } \nonumber\\
&& \times \int_{-1}^1 {\rm d}(\cos{\theta_r}) \:\dspexp{im_- P w \cos{\theta_r}
\cos{\theta_w} +im_+ Pr\cos{\theta_r} } \nonumber\\
&& \times \int_0^{2\pi} {\rm d}\phi_r \:
{\cal Y}^m_1\!\!\left(\left[\left(P\hat{z}+i\vec{\nabla}_{\vec{r}_A}\right)
\psi_{n_A L_A M_{L_A}}\!(\vec{r}_A)\right]_{\vec{r}_A=-2\vec{r}}\right) 
\nonumber \\
&& \times \int_0^{2\pi} {\rm d}\phi_w \:\dspexp{im_- Pw \sin{\theta_r} 
\sin{\theta_w} \cos{\phi_w} } \;
\psi^*_{n_B L_B M_{L_B}}\!(-R^{-1}(\phi_r,\theta_r,\pi)\:\vec{w}-\vec{r})
\nonumber \\
&& \;\;\;\;\;\;\;\;\;\;\;\;\;\;\,\times \psi^*_{n_C L_C M_{L_C}}\!(R^{-1}
(\phi_r,\theta_r,\pi)\:\vec{w}-\vec{r}).
\end{eqnarray}

We can now carry out the $\theta_r$, $\phi_r$ and $\phi_w$ integrations
analytically, in principle.  In practice, we will only do those for $\phi_r$
and $\phi_w$.  The integral over $\phi_w$ is done by means of the residue
theorem.  The integral over $\phi_r$ depends upon the form of the wavefunction
of meson $A$ -- we will again assume that the radial portions of the meson
wavefunctions are expressed as linear combinations of the first $N+1$ SHO
radial wavefunctions, as in
Eq.~\ref{shoexpansion}.  Once this assumption is made, the $\phi_r$ integral
does not require any special techniques; the final result after doing it is
rather complicated but we give it here for completeness:
\begin{eqnarray}
\lefteqn{M^{S L}(P) = \int_0^\infty {\rm d}r \int_0^\infty {\rm d}w 
\int_0^\pi {\rm d}\theta_r \int_0^\pi {\rm d}\theta_w  \;\,
g_1(r,w,\theta_r,\theta_w) \,\sqrt{2L+1}} \nonumber \\
&&\times \sum_{M_{J_B},M_{J_C}}
\langle L 0 S (M_{J_B}\!+\!M_{J_C})|J_A (M_{J_B}\!+\!M_{J_C})\rangle \:
\langle J_B M_{J_B} J_C M_{J_C} | S (M_{J_B}\!+\!M_{J_C}) \rangle \nonumber \\
&&\times \sum_{M_{L_A},M_{L_B},M_{L_C}} 
\langle L_A M_{L_A} S_A (M_{J_B}\!+\!M_{J_C}\!-\!M_{L_A}) | 
J_A (M_{J_B}\!+\!M_{J_C})\rangle \nonumber \\
&&\times \langle L_B M_{L_B} S_B (M_{J_B}\!-\!M_{L_B}) | J_B M_{J_B} \rangle \:
\langle L_C M_{L_C} S_C (M_{J_C}\!-\!M_{L_C}) | J_C M_{J_C} \rangle \nonumber\\
&&\times \langle 1(M_{L_B}\!+\!M_{L_C}\!-\!M_{L_A})1(M_{L_A}\!-\!M_{L_B}\!-
\!M_{L_C})|00 \rangle \nonumber\\
&&\times \langle \chi^{1 4}_{S_B (M_{J_B}\!-\!M_{L_B})} 
\chi^{3 2}_{S_C (M_{J_C}\!-\!M_{L_C})} |
\chi^{1 2}_{S_A (M_{J_B}\!+\!M_{J_C}\!-\!M_{L_A})} 
\chi^{3 4}_{1 (M_{L_A}\!-\!M_{L_B}\!-\!M_{L_C})} \rangle \nonumber\\
&&\times \sqrt{\frac{(L_A-M_{L_A})!}{(L_A+M_{L_A})!}}\:
\sum_{l_1=0}^{L_B}\,\sum_{l_2=0}^{L_C} \,(-1)^{L_C-l_2}\: C_{l_1}^{L_B}\: 
C_{l_2}^{L_C}\:r^{L_B+L_C-l_1-l_2}\: w^{l_1+l_2} \nonumber\\
&&\times \sum_{m_{l_1}}\sum_{m_{l_2}} 
\,\langle l_1 m_{l_1} (L_B\!-\!l_1)(M_{L_B}\!-\!m_{l_1})|L_B M_{L_B} \rangle \:
\langle l_2 m_{l_2} (L_C\!-\!l_2)(M_{L_C}\!-\!m_{l_2})|L_C M_{L_C} \rangle
\nonumber\\
&&\times \sqrt{[2(L_B-l_1)+1]\,[2(L_C-l_2)+1]\:
\frac{(L_B\!-\!l_1\!-\!M_{L_B}\!+\!m_{l_1})!\,
(L_C\!-\!l_2\!-\!M_{L_C}\!+\!m_{l_2})!}
{(L_B\!-\!l_1\!+\!M_{L_B}\!-\!m_{l_1})!\,
(L_C\!-\!l_2\!+\!M_{L_C}\!-\!m_{l_2})!}} \nonumber\\
&&\times P_{L_B-l_1}^{M_{L_B}-m_{l_1}}(\cos\theta_r)\;
P_{L_C-l_2}^{M_{L_C}-m_{l_2}}(\cos\theta_r) \nonumber\\
&&\times \sum_{m'=-l_1}^{l_1}\sum_{m''=-l_2}^{l_2} 
\sqrt{(2l_1+1)\,(2l_2+1)\,\frac{(l_1-m')! \,(l_2-m'')!}
{(l_1+m')!\,(l_2+m'')!}}\; 
(-1)^{m'+m''}  \nonumber\\
&&\times d^{l_1}_{m_{l_1}m'}(\theta_r)\; d^{l_2}_{m_{l_2}m''}(\theta_r)\;
P_{l_1}^{m'}(\cos\theta_w)\;P_{l_2}^{m''}(\cos\theta_w)\;
\dspexp{im_+ Pr\cos\theta_r} \nonumber\\
&&\times \left[i g_2(M_{L_A},M_{L_B}\!+\!M_{L_C},r,\theta_r) +
g_3(M_{L_A},M_{L_B}\!+\!M_{L_C},r,\theta_r)\right]\nonumber\\
&&\times \left[
\langle \phi^{1 4}_B \phi^{3 2}_C | \phi^{1 2}_A \phi^{3 4}_0 \rangle\:
\dspexp{im_- Pw\cos\theta_r\cos\theta_w}\; 
i^{|m'+m''|}\; {\rm sign}^{|m'+m''|}(m_- Pw\sin\theta_r \sin\theta_w)\right. 
\nonumber\\
&&\times J_{|m'+m''|}(|m_-Pw\sin\theta_r \sin\theta_w|)\nonumber\\
&& + (-1)^{L_A+L_B+L_C+S_A+S_B+S_C}\:
\langle \phi^{3 2}_B \phi^{1 4}_C | \phi^{1 2}_A \phi^{3 4}_0 \rangle\:
\dspexp{-im_- Pw\cos\theta_r\cos\theta_w}\;
i^{|m'+m''|}\nonumber\\
&& \times \left. {\rm sign}^{|m'+m''|}(-m_- Pw\sin\theta_r \sin\theta_w)\;
J_{|m'+m''|}(|m_-Pw\sin\theta_r \sin\theta_w|) \rule[-.3cm]{0cm}{0.6cm}
\right] 
\end{eqnarray}
where $d^l_{mm'}(\theta)$ is related to the rotation matrix by
\begin{equation}
D^l_{mm'}(\phi,\theta,\chi)\equiv
e^{-im\phi}\,d^l_{mm'}(\theta)\,e^{-im'\chi}, 
\end{equation}
$P_l^m(x)$ is an associated
Legendre function, $J_l(x)$ is a Bessel function of the first kind, and
sign($x$) is $+1$ if $x\geq 0$ and $-1$ if $x<0$.  The $g$ functions are given
by
\begin{eqnarray}
\lefteqn{g_1(r,w,\theta_r,\theta_w) \equiv \frac{-\sqrt{3}}{2\pi^2} 
\frac{\sqrt{2L_A+1}}{2J_A+1} \sqrt{E_A E_B E_C} \:(-1)^{L_B}\, 
\beta_A^{L_A+\frac{3}{2}} \,\beta_B^{L_B+\frac{3}{2}} \,
\beta_A^{L_B+\frac{3}{2}}} \nonumber \\
&&\times r^2 \,w^2 \,\sin\theta_r \:\sin\theta_w \:\gamma(r,w,\theta_w)\: 
(-2r)^{L_A-1}\nonumber \\
&&\times \dspexp{-2\beta_A^2r^2}\; 
\dspexp{-\beta_B^2(r^2+w^2+2rw\cos\theta_w)/2}\;
\dspexp{-\beta_C^2(r^2+w^2-2rw\cos\theta_w)/2} \nonumber \\
&&\times \left[\sum_{b=0}^N \,d_b^B\, \sqrt{\frac{2 \:b!}
{\Gamma(b+L_B+{\scriptstyle \frac{3}{2}})}} \;
L_b^{L_B+\frac{1}{2}}\!\!\left(\beta_B^2(r^2+w^2+2rw\cos\theta_w)\right)\right]
\nonumber \\
&&\times \left[\sum_{c=0}^N \,d_c^C\, \sqrt{\frac{2 \:c!}
{\Gamma(c+L_C+{\scriptstyle \frac{3}{2}})}}\;
L_c^{L_C+\frac{1}{2}}\!\!\left(\beta_C^2(r^2+w^2-2rw\cos\theta_w)\right)
\right],
\\
\lefteqn{g_2(M_{L_A},M_{L_B}\!+\!M_{L_C},r,\theta_r) \equiv } \nonumber\\ 
&& \left[\sqrt{2}\sin\theta_r \:(\delta_{M_{L_A}\!-\!M_{L_B}\!-\!M_{L_C},1}-
\delta_{M_{L_A}\!-\!M_{L_B}\!-\!M_{L_C},-1})
+2\cos\theta_r \;\delta_{M_{L_A}\!-\!M_{L_B}\!-\!M_{L_C},0}\right]
\nonumber \\
&& \times P_{L_A}^{M_{L_A}}(\cos\theta_r) \,
\left[ \sum_{a=0}^N \,d_a^A\, \sqrt{\frac{2 \:a!}
{\Gamma(a+L_A+{\scriptstyle \frac{3}{2}})}} \left(
\begin{array}{l} \,\\\, \end{array} \!\!\!\!\!\!
(L_A-4\beta_A^2r^2)\:L_a^{L_A+\frac{1}{2}}\!\!\left(4\beta_A^2r^2\right)
\right. \right. \nonumber \\
&& \left. \left. \hspace*{4cm}-8\beta_A^2r^2  \left\{
\begin{array}{ll} 0, & {\rm if}\;\;a=0 \\
L_{a-1}^{L_A+\frac{3}{2}}\!\!\left(4\beta_A^2r^2\right), & {\rm if}\;\;a\geq 1 
\end{array} \hspace*{1cm} \right. \right) \right] \nonumber \\
&& + \left\{ \rule[-.4cm]{0cm}{0.8cm}\!\! \left[\sqrt{2}\cos\theta_r \:
(\delta_{M_{L_A}\!-\!M_{L_B}\!-\!M_{L_C},1}-
\delta_{M_{L_A}\!-\!M_{L_B}\!-\!M_{L_C},-1})
-2\sin\theta_r \;\delta_{M_{L_A}\!-\!M_{L_B}\!-\!M_{L_C},0}\right] \right.
\nonumber \\
&& \hspace*{1cm}\times \frac{1}{\sin\theta_r} \left[L_A \cos\theta_r 
\:P_{L_A}^{M_{L_A}}(\cos\theta_r) -(L_A+M_{L_A})
\:P_{L_A-1}^{M_{L_A}}(\cos\theta_r) \right] \nonumber \\
&& \left. \hspace*{0.5cm}+ \frac{\sqrt{2}M_{L_A}}{\sin\theta_r}\: 
(\delta_{M_{L_A}\!-\!M_{L_B}\!-\!M_{L_C},1}+
\delta_{M_{L_A}\!-\!M_{L_B}\!-\!M_{L_C},-1}) 
\:P_{L_A}^{M_{L_A}}(\cos\theta_r) 
\right\} \nonumber \\
&&\times \left[\sum_{a=0}^N \,d_a^A\, \sqrt{\frac{2 \:a!}
{\Gamma(a+L_A+{\scriptstyle \frac{3}{2}})}}\;
L_a^{L_A+\frac{1}{2}}\!\!\left(4\beta_A^2r^2\right)\right], \\
\lefteqn{g_3(M_{L_A},M_{L_B}\!+\!M_{L_C},r,\theta_r) \equiv -4Pr\; 
\delta_{M_{L_A}\!-\!M_{L_B}\!-\!M_{L_C},0}\;
P_{L_A}^{M_{L_A}}(\cos\theta_r)} \nonumber \\
&&\times \left[\sum_{a=0}^N \,d_a^A\, \sqrt{\frac{2 \:a!}
{\Gamma(a+L_A+{\scriptstyle \frac{3}{2}})}}\;
L_a^{L_A+\frac{1}{2}}\!\!\left(4\beta_A^2r^2\right)\right],
\end{eqnarray}
where $L_n^k(x)$ is an associated Laguerre polynomial.

We have automated the calculation of partial wave amplitudes with these
expressions, using routines written for Mathematica
\cite{wolfram91:mathematica} and FORTRAN.  In the 
Mathematica code, an integrand for each partial wave amplitude is prepared
symbolically and converted to FORTRAN code.  The FORTRAN program carries out
the integrations numerically using either adaptive Monte Carlo (VEGAS
\cite{lepage78:vegas}) or a combination of adaptive Gaussian quadrature
routines.  Again, these routines are usable for any meson decay where the
radial portion of the wavefunctions can be expanded in terms of SHO
wavefunctions, and are limited only by the size of the problem and the
available computer resources.  The overlaps of the colour, flavour and spin
wavefunctions of the mesons and the created pair are calculated using the
techniques of Appendix~\ref{a:overlaps}.

\chapter{Models of Meson Decay:  Two Applications}
\markright{Chapter 3.  Models of Meson Decay:  Two Applications}
\label{3:mesonapp}

\section{Setting the Parameters of the Models}
\label{3:parameters}

In order to use the models of meson decay introduced in Chapter~\ref{2:models}
we must fit the undetermined parameter $\gamma$ ($\gamma_0$ for the flux-tube
breaking model), and we must decide which meson space wavefunctions to use.  In
addition, we must decide which phase space/normalization is best suited to our
purposes -- we discuss this below.

In this section we consider three types of space wavefunctions, described in
some detail in Appendix~\ref{a:space}: SHO wavefunctions with a common
oscillator parameter ($\beta$), wavefunctions from the relativized quark model
(RQM) \cite{godfrey85:mesons} of Godfrey and Isgur described in
Section~\ref{1:quarkmodel}, and SHO wavefunctions with the effective $\beta$'s
of Kokoski and Isgur \cite{kokoski87:meson}.  The RQM wavefunctions and SHO
wavefunctions with effective $\beta$'s are fully determined and have no
parameters left to choose.  For the SHO wavefunctions with a common $\beta$, we
take $\beta = 400$~MeV, the value used by Kokoski and Isgur which lies safely
within the range of the effective values fitted by them for the various mesons.
We also perform another fit of both $\gamma$ and $\beta$ simultaneously.

We must also decide how to normalize the mock meson wavefunctions and what kind
of phase space to use to calculate the decay widths
\cite{geiger94:distinguishing}.  We investigate three choices.  
In Chapter~\ref{2:models} we normalized the mock meson wavefunctions
relativistically to $2 E$ (Eq.~\ref{mockmesonnorm}) and used relativistic phase
space (leading to Eq.~\ref{width}), which leads to a factor of $E_B E_C/M_A$ in
the final expression for a decay width in the CM frame ($E$ is energy, $M$ is
mass).  We will refer to this as relativistic phase space/normalization (RPSN).
One could also argue that since our decay models are primarily nonrelativistic,
it makes sense to use a nonrelativistic normalization and nonrelativistic phase
space, which would replace the factor by $M_B M_C/M_A$.  We will refer to this
as nonrelativistic phase space/normalization (NRPSN).  However, there are also
arguments \cite{isgur95:personal} that heavy quark effective theory fixes the
assumptions in the mock meson prescription and suggests that the factor be
replaced by $\widetilde{M}_B \widetilde{M}_C/\widetilde{M}_A$, where the
$\widetilde{M}_i$ are the calculated masses of the meson $i$ in a
spin-independent quark-antiquark potential
\cite{kokoski87:meson}.\footnote{In other words $\widetilde{M}_i$ is
given by the hyperfine averaged mass that is equal to the centre of gravity of
the triplet and singlet masses of a meson multiplet of given orbital angular
momentum $L$.  The main effect of KIPSN is that the widths of decays with
pseudoscalars in the final state are increased.}  We will refer to this as the
Kokoski-Isgur phase space/normalization (KIPSN).

Some comments about the details of the calculations made in this chapter are in
order.  For both types of SHO wavefunctions we use quark masses in the ratio
$m_u:m_d:m_s = 3:3:5$ -- this differs from the calculations of Kokoski and
Isgur \cite{kokoski87:meson}, which ignored the strange-quark mass difference.
For the RQM wavefunctions the quark masses have already been fitted: $m_u
=220$~MeV, $m_d = 220$~MeV, and $m_s = 419$~MeV.  We treat all mesons as narrow
resonances.  Meson masses are taken from the Review of Particle Properties 1994
\cite{montanet94:review} if the state is included in their Meson Summary
Table\footnote{The one exception was the $1^3P_0$ $s \bar{s}$ state -- see
Table~\ref{t:3F2decays}.}.  If it is not, then the masses predicted in
Ref.~\cite{godfrey85:mesons} are used.  This includes the masses of the
$1^3F_2$, $1^3F_3$ and $1^3F_4$ $s\bar{s}$ mesons: 2240~MeV, 2230~MeV and
2200~MeV respectively.  We ignore mass differences between members of the same
isospin multiplet\footnote{The one exception was for the decay $\phi \to K^+
K^-$ where the charged and neutral kaon mass difference is significant to the
phase space.} and average over the quoted masses, if different.  Meson flavour
wavefunctions are also taken from Ref.~\cite{godfrey85:mesons}, and are
summarized in Appendix~\ref{a:flavours}\footnote{For the $\eta$ and $\eta'$ we
assumed perfect mixing, corresponding to $\theta_P=-9.7^\circ$, but we also
considered the effect that using $\theta_P=-20^\circ$ would have on the results
of Tables~\ref{t:f42050decays} through \ref{t:3F3decays}.  We found that while
individual decay widths varied by up to a factor of approximately 2, the
changes to the total widths were negligible compared to the other uncertainties
of the calculations.}.

In order to evaluate the above wavefunctions and phase space/normalization
schemes and to fit the value of $\gamma$ ($\gamma_0$) we carried out a series
of least squares fits of the model predictions to the decay widths of 28 of the
best known\footnote{We also required that the decays we selected not involve
any particles that undergo significant mixing (e.g.\ $\eta$, $\eta'$) in order
to minimize uncertainties in the calculation.} meson decays; we minimized the
quantity defined by $\chi^2 =\sum_i (\Gamma^{model}_i -
\Gamma^{exp}_i)^2/\sigma_{\Gamma_i}^2$ where $\sigma_{\Gamma_i}$ is the
experimental error.\footnote{For the calculations in the flux-tube breaking
model, a 1\% error due to the numerical integration was added in quadrature
with the experimental error.}  The 28 decays used, and the experimental
values of their widths
\cite{montanet94:review}, are shown in Table~\ref{t:28decays}.

\begin{table}[p]
\vspace{0.1cm}
\newcommand{\BA}{@{\vline}}
\newcommand{\SP}{\hspace*{0.1cm}}
\begin{tabular}{|l@{\SP\vline\SP}c@{\SP\vline}c\BA c\BA c\BA c\BA c\BA c\BA} \hline
& $\Gamma$ (MeV) & \multicolumn{6}{c|}{$\Gamma$ (MeV) from Models of Meson 
Decay} \\ \cline{3-8}
Decay & from & \multicolumn{2}{c|}{$^3P_0$} & \multicolumn{4}{c|}
{Flux-tube Breaking} \\ \cline{3-8}
& Experiment & \multicolumn{2}{c|}{SHO} & \multicolumn{2}{c|}{SHO} & 
\multicolumn{2}{c|}{RQM} \\ \cline{3-8}
& & \SP RPSN\SP & \SP KIPSN\SP & \SP RPSN\SP & \SP KIPSN\SP & \SP RPSN\SP & \SP KIPSN  \\ \hline \hline
$\rho\to \pi \pi$ & $151.2 \pm 1.2$ & 96 & 148 & 93 & 148 & 104 & 152 \\
$b_1(1235)\to \omega \pi$ & $142 \pm 8$  & 176 & 115 & 155 & 104 & 
                306 & 190 \\
$a_2(1320) \to \rho\pi$ & $75.0 \pm 4.5$ & 65  & 38 & 67 & 40 & 84 & 46 \\
$a_2(1320) \to K\bar{K}$ & $5.2 \pm 0.9$  & 11 & 8.0 & 11 & 8.5 
                & 7.3 & 5.0 \\
$\pi_2(1670) \to f_2(1270) \pi$ & $ 135 \pm 11$ & 147 & 116 & 143 & 
                117 & 327 & 246 \\
$\pi_2(1670) \to \rho \pi$ & $ 74 \pm 11$ & 232 & 74 & 226 & 74 & 
                323 & 97 \\
$\pi_2(1670) \to K^*(892) \bar{K}$ & $10.1 \pm 3.4$ & 38 & 17 & 37 & 
                17 & 49 & 21 \\
 \hspace{2.15cm}+c.c. &&&&&&& \\
$\rho_3(1690) \to \pi \pi $ & $50.7 \pm 5.5$ & 116 & 35 & 122 & 38 & 
                68 & 19 \\
$\rho_3(1690) \to \omega \pi $ & $34 \pm 13$ & 36 & 11 & 39 & 13 & 
                45 & 13 \\
$\rho_3(1690) \to K \bar{K} $ & $3.4 \pm 0.6$  & 9.2 & 3.8 & 9.7 & 
                4.2 & 4.2 & 1.7 \\
$f_2(1270) \to \pi \pi$ & $156.8 \pm 3.2$ & 203 & 109  & 209 & 116 & 157 & 
        80 \\
$f_2(1270) \to K\bar{K}$ & $8.6 \pm 0.8 $ & 7.2 & 5.4 & 7.4 & 5.7 & 5.0 & 
        3.5 \\
$f_4 (2050) \to \omega\omega$ & $54 \pm 13 $ & 53 & 11 & 54 & 11 & 
                94  & 18 \\
$f_4 (2050) \to \pi\pi$ & $35.4\pm 3.8$ & 123 & 25 & 132 & 28 & 58 & 11 \\
$f_4 (2050) \to K\bar{K}$ & $1.4 \pm 0.7$ & 5.4 & 1.6 & 5.8 & 1.7 & 
                1.8 & 0.5 \\
$\phi \to K^+ K^-$ & $2.17 \pm 0.05$ & 2.37 & 2.83 & 2.28 & 2.80 & 
        2.30 & 2.60\\
$f_2'(1525) \to K\bar{K}$ & $61 \pm 5$  & 117 & 61 & 118 & 64 & 98 & 49 \\
$K^*(892) \to K\pi$ & $50.2 \pm 0.5 $ &  36 & 52 & 34 & 51 & 38 & 52 \\
$K_0^*(1430)\to K\pi$ & $267 \pm 36$ & 163 & 84 & 117 & 63 & 875 & 430 \\
$K_2^*(1430)\to K\pi$ & $48.9 \pm 1.7 $  & 108 & 56 & 112 & 60 & 88 & 43 \\
$K_2^*(1430)\to K^*(892)\pi$ & $24.8\pm 1.7$ & 27 & 16 & 27 & 17 & 31 & 18\\
$K_2^*(1430) \to K\rho$ & $8.7\pm 0.8$ & 9.3 & 4.9 & 9.6 & 5.2 & 12 & 5.8 \\
$K_2^*(1430) \to K\omega$ & $2.9 \pm 0.8$ & 2.6 & 1.4 & 2.6 & 1.4 & 
        3.2 & 1.6 \\
$K_3^*(1780)\to K \rho$ & $74 \pm 10$ & 24 & 7.7 & 25 & 8.4 & 28 & 8.7 \\
$K_3^*(1780)\to K^*(892)\pi$ & $45 \pm 7$ & 33 & 11 & 34 & 12 & 37 & 12 \\
$K_3^*(1780)\to K\pi$ & $31.7\pm 3.7$ & 87 & 28 & 92 & 30 & 54 & 16 \\
$K_4^*(2045)\to K\pi$ & $19.6\pm 3.8$ & 55 & 13 & 59 & 14 & 28 & 6.2 \\
$K_4^*(2045)\to K^*(892) \phi$ & $2.8 \pm 1.4$ & 3.2 & 1.0 & 3.3 & 
                1.1 & 4.7 & 1.4 \\ \hline
\end{tabular}
\caption[The 28 meson decays used in our fits of the decay models' 
parameters]{The 28 meson decays used in our fits of the decay models' 
parameters.  The experimental widths are shown, as are the model results for 
six combinations of decay model, wavefunctions, and phase space/normalization 
(specified in the second, third and fourth rows of the heading, respectively).}
\label{t:28decays}
\end{table}

\begin{figure}[t]
\vspace{-1.1cm}
\makebox{\epsfxsize=6.5in\epsffile{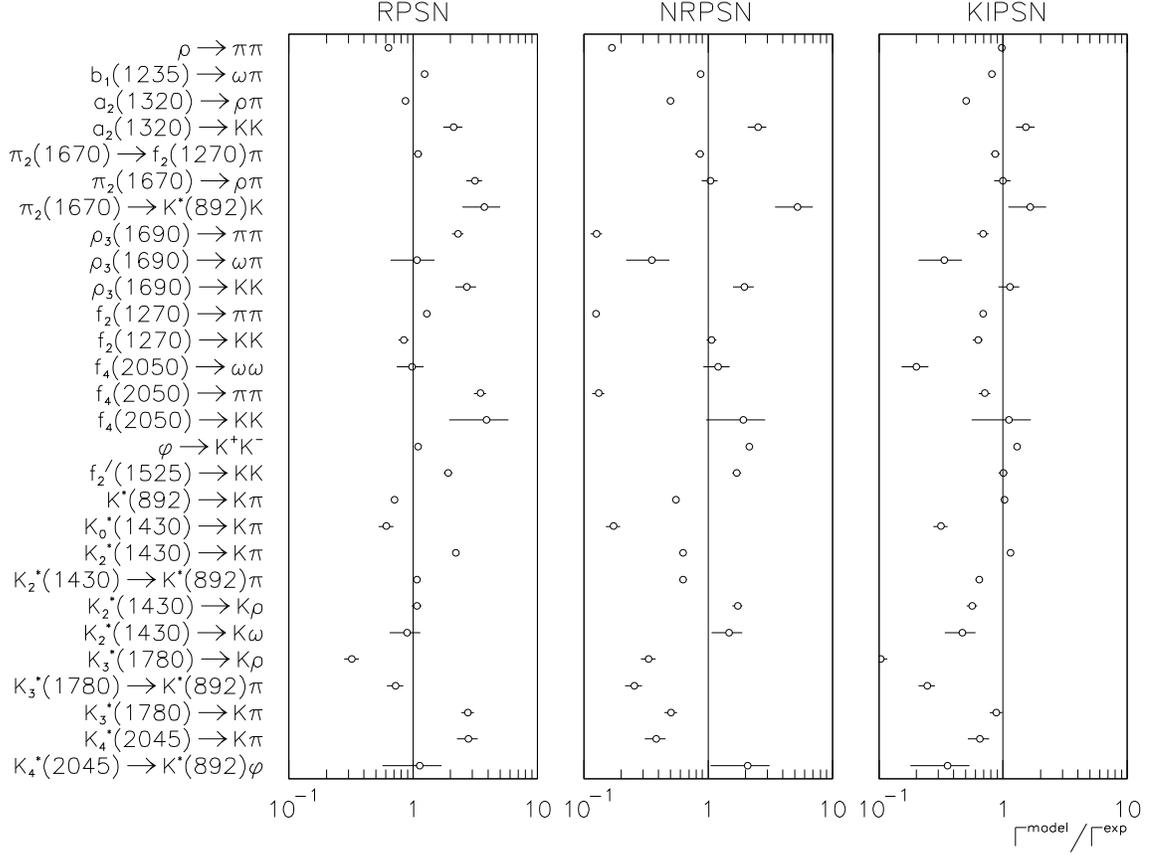}}
\vspace{-1.3cm}
\caption[Ratios of the predictions of the $^3P_0$ model of meson decay 
to the experimental values, for the widths of the 28 decays used in the fits.  
In this figure
SHO wavefunctions with a common $\beta$ were used]{Ratios of the predictions 
of the $^3P_0$ model of meson decay 
to the experimental values, for the widths of the 28 decays used in the fits.  
In this figure
SHO wavefunctions with a common $\beta$ were used.  The three graphs use 
different phase space/normalization schemes.}
\label{f:fits1}
\end{figure}

\begin{figure}[t]
\vspace{-1.1cm}
\makebox{\epsfxsize=6.5in\epsffile{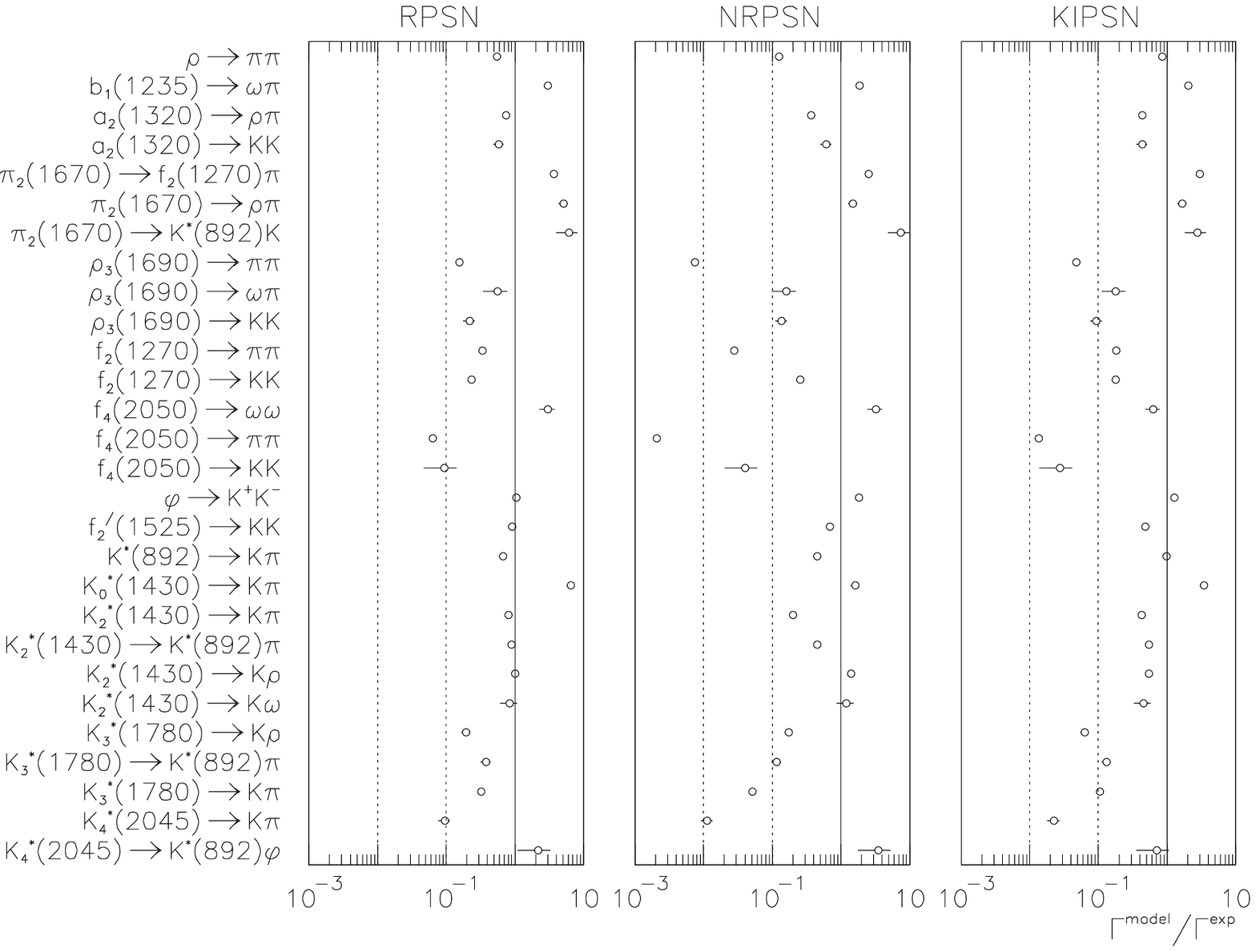}}
\vspace{-1.3cm}
\caption[Ratios of the predictions of the $^3P_0$ model of meson decay 
to the experimental values, for the widths of the 28 decays used in the fits.  
In this figure
SHO wavefunctions with effective $\beta$'s were used]{Ratios of the 
predictions of the $^3P_0$ model of meson decay 
to the experimental values, for the widths of the 28 decays used in the fits.  
In this figure
SHO wavefunctions with effective $\beta$'s were used.  The three graphs use 
different phase space/normalization schemes.}
\label{f:fits2}
\end{figure}

\begin{figure}[t]
\vspace{-1.1cm}
\makebox{\epsfxsize=6.5in\epsffile{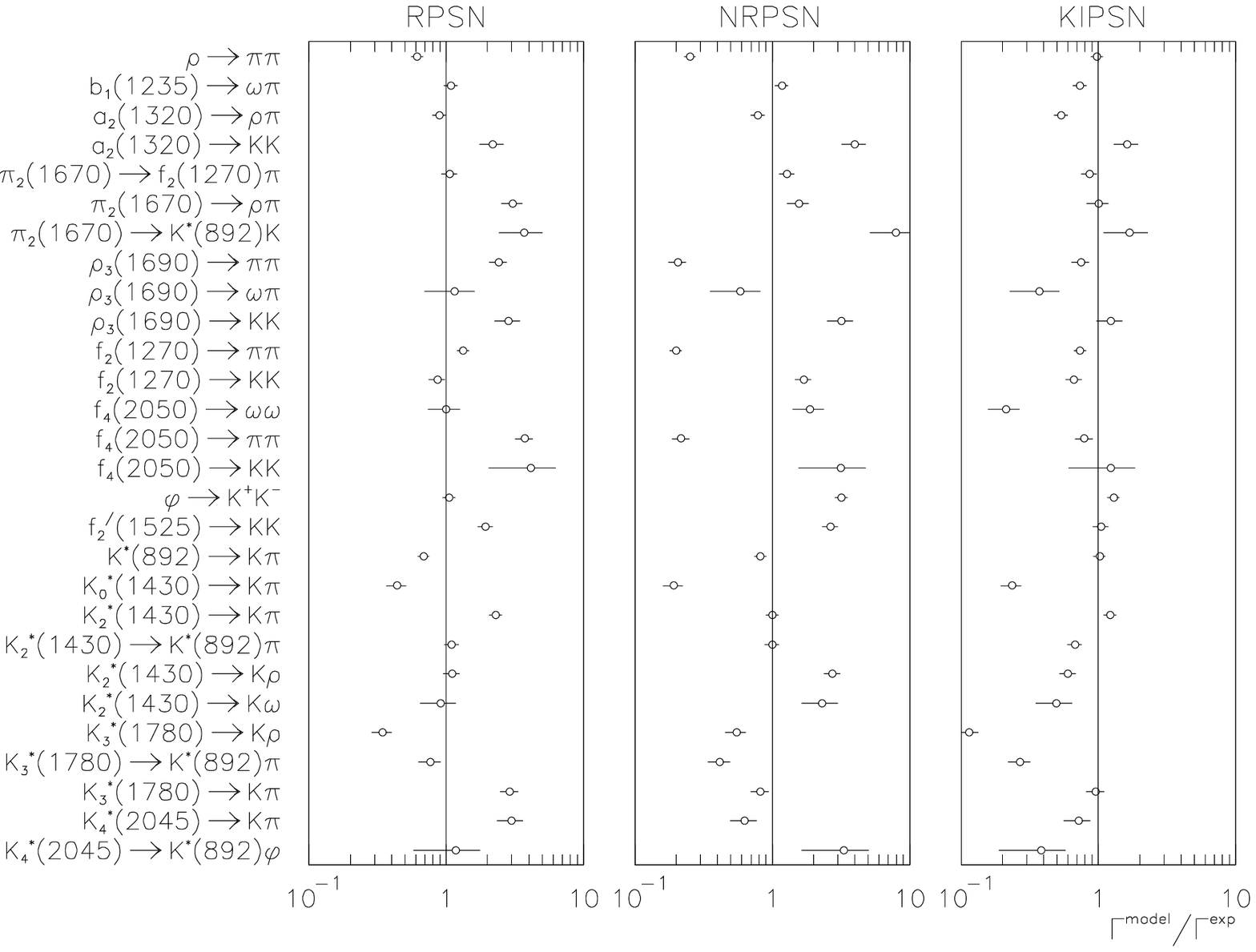}}
\vspace{-1.3cm}
\caption[Ratios of the predictions of the flux-tube breaking 
model of meson decay to the experimental values, for the widths of the 28
decays used in the fits.  In this figure SHO wavefunctions with a common
$\beta$ were\protect\\ used]{Ratios of the predictions of the flux-tube
breaking model of meson decay to the experimental values, for the widths of the
28 decays used in the fits.  In this figure SHO wavefunctions with a common
$\beta$ were used.  The three graphs use different phase space/normalization
schemes.}
\label{f:fits3}
\end{figure}

\begin{figure}[t]
\vspace{-1.1cm}
\makebox{\epsfxsize=6.5in\epsffile{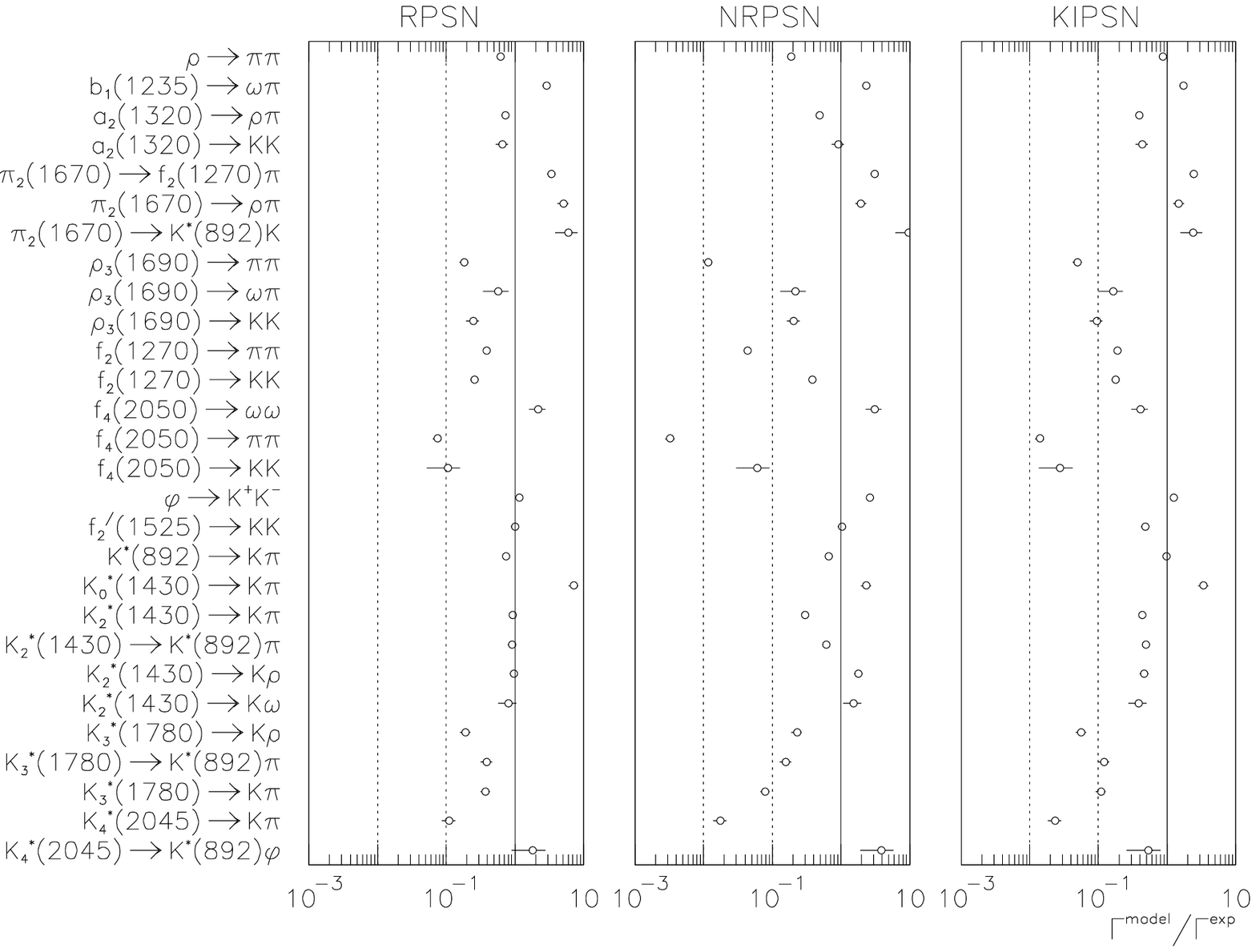}}
\vspace{-1.3cm}
\caption[Ratios of the predictions of the flux-tube breaking 
model of meson decay to the experimental values, for the widths of the 28
decays used in the fits.  In this figure SHO wavefunctions with effective
$\beta$'s were\protect\\ used] {Ratios of the predictions of the flux-tube
breaking model of meson decay to the experimental values, for the widths of the
28 decays used in the fits.  In this figure SHO wavefunctions with effective
$\beta$'s were used.  The three graphs use different phase space/normalization
schemes.}
\label{f:fits4}
\end{figure}

\begin{figure}[t]
\vspace{-1.1cm}
\makebox{\epsfxsize=6.5in\epsffile{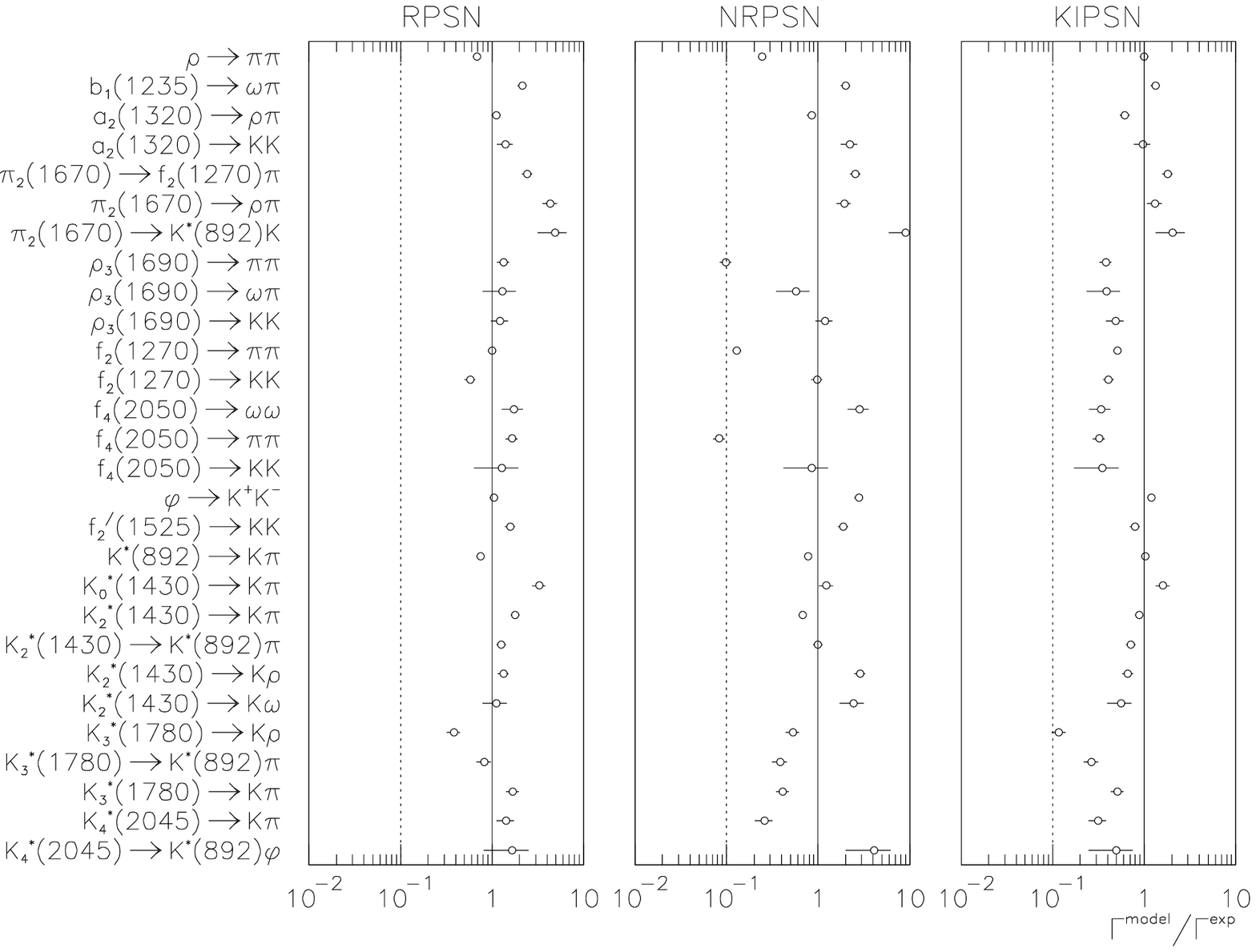}}
\vspace{-1.3cm}
\caption[Ratios of the predictions of the flux-tube breaking 
model of meson decay 
to the experimental values, for the widths of the 28 decays used in the fits.  
In this figure
RQM wavefunctions were used]{Ratios of the predictions 
of the flux-tube breaking model of meson decay 
to the experimental values, for the widths of the 28 decays used in the fits.  
In this figure
RQM wavefunctions were used.  The three graphs use 
different phase space/normalization schemes.}
\label{f:fits5}
\end{figure}

The ratios of the model predictions to the experimental values for the widths
of the 28 decays are plotted in Figures~\ref{f:fits1} to \ref{f:fits5}, on
logarithmic scales.  The fitted values of $\gamma$ ($\gamma_0$) and the
resulting $\chi^2$ per degree of freedom (dof) are listed in
Table~\ref{t:chi2results}.  We do not quote errors on the fitted values of
$\gamma$ ($\gamma_0$) because they are negligible (being at most a change in
the last digit quoted) compared to the inherent uncertainties of the models.
We have performed fits for the $^3P_0$ model with both types of SHO
wavefunctions, and for the flux-tube breaking model with all three types of
wavefunctions.  We did not use the RQM wavefunctions with the $^3P_0$ model
because the large number (up to 13) of radial SHO wavefunctions in the linear
combination (see Eq.~\ref{shoexpansion}) exceeds the capacity of our symbolic
routines on the available computer hardware\footnote{This problem arises for
the $^3P_0$ model routines, which are symbolic (and hence are slower and
require more memory), but not for the flux-tube breaking model routines, where
the brunt of the work is done numerically.}.  Because the results of the
$^3P_0$ and flux-tube breaking models appear to be very similar, we chose not
to write a numerical version of the $^3P_0$ routines to enable us to use the
RQM wavefunctions.  For each decay model and wavefunction combination, we tried
all three phase space/normalization schemes.

\begin{table}[t]
\vspace{-0.3cm}
\begin{center}
\begin{tabular}{|c|c|c|c|c|c|} \hline
Meson Decay & Wavefunctions & Phase Space/ & $\beta$ & $\gamma$ ($\gamma_0$)& 
$\chi^2/{\rm dof}$ \\ 
Model & & Normalization & & & \\ \hline \hline
 & & RPSN & & 9.73 & 223 \\ 
$^3P_0$ & SHO (common $\beta$) & NRPSN & 400 MeV & 14.1 & 675 \\
 & & KIPSN & & 6.25 & 26.9 \\ \hline
 & & RPSN & & 17.6 & 410 \\
$^3P_0$ & SHO (effective $\beta$'s) & NRPSN & varies & 23.8 & 776 \\
 & & KIPSN & & 11.4 & 161 \\ \hline
Flux-tube & & RPSN & & 16.0 & 185 \\
Breaking & SHO (common $\beta$) & NRPSN & 400 MeV & 28.7 & 546 \\
 & & KIPSN & & 10.4 & 22.0 \\ \hline
Flux-tube & & RPSN & & 22.1 & 271 \\
Breaking & SHO (effective $\beta$'s) & NRPSN & varies & 34.4 & 663 \\
 & & KIPSN & & 13.6 & 130 \\ \hline
Flux-tube & & RPSN & & 20.5 & 128 \\
Breaking & RQM & NRPSN & varies & 34.4 & 556 \\
 & & KIPSN & & 12.8 & 40.3 \\ \hline \hline
 & & RPSN & 481 MeV & 13.4 & 158 \\
$^3P_0$ & SHO (common $\beta$) & NRPSN & 444 MeV & 16.5 & 696 \\
 & & KIPSN & 371 MeV & 5.60 & 26.0 \\
\hline
\end{tabular}
\end{center}
\vspace{-0.4cm}
\caption[The details of the various fits of the decay models' 
parameters, and the resulting $\chi^2/{\rm dof}$]{The details of the various 
fits of the decay models' 
parameters, and the resulting $\chi^2/{\rm dof}$.  For the last three fits, 
both $\gamma$ and $\beta$ were fit simultaneously.}
\label{t:chi2results}
\end{table}

\begin{table}[t]
\vspace{-0.3cm}
\begin{center}
\begin{tabular}{|c|c|c|c|c|c|c|} \hline
Meson Decay & Wavefunctions & Phase Space/ & $\beta$ & $\bar{\gamma}$ & 
$\sigma_\gamma$ & $\sigma_\gamma/\bar{\gamma}$ \\ 
Model & & Normalization & (MeV) & & & \\ \hline \hline
 & & RPSN & & 8.73 & 2.84 & 0.325 \\ 
$^3P_0$ & SHO (common $\beta$) & NRPSN & 400 & 19.3 & 10.1 & 0.527 \\
 & & KIPSN & & 8.17 & 3.15  & 0.386 \\ \hline
 & & RPSN & & 25.7 & 16.3 & 0.634 \\
$^3P_0$ & SHO (effective $\beta$'s) & NRPSN & varies & 76.4 & 108 & 1.41 \\
 & & KIPSN & & 26.5 & 23.0 & 0.868 \\ \hline
Flux-tube & & RPSN & & 14.3 & 4.86 & 0.339 \\
Breaking & SHO (common $\beta$) & NRPSN & 400 & 31.6 & 16.9 & 0.537 \\
 & & KIPSN & & 13.4 & 5.13 & 0.384 \\ \hline
Flux-tube & & RPSN & & 31.0 & 18.7 & 0.604 \\
Breaking & SHO (effective $\beta$'s) & NRPSN & varies & 90.5 & 124 & 1.37 \\
 & & KIPSN & & 31.9 & 26.6 & 0.834 \\ \hline
Flux-tube & & RPSN & & 18.1 & 5.02 & 0.278 \\
Breaking & RQM & NRPSN & varies & 42.1 & 27.6 & 0.656 \\
 & & KIPSN & & 17.1 & 6.01 & 0.351 \\ 
\hline
\end{tabular}
\end{center}
\vspace{-0.4cm}
\caption[The values of $\bar{\gamma}$,
$\sigma_\gamma$ and $\sigma_\gamma/\bar{\gamma}$ for the 28 decays used in the
fits, where the $\gamma$ ($\gamma_0$) value for a decay matches the model
prediction to the experimental value]{The values of $\bar{\gamma}$,
$\sigma_\gamma$ and $\sigma_\gamma/\bar{\gamma}$ for the 28 decays used in the
fits, where the $\gamma$ ($\gamma_0$) value for a decay matches the model
prediction to the experimental value.  Thus $\sigma_\gamma/\bar{\gamma}$
provides a measure of the spread of the results that ignores the experimental
errors.}
\label{t:avgingresults}
\end{table}

When examining Figures~\ref{f:fits1} to \ref{f:fits5}, note that if a
particular model described the data exactly, all the points for it would lie
along the vertical solid line at $\Gamma^{\rm model}/\Gamma^{\rm exp}$~=~1.
Because $\gamma$ ($\gamma_0$) is an overall factor in the amplitude, changing
it would move all of the points to the left or right by the same amount.  Some
distributions appear lopsided about $\Gamma^{\rm model}/\Gamma^{\rm exp}=1$,
because the experimentally well known decays dominate the fit.

The first thing to notice from the figures and the $\chi^2$ values of
Table~\ref{t:chi2results} is that the results with NRPSN are significantly
worse than the ones using the other two choices of phase space/normalization.
We will hence not use NRPSN for our calculations.  Similarly, the SHO
wavefunctions with effective $\beta$'s do worse than either of the other two
types of wavefunctions, and we will not use them either.  That leaves us with
six different combinations of decay model, wavefunction, and phase
space/normalization -- these are the combinations given in
Table~\ref{t:28decays}.  Some general comments are in order.

First, the $\chi^2$ values of Table~\ref{t:chi2results} make it clear that
these models are not very accurate\footnote{It should be noted that past
practice has often been to look for agreement in the amplitudes, rather than
the widths, which tends to make the situation look better.} -- the best they
can hope for is to predict a decay width to within a factor of 2, and even
larger deviations are common.  Since they are coarse models of a complicated
theory, this is not surprising.  In addition, the six combinations can give
rather different results.  We will use this to our advantage by taking the
spread in the predictions of the different combinations as a rough guide to the
trustworthiness of the calculations.

From Table~\ref{t:28decays} and Figures~\ref{f:fits1} through \ref{f:fits5} one
can see that the results for the $^3P_0$ and flux-tube breaking models for the
SHO wavefunctions are very similar, as previously noted.\footnote{The one
exception to this is the S-wave decay $K^*_0(1430) \to K\pi$ which seems
particularly sensitive to the model.}  This is in agreement with the
observations of Kokoski and Isgur \cite{kokoski87:meson}, who argued that the
overlap of the space wavefunctions in the $^3P_0$ model makes it unlikely that
the decay will proceed if the quark-antiquark pair is created far from a
straight line connecting the two mesons.  This would have the same effect as
the functional form of $\gamma(\vec{r},\vec{w})$ in the flux-tube breaking
model.  Despite the similarity in the individual decays, the $\chi^2$ values of
the fits are better for the flux-tube breaking model, indicating at least some
preference for this model.

In comparing the SHO wavefunctions (we will henceforth only speak of the SHO
wavefunctions with a common $\beta$) to the RQM wavefunctions, we note that the
RQM wavefunctions do better for RPSN, but worse for KIPSN, showing no clear
preference.

KIPSN gives better overall fits to the data than RPSN.  However, certain
decays, $K_3^*(1780) \to K\rho$ and $f_4(2050) \to \omega\omega$ for example,
are fit better using RPSN.  Also, if we wished to remove the dominant
effect that the experimentally well known decays have on the fit, we could
follow Geiger and Swanson \cite{geiger94:distinguishing} and look at the spread
of the values of $\gamma$ needed to match the model predictions to the
experimental values exactly; if $\bar{\gamma}$ is the mean of these $\gamma$
values, and $\sigma_\gamma$ the standard deviation of the distribution, then
$\sigma_\gamma/\bar{\gamma}$ provides a measure of the spread of the results
that ignores the experimental errors.  Using this measure, RPSN is consistently
better than KIPSN, as can be seen in Table~\ref{t:avgingresults}.

We can make more specific observations using the SHO wavefunctions: Decays with
two pseudoscalars in the final state tend to do better with KIPSN, but KIPSN
generally underestimates decays of high $L$ mesons with vector mesons in the
final state.  On the other hand RPSN tends to overestimate decays with two
pseudoscalars in the final state.  Trends can also be observed with the
flux-tube breaking model using the RQM wavefunctions.  Having said all this we
stress that these are only general observations and exceptions can be found to
any of them in Table \ref{t:28decays}.  One must therefore be very careful not
to take any predictions of these models at face value but should try if
possible to compare the predicted decay to a similar one that is experimentally
well known.

\begin{figure}[t]
\vspace{-1.1cm}
\makebox{\epsfxsize=6.5in\epsffile{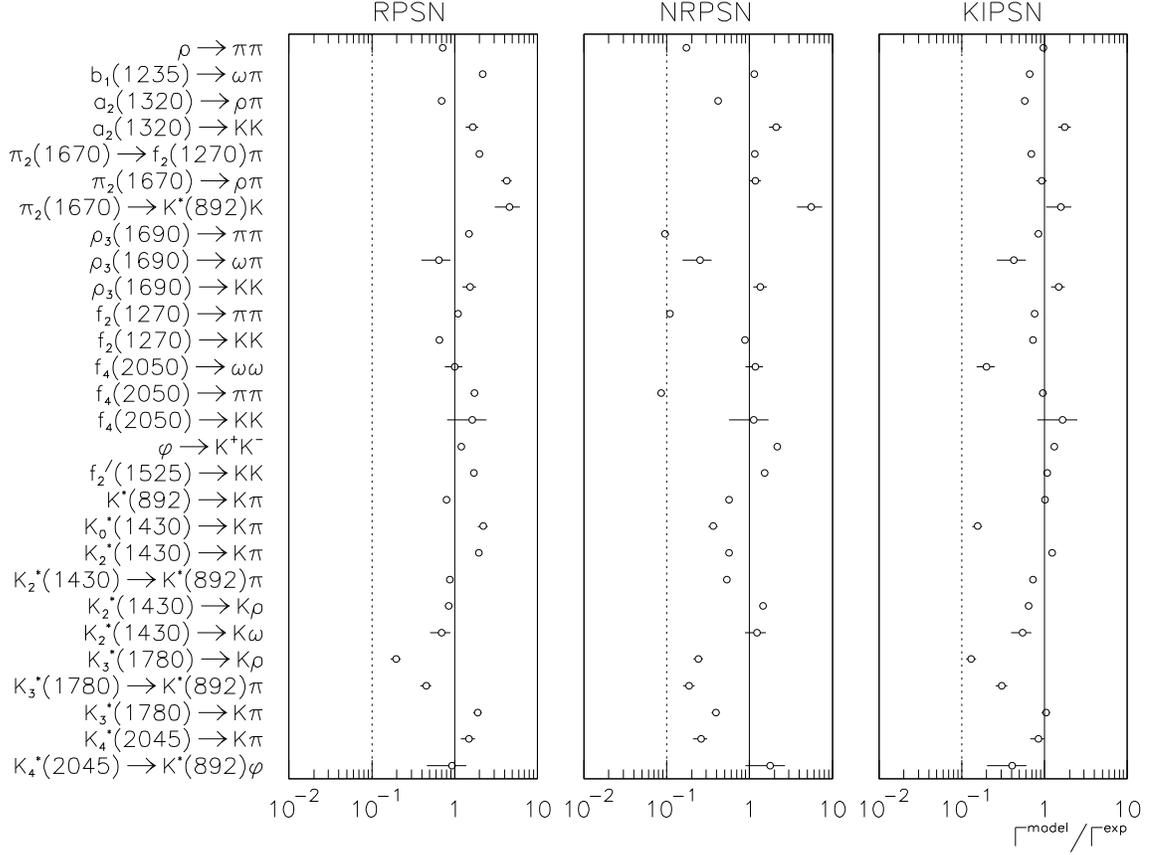}}
\vspace{-1.3cm}
\caption[Ratios of the predictions of the $^3P_0$ model of meson decay 
to the experimental values, for the widths of the 28 decays used in the fits.  
In this figure
SHO wavefunctions with a common $\beta$ were used, but both $\beta$ and 
$\gamma$ were fit simultaneously]{Ratios of the predictions 
of the $^3P_0$ model of meson decay 
to the experimental values, for the widths of the 28 decays used in the fits.  
In this figure
SHO wavefunctions with a common $\beta$ were used, and both $\beta$ and 
$\gamma$ were fit simultaneously.  The three graphs use 
different phase space/normalization schemes.}
\label{f:fits6}
\end{figure}

Finally, we consider the sensitivity of our results to $\beta$, which we must
choose for the SHO wavefunctions.  Different quark models
predict different values of $\beta$, so there is no clear theoretical
prediction to draw upon.  It should be noted that although meson decay widths
can depend strongly on the value of $\beta$, to some extent we alleviate this
effect because we fit $\gamma$ to experimental data after selecting a value of
$\beta$.  In the fits discussed above, we took $\beta = 400$~MeV which is the
value used by Kokoski and Isgur \cite{kokoski87:meson}.  In addition to these
fits, we performed simultaneous fits of both $\gamma$ and $\beta$ to the 28
decay widths -- see Figure~\ref{f:fits6}.  The overall fits improved, but as
usual some widths were in better agreement and some in worse agreement with
experiment when compared to the fits for $\beta=400$~MeV.  In general we found
that the deviation in the results lies within the overall uncertainty we assign
to the models, and so we choose to follow the literature and use $\beta =
400$~MeV.

\section{The $f_4(2220)$ Reexamined - What Is it Really?}
\label{3:xi}

\begin{figure}[t]
\vspace{-1.5cm}
\makebox{\epsfxsize=6.5in\epsffile{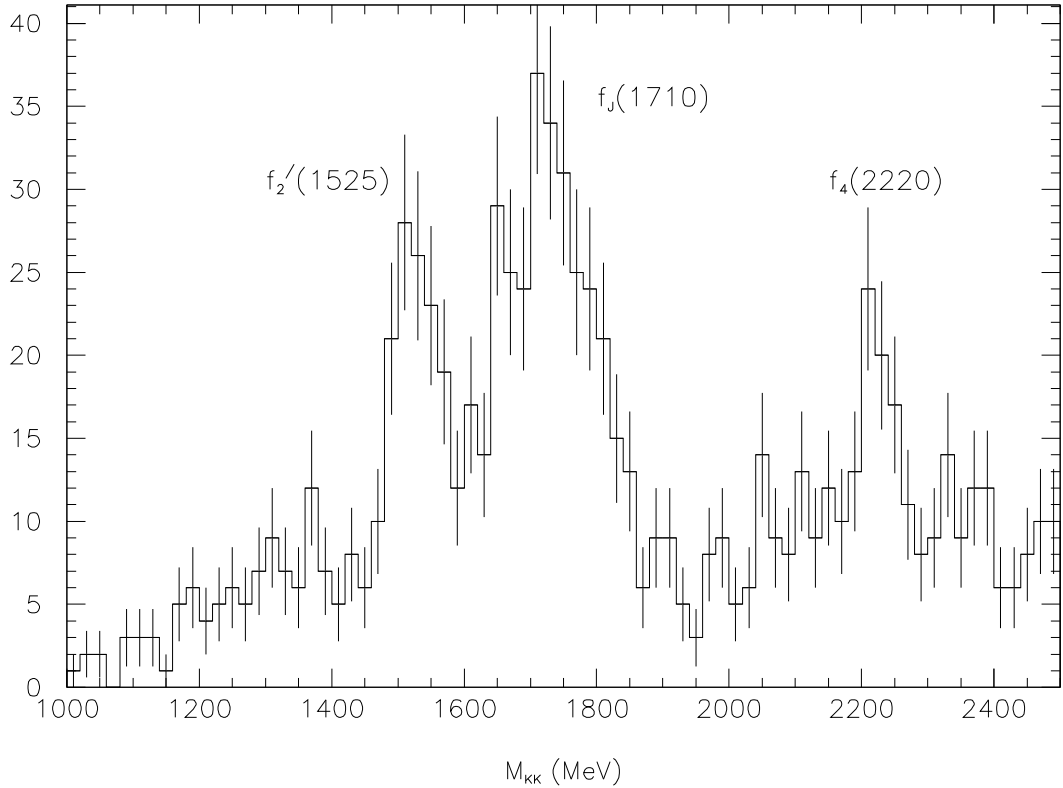}}
\vspace{-1cm}
\caption[The original data from the discovery of the $f_4(2220)$ by the 
MARK~III Collaboration.  The distribution of the $K^+K^-$ invariant mass in
$e^+e^- \to J/\psi \to \,\gamma K^+K^-$ is shown, revealing three resonances]
{The original data from the discovery of the $f_4(2220)$ by the MARK~III
Collaboration \cite{baltrusaitis86:observation}.  The distribution of the
$K^+K^-$ invariant mass in $e^+e^- \to J/\psi \to \gamma \,K^+K^-$ is shown,
revealing three resonances.  The errors shown are purely statistical.}
\label{f:xidata}
\end{figure}

The $f_4(2220)$ (formerly known as the $\xi(2220)$) state was discovered by the
Mark III Collaboration \cite{baltrusaitis86:observation} in 1983, in the decay
$e^+e^- \to J/\psi \to \gamma \,f_4(2220)\to\gamma \,K^+K^-$ -- see
Figure~\ref{f:xidata}.  It attracted considerable attention because of its
narrow width of approximately 30~MeV, which was thought to be too small for a
conventional meson of that mass.  Over the years this has led to speculation
that it might be a Higgs boson \cite{barnett84:implications}, a bound state of
coloured scalars \cite{shatz84:can}, a four quark state
\cite{chao84:possible,pakvasa84:xi}, a $\Lambda \bar{\Lambda}$ bound state
\cite{ono87:xi}, or a meson hybrid or a glueball \cite{chanowitz83:glueballs},
among others (we only reference early works on each suggestion).

The $f_4(2220)$ has since been seen by other experiments, the results of which
are summarized in Table~\ref{t:ximeasurements}.  The rather broad state seen by
DM2 is unlikely to be the same state seen by the other experiments.  The recent
measurements by the BES Collaboration are of particular interest, because the
previously unseen $\pi^+\pi^-$ and $p\bar{p}$ decay modes indicate a decay
pattern that is roughly flavour-symmetric, in keeping with the glueball
interpretation (see for example, the recent paper by Chao \cite{chao95:xi}).
The $f_4(2220)$ has not been seen by $p\bar{p}$ experiments, but the limits set
are not stringent enough to exclude the meson identification (for example see
Barnes {\it et al.}\ \cite{barnes93:measurement}).

\begin{table}[t]
\newcommand{\BA}{@{\vline}}
\newcommand{\SP}{\hspace*{0.1cm}}
\vspace{-0.3cm}
\begin{center}
\begin{tabular}{|l|l|l|l|c|} \hline
\hspace{0.3cm}Experiment & \hspace{2cm}Process & \hspace{0.8cm}Mass & 
\hspace{0.45cm}Width & J  \\
 & & \hspace{0.65cm}(MeV) & \hspace{0.45cm}(MeV) & \\ \hline \hline 
SLAC-SP-032 & $e^+e^- \to J/\psi \to \gamma \,f_4(2220)$ & & & \\
(Mark III)  $^{\rm a}$ & \hspace{3.4cm}$\hookrightarrow K^+K^-$ & 
$2230\pm 6\pm 14$ & $26^{+20}_{-16} \pm 17$ & \\
 & \hspace{3.4cm}$\hookrightarrow K^0_S\bar{K}^0_S$ & $2232\pm 7\pm 7$ & 
$18^{+23}_{-15} \pm 10$ & \\ \hline
(GAMS-2000, & $\pi^-p\to f_4(2220) \,n$ & & & \\
GAMS-4000) $^{\rm b}$ & \hspace{1.6cm}$\hookrightarrow \eta \eta'$ & 
$2220\pm 10$ & & $\geq 2$ \\ \hline
SERP-E-147 & $\pi^-p\to f_4(2220) \,n$ & & & \\
(MIS) $^{\rm c}$ & \hspace{1.6cm}$\hookrightarrow K^0_S\bar{K}^0_S$ & 
$2230\pm 20$ & 
$80\pm 30$ & 2 \\ \hline
SLAC-E-135 & $K^- p\to f_4(2220)\,\Lambda $ & & & \\
(LASS) $^{\rm d}$ & \hspace{1.7cm}$\hookrightarrow K^+K^-$ & $2209^{+17}_{-15}\pm 10$ & 
$60^{+107}_{-57}$ & 4 \\ \hline \hline
RPP Average $^{\rm e}$ & & $2225\pm 6$ & $38^{+15}_{-13}$ & 4? \\ \hline 
\hline
(DM2) $^{\rm f}$ & $e^+e^- \to J/\psi \to \gamma\,$? & & & \\
 & \hspace{3.3cm}$\hookrightarrow K^0_S\bar{K}^0_S$ & $2197\pm 17$ & 
$201\pm 51$ & \\ \hline
(BES) $^{\rm g}$ & $e^+e^- \to J/\psi \to \gamma \,f_4(2220)$ & & & \\
 & \hspace{3.4cm}$\hookrightarrow K^+K^-$ & $2230^{+6}_{-7}\pm 12$ & 
$20^{+20}_{-15}\pm 12$ & \\
 & \hspace{3.4cm}$\hookrightarrow K^0_S\bar{K}^0_S$ & $2232^{+8}_{-7}\pm 15$ & 
$20^{+25}_{-16}\pm 10$ & \\
 & \hspace{3.4cm}$\hookrightarrow p \bar{p}$ & $2235\pm 4\pm 5$ & 
$15^{+12}_{-9}\pm 9$ & \\
 & \hspace{3.4cm}$\hookrightarrow \pi^+ \pi^-$ & $2235\pm 4\pm 6$ & 
$19^{+13}_{-11}\pm 12$ & \\ \hline
\end{tabular}
\end{center}
\vspace{-0.4cm}
\caption[Summary of measurements of the $f_4(2220)$]
{Summary of measurements of the $f_4(2220)$.  The RPP average is for the first
four entries in the table only; they feel that the $f_4(2220)$ still needs
confirmation.  In the case of the DM2 results, the resonance
seen was not identified with the $f_4(2220)$, and so it is not included in
the RPP average; the BES data is too recent.\\
$^{\rm a}$ Reference~\cite{baltrusaitis86:observation}. \\
$^{\rm b}$ Reference~\cite{alde86:2.22}. \\
$^{\rm c}$ Reference~\cite{bolonkin88:study}. \\
$^{\rm d}$ Reference~\cite{aston88:evidence}. \\
$^{\rm e}$ Reference~\cite{montanet94:review}. \\
$^{\rm f}$ Reference~\cite{augustin88:measurement}. \\
$^{\rm g}$ Reference~\cite{bes95:study}. 
}
\label{t:ximeasurements}
\end{table}

The intermediate state $J/\psi$ and the final observed states indicate that the
parity and charge conjugation quantum numbers of the $f_4(2220)$ are $^{++}$
and that its spin $J$ must be even, although as can been seen from
Table~\ref{t:ximeasurements} the actual value of $J$ is rather uncertain.  The
Particle Data Group \cite{montanet94:review} has tentatively identified the
$f_4(2220)$ as the $^3F_4$ $s\bar{s}$ meson (hence its presence in
Table~\ref{t:mesonspectrum}), but notes that this needs confirmation.

In order to determine whether the $f_4(2220)$ is a high-spin meson or something
new and much more interesting, we need theoretical predictions of the
properties of the various options in order to make the identification.  In this
work we carefully examine the decays of the $^3F_2$ and $^3F_4$ $s\bar{s}$
mesons in order to see if the meson identification is tenable.  These are the
best meson candidates because of their $J^{\cal PC}$'s of $2^{++}$ and $4^{++}$
respectively, and because their masses have been predicted to lie in the right
range (Godfrey and Isgur \cite{godfrey85:mesons} calculated masses of 
2240~MeV and 2200~MeV respectively).

The decay widths of these mesons have been calculated previously.  The first
was by Godfrey, Kokoski and Isgur \cite{godfrey84:xi} using the flux-tube
breaking model (and somewhat different masses: 2270~MeV and 2210~MeV
respectively).  Despite the fact that one na\"{\i}vely expects heavy mesons to
be broad, since more phase space and decay channels lead to larger widths, they
found the total widths to be approximately 60~MeV and 50~MeV respectively.
Based on production rate arguments, they concluded that the $f_4(2220)$ was
most likely to be the $^3F_2$ $s\bar{s}$ state.  However, the analysis was not
exhaustive in that it did not calculate the partial widths to all possible
final states.  In particular it made the assumption, which we will see to be
incorrect, that the decays to an $L=1$ meson and a $K$ or $\eta$ were small on
the basis of phase space arguments alone.\footnote{In addition, they note that
the decay to $K^*(892)\bar{K}^*(892)$ is dominated by the $1,D$ partial wave
(using the $S,L$ notation of Table~\ref{t:f42050decays}).  Unfortunately, $S=1$
is forbidden by conservation of parity and charge conjugation, making the
result for this decay questionable.}

In addition, Pakvasa, Suzuki and Tuan \cite{pakvasa84:xi} have used Regge
theory to estimate the partial width of the $^3F_4$ $s\bar{s}$ state decaying
to two $^1S_0$ mesons as approximately 40~MeV, with the total width expected to
be a few times larger.  And finally, Ono, P\`{e}ne and Sch\"{o}berl
\cite{ono85:quarkonium} have used the $^3P_0$ model to calculate the partial 
width of the $^3F_2$ $s\bar{s}$ state (assuming a mass of 2200~MeV) decaying to
$^1S_0$ and $^3S_1$ states to be 417~MeV (with the width to $K^*(892)
\bar{K}^*(892)$ dominant at 348~MeV).  In light of the new data, and these
contradictory theoretical results, a new, complete calculation is required.

\begin{table}[t]
\vspace{-0.3cm}
\begin{center}
\begin{tabular}{|l|c|c|c|} \hline
& $\Gamma$ (MeV) & \multicolumn{2}{c|}{$\Gamma$ (MeV) from Models} \\ 
\cline{3-4}
\hspace{0.5cm} Decay & from & \multicolumn{2}{c|}{$^3P_0$} \\ \cline{3-4}
& Experiment & \multicolumn{2}{c|}{SHO} \\ \cline{3-4}
$f_4(2050) \to$ & & \hspace{0.25cm} RPSN \hspace{0.25cm} & KIPSN\\ \hline\hline
\hspace{0.5cm} $[\pi \pi]_{0,G}$ & $35.4\pm 3.8$ & 123 & 25 \\
\hspace{0.5cm} $[\pi \pi(1300)]_{0,G}$ & & 3.9 & 1.9 \\
\hspace{0.5cm} $[\pi a_1(1260)]_{1,F}$ & & 18 &  7.5 \\
\hspace{0.5cm} $[\pi a_2(1320)]_{2,F}$ & & 44 & 19 \\
\hspace{0.5cm} $[\pi \pi_2(1670)]_{2,D}$ &  & 2.1 & 1.8 \\
\hspace{0.5cm} $[\rho \rho]_{0,G}$ & & 1.9 & 0.4 \\
\hspace{0.5cm} $[\rho \rho]_{2,D}$ & & 159 & 33 \\
\hspace{0.5cm} $[\rho \rho]_{2,G}$ & & 7.3 & 1.5 \\
\hspace{0.5cm} $[\eta\eta]_{0,G}$ & $0.4\pm 0.2$& 3.2 & 0.9 \\
\hspace{0.5cm} $[\eta\eta']_{0,G}$ & & 1.0 & 0.3 \\
\hspace{0.5cm} $[\eta f_2(1270)]_{2,F}$ & & 1.1 &  0.5 \\
\hspace{0.5cm} $\left. \begin{array}{l}
\!\!\![\omega\omega]_{2,D} \\
\!\!\![\omega\omega]_{2,G}
\end{array} \right\} $ &
$54\pm 13$ $^{\rm a}$ &
$\begin{array}{c}
50 \\
2.0
\end{array} $ &
$\begin{array}{c}
10 \\
0.4
\end{array} $ \\
\hspace{0.5cm} $[K\bar{K}]_{0,G} $ & $1.4^{+0.7}_{-0.4}$ & 5.4 & 1.6 \\
\hspace{0.5cm} $[K\bar{K}^*(892)+{\rm c.c.}]_{1,G} $ & & 2.7 & 0.8 \\
\hspace{0.5cm} $[K \bar{K}_1(1270)+{\rm c.c.}]_{1,F}$ & & 2.3  & 1.2 \\
\hspace{0.5cm} $[K^*(892) \bar{K}^*(892)]_{2,D}$ & & 7.3 & 2.1 \\ \hline
Total / $\sum_i\Gamma_i$ & $208 \pm 13$ & 435 & 109 \\ \hline
\end{tabular}
\end{center}
\vspace{-0.4cm}
\caption[The calculated partial decay widths and total width of the 
$f_4(2050)$ meson.  The available experimental results
are also shown]{The calculated partial decay widths and total width of the 
$f_4(2050)$ meson.  The available experimental results
\cite{montanet94:review} are also shown.  The model results are given
for two combinations of decay model, wavefunctions, and phase
space/normalization (specified in the second, third and fourth rows of the
heading, respectively).  We have calculated the partial widths of all
kinematically-allowed, OZI-allowed, two-body strong decays, but only show those
partial widths that are $\geq 1$~MeV in at least one model.  For this reason
the total widths may not equal the sum of the partial widths shown.  The
subscripts on the decays refer to the $S$ and $L$ (see Appendix
\ref{a:partialwaves}) of the given partial wave -- the $L$ is in
spectroscopic notation ($S, P, D, F, G, H$).  We have used the meson
mixings given in Appendix~\ref{a:mixings} where applicable. \\ 
$^{\rm a}$ This number is the total for all partial waves.}
\label{t:f42050decays}
\end{table}

We have calculated the $^3F_2$ and $^3F_4$ $s\bar{s}$ meson decay widths using
the six combinations of model, wavefunction, and phase space/normalization
discussed in Section~\ref{3:parameters}.  In order to judge the reliability of
the results, it is useful to examine how well our calculations do for similar
mesons.  We do not have any good candidates for the $^3F_2$, but the
$f_4(2050)$ and $K_4^*(2045)$ differ only in flavour from the $^3F_4$
$s\bar{s}$.  Examining Figures~\ref{f:fits1}, \ref{f:fits3} and \ref{f:fits5},
we see that for the five decays in question, the choice of phase
space/normalization makes a significant difference: RPSN tends to overestimate
the partial widths, while KIPSN tends to underestimate them.  For the
$K_3^*(1780)$, which one would expect to behave similarly to the $^3F_4$ states
(since both are $^3L_{L+1}$), the situation is unfortunately less clear.  We
can also examine the total width of the $f_4(2050)$ and $K_4^*(2045)$ states.
In Tables~\ref{t:f42050decays} and \ref{t:K42045decays} we give the partial
widths for all significant, kinematically-allowed decays and their sums, in
the $^3P_0$ model and with SHO wavefunctions only.  It can be seen that once
again the RPSN results are high, and the KIPSN results low.  The main
conclusion we can draw from these results is that the widths of the $^3F_4$
$s\bar{s}$ meson probably lie between the RPSN and KIPSN estimates but it is
difficult to guess if they are closer to the lower or upper value.

\begin{table}[t]
\vspace{-0.3cm}
\begin{center}
\begin{tabular}{|l|c|c|c|} \hline
& $\Gamma$ (MeV) & \multicolumn{2}{c|}{$\Gamma$ (MeV) from Models} \\ 
\cline{3-4}
\hspace{0.5cm} Decay & from & \multicolumn{2}{c|}{$^3P_0$} \\ \cline{3-4}
& Experiment & \multicolumn{2}{c|}{SHO} \\ \cline{3-4}
$K_4^*(2045) \to$ & & \hspace{0.25cm} RPSN \hspace{0.25cm} & KIPSN\\ \hline\hline
\hspace{0.5cm} $[K \pi]_{0,G}$ & $19.6 \pm 3.8 $& 55 & 13 \\
\hspace{0.5cm} $[K \rho]_{1,G}$ & & 19 &  4.4 \\
\hspace{0.5cm} $[K b_1(1235)]_{1,F}$ & & 4.9 & 2.2 \\
\hspace{0.5cm} $[K a_1(1260)]_{1,F}$ & & 1.3 &  0.6 \\
\hspace{0.5cm} $[K a_2(1320)]_{2,F}$ & & 2.2 &  1.0 \\
\hspace{0.5cm} $[K^*(892) \pi]_{1,G}$ & & 23 & 5.5 \\
\hspace{0.5cm} $\left. \begin{array}{l}
\!\!\![K^*(892) \rho]_{2,D} \\
\!\!\![K^*(892) \rho]_{2,G} 
\end{array} \right\} $ &
$18 \pm 10$ $^{\rm a}$ &
$\begin{array}{c}
76 \\
2.1
\end{array} $ &
$\begin{array}{c}
18 \\
0.5
\end{array}$ \\
\hspace{0.5cm} $[K_1(1270) \pi]_{1,F}$ & & 1.6 & 0.7 \\
\hspace{0.5cm} $[K_1(1400) \pi]_{1,F}$ & & 5.3 & 2.6 \\
\hspace{0.5cm} $[K_2^*(1430) \pi]_{2,F}$ & & 5.2  & 2.6 \\
\hspace{0.5cm} $[K\eta']_{0,G}$ & & 3.3  & 0.9 \\
\hspace{0.5cm} $[K\omega]_{1,G}$ & & 6.0  & 1.4 \\
\hspace{0.5cm} $[K\phi]_{1,G}$ & & 1.1  & 0.4 \\
\hspace{0.5cm} $[K h_1(1170)]_{1,F}$ & & 2.9 & 1.3 \\
\hspace{0.5cm} $[K f_2(1270)]_{2,F}$ & & 1.3 & 0.6 \\
\hspace{0.5cm} $[K^*(892) \eta]_{1,G}$ & & 4.9 & 1.4 \\
\hspace{0.5cm} $[K^*(892) \omega]_{2,D}$ & & 24 & 5.7 \\
\hspace{0.5cm} $[K^*(892) \phi]_{2,D}$ & $2.8 \pm 1.4$ & 3.2 & 1.0 \\ \hline
Total / $\sum_i\Gamma_i$ & $198 \pm 30$ & 247 & 65 \\ \hline
\end{tabular}
\end{center}
\vspace{-0.4cm}
\caption[The calculated partial decay widths and total width of the 
$K_4^*(2045)$ meson.  The available experimental results
are also shown]{The calculated partial decay widths and total width of the 
$K_4^*(2045)$ meson.  The available experimental results
\cite{montanet94:review} are also shown.  For additional comments, see 
Table~\ref{t:f42050decays}. \\
$^{\rm a}$ This number is actually for the final state $K^*(892) \pi\pi$, 
and is the total for all partial waves.}
\label{t:K42045decays}
\end{table}

\begin{table}[p]
\newcommand{\BA}{@{\vline}}
\newcommand{\SP}{\hspace*{0.15cm}}
\vspace{-0.3cm}
\begin{center}
\begin{tabular}{|l\BA c\BA c\BA c\BA c\BA c\BA c\BA} \hline
& \multicolumn{6}{c|}{$\Gamma$ (MeV) from Models of Meson Decay} \\ \cline{2-7}
\hspace{0.5cm} Decay & \multicolumn{2}{c|}{$^3P_0$} & \multicolumn{4}{c|}
{Flux-tube Breaking} \\ \cline{2-7}
& \multicolumn{2}{c|}{SHO} & \multicolumn{2}{c|}{SHO} & 
\multicolumn{2}{c|}{RQM} \\ \cline{2-7}
$^3F_2 \;\;s\bar{s} \to$ & \SP RPSN\SP & \SP KIPSN\SP & \SP RPSN\SP & 
\SP KIPSN\SP & \SP RPSN\SP & \SP KIPSN \\ \hline \hline
\hspace{0.5cm} $[K \bar{K}]_{0,D} $ & 51 & 12 & 47 & 12 & 101 & 23 \\
\hspace{0.5cm} $[K_r \bar{K}+{\rm c.c.}]_{0,D} $ $^{\rm a}$ & 2.9 & 1.5 & 0.9 
& 0.5 & 25 & 12 \\
\hspace{0.5cm} $[K^*(892) \bar{K} +{\rm c.c.}]_{1,D} $ & 108 & 26 & 107 & 26 
& 165 & 38 \\
\hspace{0.5cm} $[K^*(1410) \bar{K}+{\rm c.c.}]_{1,D} $& 2.6 & 1.3 & 0.6 & 
0.3 & 4.0 & 1.9 \\
\hspace{0.5cm} $[K_1(1270) \bar{K}+{\rm c.c.}]_{1,P} $ & 445 & 187 & 449 & 194 
& 1072 & 426 \\
\hspace{0.5cm} $[K_1(1270) \bar{K}+{\rm c.c.}]_{1,F} $ & 25 & 11 & 27 & 12 & 
41 & 16 \\
\hspace{0.5cm} $[K_1(1400) \bar{K}+{\rm c.c.}]_{1,P} $ & 14 & 6.3 & 15 & 6.9 & 
29 & 12 \\
\hspace{0.5cm} $[K_1(1400) \bar{K}+{\rm c.c.}]_{1,F} $ & 0.8 & 0.4 & 1.0 & 0.4 
& $\sim 0$ & $\sim 0$ \\
\hspace{0.5cm} $[K_2^*(1430) \bar{K}+{\rm c.c.}]_{2,P} $ & 54 & 24 & 55 & 25 & 
112 & 47 \\
\hspace{0.5cm} $[K_2^*(1430) \bar{K}+{\rm c.c.}]_{2,F}$ & 9.6 & 4.3 & 10 & 4.7
& 22 & 9.1 \\
\hspace{0.5cm} $[K^*(892) \bar{K}^*(892)]_{0,D}$ & 24 & 5.7 & 24 & 5.9 & 39 
& 8.9 \\
\hspace{0.5cm} $[K^*(892) \bar{K}^*(892)]_{2,D}$ & 14 & 3.3 & 14 & 3.4 & 23 
& 5.1 \\
\hspace{0.5cm} $[K^*(892) \bar{K}^*(892)]_{2,G}$ & 48 & 12 & 52 & 13 & 83 & 
19 \\
\hspace{0.5cm} $[K_1(1270) \bar{K}^*(892) +{\rm c.c.}]_{1,P}$ \SP & 99 & 40 & 
102 & 42 & 209 & 79 \\
\hspace{0.5cm} $[K_1(1270) \bar{K}^*(892) +{\rm c.c.}]_{1,F}$ & 0.5 & 0.2 & 
0.6 & 0.2 & 1.1 & 0.4 \\
\hspace{0.5cm} $[K_1(1270) \bar{K}^*(892) +{\rm c.c.}]_{2,P}$ & 33 & 13 & 34 
& 14 & 70 & 26 \\
\hspace{0.5cm} $[K_1(1270) \bar{K}^*(892) +{\rm c.c.}]_{2,F}$ & 0.8 & 0.3 & 
0.9 & 0.4 & 1.8 & 0.7 \\
\hspace{0.5cm} $[\eta \eta]_{0,D}$ & 14 & 3.3 & 13 & 3.2 & 20 & 4.4 \\
\hspace{0.5cm} $[\eta' \eta]_{0,D}$ & 29 & 7.0 & 29 & 7.2 & 29 & 6.6 \\
\hspace{0.5cm} $[f_1(1510) \eta]_{1,P}$ & 45 & 22 & 46 & 24 & 92 & 43 \\
\hspace{0.5cm} $[f_2'(1525) \eta]_{2,P}$ & 14 & 6.9 & 14 & 7.3 & 29 & 14  \\
\hspace{0.5cm} $[\eta' \eta']_{0,D}$ & 6.6 & 1.6 & 6.7 & 1.7 & 4.9 & 1.1 \\
\hspace{0.5cm} $[\phi \phi ]_{0,D}$ & 3.9 & 1.2 & 3.9 & 1.3 & 5.5 & 1.6 \\
\hspace{0.5cm} $[\phi \phi ]_{2,D}$ & 2.2 & 0.7 & 2.3 & 0.7 & 3.1 & 0.9 \\
\hspace{0.5cm} $[\phi \phi ]_{2,G}$ & 1.0 & 0.3 & 1.0 & 0.3 & 1.1 & 0.3 \\ 
\hline
$\sum_i \Gamma_i$ & 1046 & 391 & 1058 & 406 & 2181 & 797 \\ \hline
\end{tabular}
\end{center}
\vspace{-0.4cm}
\caption[The calculated partial decay widths and total width of the 
$^3F_2 \;\;s\bar{s}$\protect\\ meson]{The calculated partial decay widths and
total width of the $^3F_2 \;\;s\bar{s}$ meson.  We do not include a decay to
$f_0(980) f_0(980)$ because we question its assignment as a $^3P_0$ meson.  At
a more likely mass for the $^3P_0$ $s \bar{s}$ meson, this decay is
kinematically inaccessible.  For additional comments, see
Table~\ref{t:f42050decays}. \\ $^{\rm a}$ $K_r$ is our notation for the first
radial excitation ($2^1S_0$) of the $K$.}
\label{t:3F2decays}
\end{table}

\begin{table}[t]
\newcommand{\BA}{@{\vline}}
\newcommand{\SP}{\hspace*{0.15cm}}
\vspace{-0.3cm}
\begin{center}
\begin{tabular}{|l\BA c\BA c\BA c\BA c\BA c\BA c\BA} \hline
& \multicolumn{6}{c|}{$\Gamma$ (MeV) from Models of Meson Decay} \\ \cline{2-7}
\hspace{0.5cm} Decay & \multicolumn{2}{c|}{$^3P_0$} & \multicolumn{4}{c|}
{Flux-tube Breaking} \\ \cline{2-7}
& \multicolumn{2}{c|}{SHO} & \multicolumn{2}{c|}{SHO} & 
\multicolumn{2}{c|}{RQM} \\ \cline{2-7}
$^3F_4 \;\;s\bar{s} \to$ & \SP RPSN\SP & \SP KIPSN\SP & \SP RPSN\SP & 
\SP KIPSN\SP & \SP RPSN\SP & \SP KIPSN \\ \hline \hline
\hspace{0.5cm} $[K \bar{K}]_{0,G} $ & 118 & 29 & 125 & 31 & 62 & 14 \\
\hspace{0.5cm} $[K_r \bar{K}+{\rm c.c.}]_{0,G} $ & 0.7 & 0.4 & 0.4 & 0.2 &
        2.4 & 1.2 \\
\hspace{0.5cm} $[K^*(892) \bar{K} +{\rm c.c.}]_{1,G}$ & 107 & 27 & 115 & 29 & 
112 & 26 \\
\hspace{0.5cm} $[K^* (1410) \bar{K} +{\rm c.c.}]_{1,G}$\SP & 1.7 & 0.9 & 
        0.8 & 0.4 & 5.0 & 2.4\\
\hspace{0.5cm} $[K_1(1270) \bar{K}+{\rm c.c.}]_{1,F}$ & 6.4 & 2.8 & 7.0 &
        3.1 & 10 & 4.2 \\
\hspace{0.5cm} $[K_1(1270) \bar{K}+{\rm c.c.}]_{1,H}$ & 1.3 & 0.6 & 
        1.4 & 0.6 & 3.7 & 1.5 \\
\hspace{0.5cm} $[K_1(1400) \bar{K}+{\rm c.c.}]_{1,F}$ & 14 & 6.4 & 15 & 
         7.0 & 29 & 12 \\
\hspace{0.5cm} $[K_2^*(1430) \bar{K}+{\rm c.c.}]_{2,F}$ & 15 & 7.0 & 16 & 7.7 
& 35 & 15 \\
\hspace{0.5cm} $[K^*(892) \bar{K}^*(892)]_{0,G}$ & 2.1 & 0.5 & 2.3 & 0.6 & 
4.3 & 1.0 \\
\hspace{0.5cm} $[K^*(892) \bar{K}^*(892)]_{2,D} $ & 181 & 44 & 184 & 46 & 312 
& 72 \\
\hspace{0.5cm} $[K^*(892) \bar{K}^*(892)]_{2,G} $ & 8.2 & 2.0 & 8.9 & 2.2 & 
17 & 3.9 \\
\hspace{0.5cm} $[\eta \eta]_{0,G} $ & 14 & 3.5 & 15 & 3.9 & 5.0 & 1.2 \\
\hspace{0.5cm} $[\eta' \eta]_{0,G} $ & 6.9 & 1.7 & 7.5 & 1.9 & 2.4 & 0.6 \\
\hspace{0.5cm} $[\phi \phi ]_{2,D} $ & 20 & 6.6 & 21 & 7.1 & 31 & 9.5 \\ \hline
$\sum_i \Gamma_i$ & 498 & 132 & 522 & 142 & 633 & 166 \\ \hline
\end{tabular}
\end{center}
\vspace{-0.4cm}
\caption[The calculated partial decay widths and total width of the 
$^3F_4 \;\;s\bar{s}$\protect\\ meson]{The calculated partial decay widths and
total width of the $^3F_4 \;\;s\bar{s}$ meson.  For additional comments, see
Tables~\ref{t:f42050decays} and \ref{t:3F2decays}. }
\label{t:3F4decays}
\end{table}

\begin{table}[t]
\vspace{-0.3cm}
\begin{center}
\begin{tabular}{|l|c|c|} \hline
& \multicolumn{2}{c|}{$\Gamma$ (MeV) from Models} \\ 
\cline{2-3}
\hspace{0.5cm} Decay & \multicolumn{2}{c|}{$^3P_0$} \\ \cline{2-3}
& \multicolumn{2}{c|}{SHO} \\ \cline{2-3}
$^3F_3 \;\;s\bar{s} \to$ & \hspace{0.25cm} RPSN \hspace{0.25cm} & KIPSN\\ 
\hline\hline
\hspace{0.5cm} $[K^*(892) \bar{K} +{\rm c.c.}]_{1,D}$ & 154 & 38 \\
\hspace{0.5cm} $[K^*(892) \bar{K} +{\rm c.c.}]_{1,G}$ & 98 & 24 \\
\hspace{0.5cm} $[K^* (1410) \bar{K} +{\rm c.c.}]_{1,D}$ & 2.5 & 1.3 \\ 
\hspace{0.5cm} $[K^* (1410) \bar{K} +{\rm c.c.}]_{1,G}$ & 2.2 & 1.1 \\ 
\hspace{0.5cm} $[K_0^*(1430) \bar{K}+{\rm c.c.}]_{0,F}$ & 5.3 & 2.4 \\ 
\hspace{0.5cm} $[K_1(1270) \bar{K}+{\rm c.c.}]_{1,F}$ & 71 & 30 \\ 
\hspace{0.5cm} $[K_2^*(1430) \bar{K}+{\rm c.c.}]_{2,P}$ & 305 & 137 \\ 
\hspace{0.5cm} $[K^*(892) \bar{K}^*(892)]_{2,D}$ & 66 & 16 \\ 
\hspace{0.5cm} $[K^*(892) \bar{K}^*(892)]_{2,G}$ & 27 & 6.6 \\ 
\hspace{0.5cm} $[K_1(1270) \bar{K}^*(892) +{\rm {\rm c.c.}}]_{2,P}$ & 157 & 
63\\ 
\hspace{0.5cm} $[f_2'(1525) \eta]_{2,P}$ & 73 & 37 \\ 
\hspace{0.5cm} $[\phi \phi ]_{2,D}$ & 10 & 3.2\\ \hline
$\sum_i \Gamma_i$ & 974 & 360 \\ \hline
\end{tabular}
\end{center}
\vspace{-0.4cm}
\caption[The calculated partial decay widths and total width of the 
$^3F_3 \;\;s\bar{s}$\protect\\ meson]{The calculated partial decay widths and
total width of the $^3F_3 \;\;s\bar{s}$ meson.  We do not include decays to
$f_0(980)
\eta$ or $f_0(980)\eta'$ because we question 
the assignment of the $f_0(980)$ as a $^3P_0$ meson.  At a more 
likely mass for the 
$^3P_0$ $s \bar{s}$ meson, this 
decay is kinematically inaccessible.  For additional comments, see 
Table~\ref{t:f42050decays}. }
\label{t:3F3decays}
\end{table}

Turning now to our results for the $^3F_2$ and $^3F_4$ $s\bar{s}$ mesons, in
Tables~\ref{t:3F2decays} and \ref{t:3F4decays} we give the partial widths for
all significant, kinematically-allowed decays and their sums, for all six
combinations of decay model, wavefunction, and phase space/normalization.  For
a degree of completeness, in Table~\ref{t:3F3decays} we also give the results
for the other $L=3$, spin-triplet $s\bar{s}$ meson, the $^3F_3$, in the $^3P_0$
model and with SHO wavefunctions only.

For the $^3F_2$ $s\bar{s}$ state the main decay modes are, in descending order,
including charge conjugate pairs:
\begin{eqnarray*}
&&K_1(1270)\bar{K},\; K_1(1270)\bar{K}^*(892),\;  K^*(892)\bar{K},\;  
K^*(892)\bar{K}^*(892), \\  
&&K_2^*(1430)\bar{K},\;  K\bar{K},\; f_1(1510)\eta.
\end{eqnarray*}
It is clearly a mistake to neglect the decays to an $L=1$ meson and a $K$ or
$\eta$; by far the largest partial width is for the decay to
$K_1(1270)\bar{K}$, and many of the other decays involving $L=1$ mesons are
significant.  The results using RQM wavefunctions are surprisingly large --
about double those using SHO wavefunctions.  Turning to the results with SHO
wavefunctions, we see that if we believe that the KIPSN results give a lower
bound for the $^3F_2$ as well as the $^3F_4$, we can expect a total width
\approxge~400~MeV, in keeping with na\"{\i}ve expectations.
Even if this width is too large by a factor of 2, the $f_4(2220)$ cannot be the
$^3F_2$ $s\bar{s}$ meson.  Indeed, the $^3F_2$ $s\bar{s}$ may well be too broad
to be found experimentally.

For the $^3F_4$ $s\bar{s}$ state the main decay modes are, in descending order,
including charge conjugate pairs:
\[
K^*(892)\bar{K}^*(892), \; K\bar{K}, \; K^*(892)\bar{K}, \; \phi\phi.
\]
Note that the total width of the $^3F_4$ $s\bar{s}$ is quite a bit less than
that of the $^3F_2$ $s\bar{s}$ .  This can be understood in terms of the
orbital angular momentum between the mesons in the final state.  Because of the
higher spin of the $^3F_4$, it tends to yield final states with higher orbital
angular momentum in order to conserve angular momentum.  These decays are then
suppressed by the higher angular momentum barrier.  Examining
Table~\ref{t:3F2decays}, we see that the lowest orbital angular momentum final
states in $^3F_2$ $s\bar{s}$ decay are in P-waves.  All of these decays are
relatively broad but the decay to $K_1(1270)\bar{K}$ is the P-wave decay with
the largest available phase space.  In fact, one could almost order the P-wave
decays using phase space alone.  The analogous decay of the $^3F_4$ $s\bar{s}$
is in an F-wave and therefore is suppressed by a larger angular momentum
barrier.  The lowest angular momentum partial wave for $^3F_4$ $s\bar{s}$
decays is a D-wave which, although it has the largest partial width of all
$^3F_4$ $s\bar{s}$ decays, is still smaller than the P-wave $^3F_2$ $s\bar{s}$
decay.

Looking at the total width of the $^3F_4$ $s\bar{s}$ state, if we accept that
the true result lies between those obtained from RPSN and KIPSN, then we would
expect the width to be \approxge~140~MeV and \approxle~600~MeV.  While this
is not very specific, we are really interested in the lower bound, and whether
this could be the $f_4(2220)$.  It is possible to look at the behaviour of the
similar $f_4(2050)$ and $K_4^*(2045)$ (and perhaps $K_3^*(1780)$) decays in
selected model, wavefunction and phase space/normalization combinations.  For
example, in the $^3P_0$ model with SHO wavefunctions and RPSN, the decays to
two $^3S_1$ mesons are well predicted, indicating that the largest partial
width of the $^3F_4$ $s\bar{s}$ (to $K^*(892)\bar{K}^*(892)$) may well be close
to the calculated value.  On the other hand, the partial widths to two $^1S_0$
mesons may be less than predicted, and to a $^1S_0$ and a $^3S_1$ slightly
more.  Based on this, we think it possible, but unlikely, that the calculated
$^3F_4$ $s\bar{s}$ width is high by a factor of 2.  Indeed, we would expect the
actual width to be higher, i.e.\ $>~140$~MeV.  Unfortunately, the uncertainties
associated with the models we are using make it impossible to rule out the
identification of the $f_4(2220)$ as the $^3F_4$ $s\bar{s}$ meson with any
surety; however, we feel it is unlikely.

Aside from the total width of the $^3F_4$ $s\bar{s}$, the roughly flavour
symmetric nature of the $f_4(2220)$ decays observed at BES suggest that it is
not an $s\bar{s}$ meson, since production of $\pi\pi$ and $p\bar{p}$ would be
OZI-suppressed.  We thus propose a second explanation of what is being observed
in this mass region -- that two different hadron states are present: a narrow
state produced in $J/\psi$ radiative decay and a broader state seen in hadron
beam experiments.  The broader state would be identified with the $^3F_4$
$s\bar{s}$ state.  The measured width is consistent with our predictions and
the LASS Collaboration \cite{aston88:evidence} shows evidence that its quantum
numbers are $J^{PC}=4^{++}$.  We would then identify the narrow hadron state
observed in the gluon rich $J/\psi$ radiative decays as a glueball
\cite{chao95:xi}.  The narrow state is not seen in hadron beam production
because it is narrow, is produced weakly in these experiments through
intermediate gluons, and is hidden by the $s\bar{s}$ state.  Conversely, the
broader state is not seen in $J/\psi$ radiative decays because this mode
preferentially produces states with a high glue content.  Crucial to this
explanation is the experimental verification of the BES results on the flavour
symmetric couplings of the state produced in $J/\psi$ radiative decay and the
observation of other decay modes for the broader state (primarily
$K^*(892)\bar{K}^*(892)$ and $K^*(892)\bar{K}$) in addition to the theoretical
verification that the predicted tensor glueball is as narrow as the observed
width.

We conclude that the $f_4(2220)$ cannot be the $^3F_2$ $s\bar{s}$ meson, and
that taking all of the data into account, it is unlikely to be the $^3F_4$
$s\bar{s}$ meson alone.  Instead we suggest that the broad state seen in hadron
beam experiments is the $^3F_4$ $s\bar{s}$ meson, and the narrow state seen in
$J/\psi$ radiative decay is a glueball.  Further experimental data is needed to
finally identify the $f_4(2220)$.

\section{Constraining the $K_1$ Mixing Angle}
\label{3:k1}

As discussed in Appendix~\ref{a:mixings}, the $^1P_1$ and $^3P_1$ strange
mesons mix to produce the physical states $K_1(1270)$ and $K_1(1400)$.  A
possible source of the mixing can be seen in the spin-orbit portion of the
Hamiltonian of Godfrey and Isgur's relativized quark model
\cite{godfrey85:mesons}, described in Section~\ref{1:quarkmodel}.  We can
rewrite Eq.~\ref{spinorbit} as
\begin{equation}
H^{\rm SO}_{q\bar{q}} = H^{\rm SO+}_{q\bar{q}} + H^{\rm SO-}_{q\bar{q}} 
\end{equation}
where
\begin{eqnarray}
H^{\rm SO+}_{q\bar{q}} &=& \left[ -\frac{\alpha_s}{2r^3} \left(\frac{1}{m_q}+
\frac{1}{m_{\bar{q}}} \right) ^2 
\vec{F}_q \cdot \vec{F}_{\bar{q}}- \frac{1}{4r} \frac{\partial 
H_{q\bar{q}}^{\rm conf}}{\partial r}
\left(\frac{1}{m_q^2} + \frac{1}{m_{\bar{q}}^2} \right) \right]
(\vec{S}_q + \vec{S}_{\bar{q}} )  \cdot \vec{L}, \nonumber \\
H^{\rm SO-}_{q\bar{q}} &=& \left( -\frac{\alpha_s}{2r^3} 
\vec{F}_q \cdot \vec{F}_{\bar{q}}
- \frac{1}{4r} \frac{\partial 
H_{q\bar{q}}^{\rm conf}}{\partial r} \right)
\left( \frac{1}{m_q^2} - \frac{1}{m_{\bar{q}}^2} \right)
(\vec{S}_q - \vec{S}_{\bar{q}} )  \cdot \vec{L}. \label{spinorbitpm}
\end{eqnarray}

Using this quark model Hamiltonian, the mass formulae for the P-wave mesons are
\begin{eqnarray}
M(^3P_2) & = & M_0 +\frac{1}{4} \langle H^{\rm cont}_{q\bar{q}} \rangle - 
\frac{1}{10} \langle H^{\rm ten}_{q\bar{q}} \rangle + 
\langle H^{SO+}_{q\bar{q}} \rangle, \nonumber\\
\left(\!\!\begin{array}{c} M(^3P_1) \\ M(^1P_1) \end{array} \!\!\right)
& = & \left(\!\begin{array}{cc} M_0 + \frac{1}{4} \langle 
H^{\rm cont}_{q\bar{q}} 
\rangle + \frac{1}{2} \langle H^{\rm ten}_{q\bar{q}} \rangle - \langle 
H^{SO+}_{q\bar{q}} \rangle & \sqrt{2} \langle H^{SO-}_{q\bar{q}} \rangle \\
\sqrt{2} \langle H^{SO-}_{q\bar{q}} \rangle   &
M_0 -\frac{3}{4} \langle H^{\rm cont}_{q\bar{q}} \rangle \end{array} \!\right)
\left(\! \begin{array}{c} ^3P_1 \\ ^1P_1 \end{array} \!\right), \nonumber \\
M(^3P_0) & = & M_0 +\frac{1}{4} \langle H^{\rm cont}_{q\bar{q}} \rangle - 
\langle H^{\rm ten}_{q\bar{q}} \rangle -2  \langle H^{SO+}_{q\bar{q}} \rangle, 
\label{massmatrix}
\end{eqnarray}
where the $\langle H^i_{q\bar{q}} \rangle$ are the expectation values of the
spatial parts of the various terms of the Hamiltonian, $M_0$ is the center
of mass of the multiplet, and we have adopted a phase
convention corresponding to the order of coupling 
$\vec{L} \times \vec{S}_q \times \vec{S}_{\bar{q}}$.  It can be seen that $H^{\rm SO-}_{q\bar{q}}$ leads
to mixing between the $^1P_1$ and $^3P_1$ states.  However, $H^{\rm
SO-}_{q\bar{q}}$ is non-zero only if $m_q \neq m_{\bar{q}}$, so the
strange mesons are the lightest states where this mixing can occur.

If we use Eq.~\ref{mixingeg} to describe the mixing, we find that to
diagonalize the $^1P_1$--$^3P_1$ Hamiltonian we must have
\begin{equation}
\tan(2\theta_K)=-\frac{2\sqrt{2}\langle H^{SO-}_{q\bar{q}} \rangle}
{\langle H^{\rm cont}_{q\bar{q}} \rangle + 
\frac{1}{2} \langle H^{\rm ten}_{q\bar{q}} \rangle - \langle 
H^{SO+}_{q\bar{q}} \rangle }, \label{diagcond}
\end{equation}
where $\theta_K$ is the $K_1$ mixing angle.
In this particular quark model, the expectation values of the spatial parts of
the various terms for strange mesons are $M_0=1378$~MeV, $\langle H^{\rm
cont}_{q\bar{q}} \rangle=33$~MeV, $\langle H^{\rm ten}_{q\bar{q}}
\rangle=56$~MeV, $\langle H^{SO+}_{q\bar{q}} \rangle=47$~MeV and $\langle
H^{SO-}_{q\bar{q}} \rangle=-1$~MeV \cite{godfrey91:properties}.  This gives
rise to a prediction of $\theta_K=6^\circ$.

In this section we compare the predictions of the $^3P_0$ and flux-tube
breaking models of meson decay to experimental data for five decay widths and
two ratios of D to S amplitudes\footnote{By ``ratio of D to S amplitudes'' we
mean the ratio of the $L=2$ partial wave amplitude for a decay to the $L=0$
partial wave amplitude for the same decay.}, in order to determine $\theta_K$.
This mixing angle is of interest both for its own sake, and for what it can can
tell us about the quark model Hamiltonian.

The $K_1(1270)$ and $K_1(1400)$ can both decay to $K\rho$ and $K^*(892)\pi$;
the $K_1(1400)$ can also decay to $K\omega$.  All of the decays occur in the
partial waves $1,S$ and $1,D$ (in $S,L$ notation).  Although decays to other
final states are observed they lie below threshold and proceed
through the tails of the Breit-Wigner resonances making the calculations less
reliable.  In addition to these five decay widths, there are also experimental
results available for the ratio of D to S amplitudes for both $K_1(1270)$ and
$K_1(1400)$ decaying to $K^*(892)\pi$.

\begin{figure}[t]
\vspace{-2cm}
\begin{center}
\makebox{\epsfxsize=6.5in\epsffile{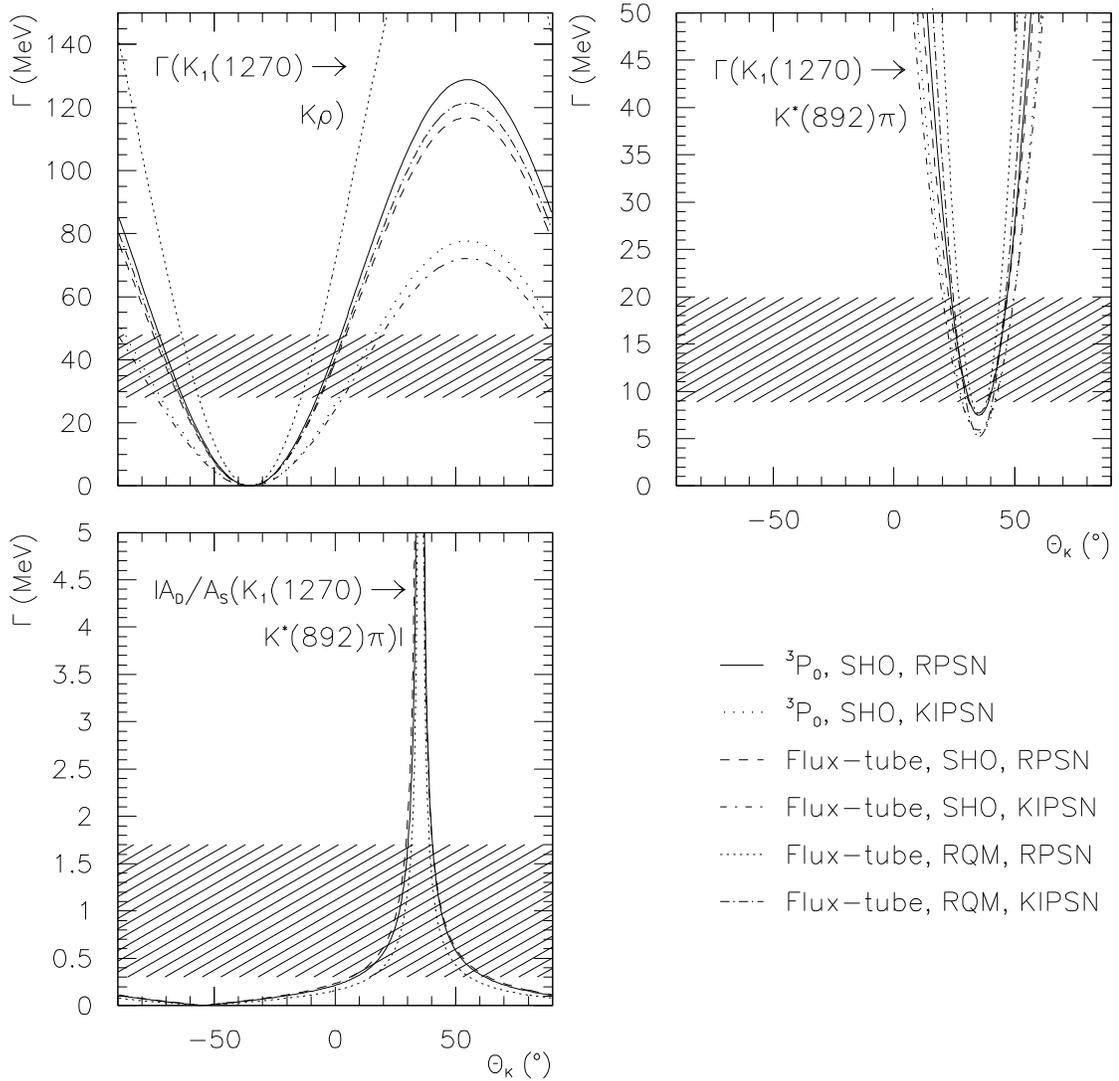}}
\end{center}
\vspace{-1.3cm}
\caption[Graphs of the $K_1(1270)$ partial decay widths and ratio of D to S 
amplitudes considered in this work, vs.\ the mixing angle $\theta_K$]{Graphs of
the $K_1(1270)$ partial decay widths and ratio of D to S amplitudes considered
in this work, vs.\ the mixing angle $\theta_K$.  The shaded regions are the
1--$\sigma$ limits on the experimental measurements of the quantities
\cite{montanet94:review}.  In the key the labels refer to the meson decay
model, the wavefunctions used, and the type of phase space/normalization used.
Only three curves are shown for $|A_D/A_S(K_1(1270)\to K^*(892)\pi)|$, because
the differences between RPSN and KIPSN cancel in the ratio of amplitudes,
making the RPSN and KIPSN curves for the same model and wavefunctions
identical.}
\label{f:k1cross1}
\end{figure}

\begin{figure}[t]
\vspace{-2cm}
\begin{center}
\makebox{\epsfxsize=6.5in\epsffile{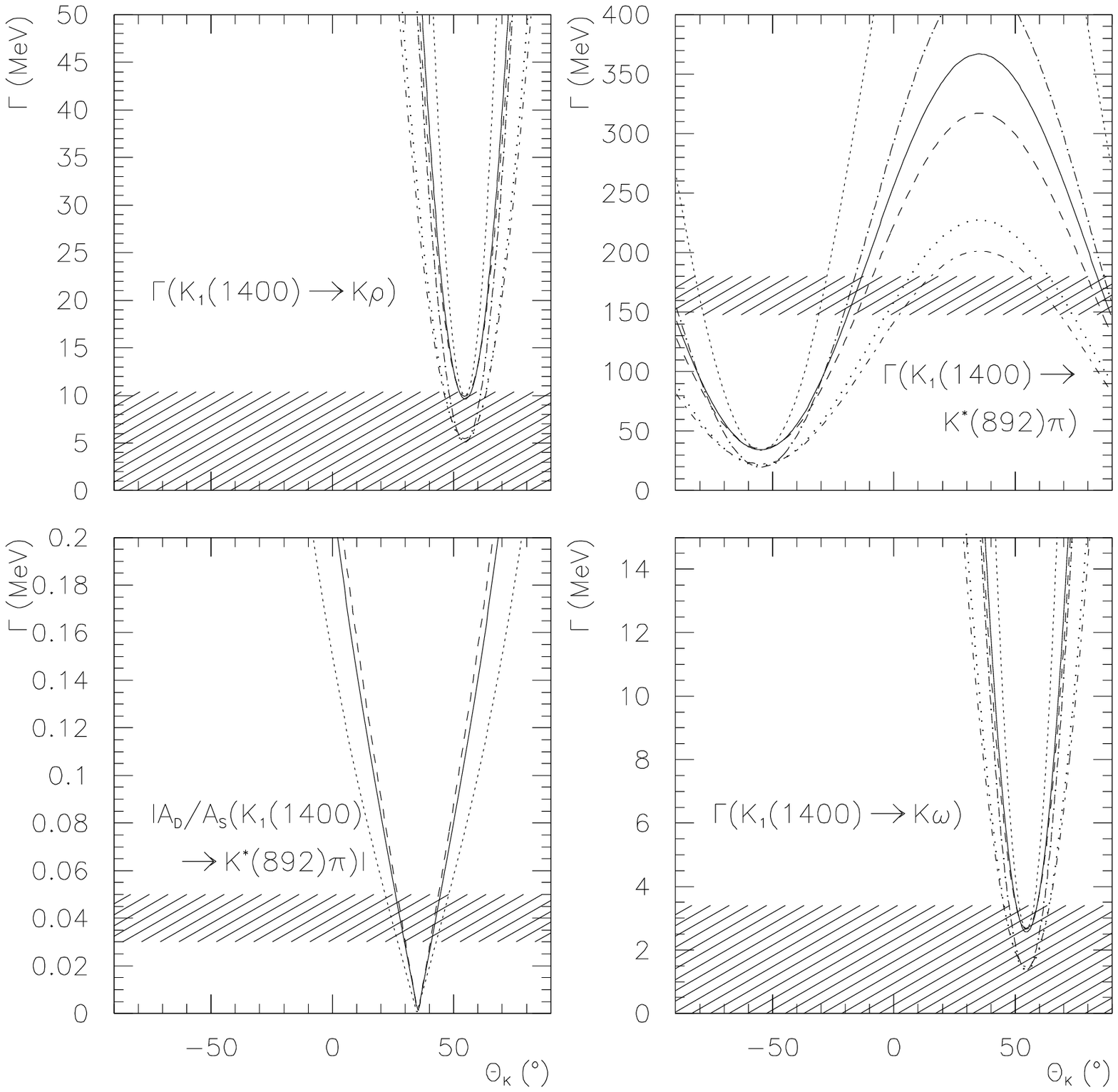}}
\end{center}
\vspace{-1.3cm}
\caption[Graphs of the $K_1(1400)$ partial decay widths and ratio of D to S 
amplitudes considered in this work, vs.\ the mixing angle $\theta_K$] {Graphs
of the $K_1(1400)$ partial decay widths and ratio of D to S amplitudes
considered in this work, vs.\ the mixing angle $\theta_K$.  See
Figure~\ref{f:k1cross1} for the explanation of the line labels and additional
notes.}
\label{f:k1cross2}
\end{figure}

We have calculated these seven quantities using the six combinations of model,
wavefunction, and phase space/normalization selected in
Section~\ref{3:parameters}.  The $K_1(1270)$ quantities are plotted as a
function of the $K_1$ mixing angle in Figure~\ref{f:k1cross1}, and the
$K_1(1400)$ quantities are plotted in Figure~\ref{f:k1cross2} -- the
experimental results are also shown in both figures.

In order to determine the best fit values of $\theta_K$, we have carried out a
least squares fit for each of the six combinations of decay model,
wavefunction, and phase space/normalization.  The best fit values of $\theta_K$
and their corresponding $\chi^2/{\rm dof}$ are shown in Table~\ref{t:k1fits},
as are the resulting values of each of the seven quantities being considered,
and their experimental values.  Examining the results, we see that as usual,
the values of $\chi^2/{\rm dof}$ are high, but not as high as were obtained in
the 28-decay fits of Section~\ref{3:parameters}.  This is not unexpected, since
the similar decay $b_1(1235)\to\omega\pi$ was described reasonably well in
those fits.  It is also clear that the SHO wavefunctions do better than the
RQM, and in terms of phase space/normalization, KIPSN does better than the
RPSN.

\begin{singlespace}
\begin{sidewaystable}[p]
\newcommand{\BA}{@{\vline}}
\newcommand{\SP}{\hspace*{0.1cm}}
\centering
\begin{tabular}{@{\vline\SP} l@{\SP\vline\SP}c@{\SP\vline}c\BA c\BA c\BA c\BA 
c\BA c\BA} \hline
& Value & \multicolumn{6}{c|}{Value Calculated from Models of Meson 
Decay} \\ \cline{3-8}
Quantity & from & \multicolumn{2}{c|}{$^3P_0$} & \multicolumn{4}{c|}
{Flux-tube Breaking} \\ \cline{3-8}
& Experiment & \multicolumn{2}{c|}{SHO} & \multicolumn{2}{c|}{SHO} & 
\multicolumn{2}{c|}{RQM} \\ \cline{3-8}
& &  RPSN &  KIPSN &  RPSN &  KIPSN &  RPSN &  KIPSN  \\ \hline \hline
Fitted mixing angle $\theta_K$ & & \SP$48.4\pm 1.1^\circ$\SP & 
\SP$45.2^{+1.4\,\circ}_{-1.3}$\SP & \SP$46.8^{+1.1\,\circ}_{-1.0}$\SP & 
\SP$44.2\pm 1.3^\circ$\SP & \SP$55.4^{+0.7\,\circ}_{-0.8}$\SP & 
\SP$50.6^{+1.3\,\circ}_{-1.2}$\SP \\ 
$\chi^2$/dof & & 38.9 & 5.3 & 26.0 & 3.3 & 231 & 51.2 \\ \hline
$\Gamma(K_1(1270) \to K \rho)$ & $38 \pm 10$
         & 127 & 76 & 115 & 70 & 213 & 121 \\
$\Gamma(K_1(1270) \to K^*(892) \pi)$ & $14.4 \pm 5.5$
         & 24 & 13 & 20 & 12 & 76 & 34 \\
$|A_D/A_S (K_1(1270) \to K^*(892) \pi)|$ & $1.0\pm 0.7$
         & 0.65 & 0.86 & 0.77 & 0.99 & 0.31 & 0.41 \\
$\Gamma(K_1(1400) \to K \rho)$ & $5.2 \pm 5.2$
         & 15 & 12 & 17 & 12 & 10 & 7.0 \\
$\Gamma(K_1(1400) \to K^*(892) \pi)$ & $164\pm 16$
         & 350 & 221 & 306 & 197 & 653 & 402 \\
$|A_D/A_S(K_1(1400) \to K^*(892) \pi)|$ & $0.04 \pm 0.01$
        & 0.071 & 0.054 & 0.069 & 0.053 & 0.080 & 0.061 \\ 
$\Gamma(K_1(1400) \to K \omega)$ & $1.7\pm 1.7$
        & 4.3 & 3.5 & 5.0 & 3.8 & 2.7 & 2.0 \\ \hline
\end{tabular}
\caption[The partial decay widths and ratios of D to S amplitudes used in our 
fits of the $K_1(1270)$ and $K_1(1400)$ mixing angle $\theta_K$] {The partial
decay widths and ratios of D to S amplitudes used in our fits of the
$K_1(1270)$ and $K_1(1400)$ mixing angle $\theta_K$.  The experimental values
\cite{montanet94:review} are shown, as are the model results for the six
combinations of decay model, wavefunctions, and phase space/normalization
(specified in the second, third and fourth rows of the heading, respectively).
For each combination we also give the fitted value of $\theta_K$, and the
corresponding $\chi^2$/dof.  The partial widths are given in units of MeV.  
All errors shown are 1--$\sigma$. }
\label{t:k1fits}
\end{sidewaystable}
\end{singlespace}
The resulting values of $\theta_K$ are all fairly similar, falling in the range
$44-56^\circ$.  The lowest values of $\chi^2/{\rm dof}$ occur for fits near
$44^\circ$, and so we take that as our approximate result.  What can this tell
us about the quark model Hamiltonian?  A mixing angle of $44^\circ$ implies
$\langle H^{SO-}_{q\bar{q}} \rangle=-142$~MeV, rather different from the quark
model value.  However, the value of $\langle H^{SO-}_{q\bar{q}}\rangle$ is very
sensitive to the angle in this region (e.g.\ $\theta_K=41^\circ$ implies
$\langle H^{SO-}_{q\bar{q}} \rangle=-35$~MeV, while $\theta_K=45^\circ$ implies
$\langle H^{SO-}_{q\bar{q}} \rangle=-\infty$), so the only conclusion we can
really draw is that the quark model value is very low, {\em if $\langle
H^{SO-}_{q\bar{q}}\rangle$ is responsible for all of the mixing}.  We cannot
rule out the possibility that some other interaction is partly responsible for
the mixing, such as mixing via common decay channels.

To consider the SO Hamiltonian further, define $\bar{m}=\frac{1}{2} (m_q +
m_{\bar{q}})$ and $\Delta = (m_{\bar{q}}-m_q)$.  Next expand $1/m_q=
1/\bar{m}+\Delta/(2\bar{m}^2)+...$ and $1/m_{\bar{q}}=
1/\bar{m}-\Delta/(2\bar{m}^2)+...$ to first order in $\Delta/\bar{m}$ (each of
these truncated results has less than an 11\% error for the quark masses we
used).  In terms of these, the contributions to the spin-orbit Hamiltonian
become
\begin{eqnarray}
H^{\rm SO+}_{q\bar{q}} &\simeq& \left( -\frac{2\alpha_s}{\bar{m}^2r^3} 
\vec{F}_q \cdot \vec{F}_{\bar{q}}- \frac{1}{2\bar{m}^2r} \frac{\partial 
H_{q\bar{q}}^{\rm conf}}{\partial r}\right)
(\vec{S}_q + \vec{S}_{\bar{q}} )  \cdot \vec{L}, \nonumber \\
H^{\rm SO-}_{q\bar{q}} &\simeq& \frac{\Delta}{\bar{m}}
\left( -\frac{\alpha_s}{\bar{m}^2r^3} 
\vec{F}_q \cdot \vec{F}_{\bar{q}}
- \frac{1}{2\bar{m}^2r} \frac{\partial 
H_{q\bar{q}}^{\rm conf}}{\partial r} \right)
(\vec{S}_q - \vec{S}_{\bar{q}} )  \cdot \vec{L}. \label{spinorbitpm}
\end{eqnarray}
The $\langle (\vec{S}_q\pm\vec{S}_{\bar{q}})\cdot\vec{L}\rangle$ factors are
roughly the same size, and the $\Delta/\bar{m}$ term is not too far from 1 (it
is $\geq\frac{1}{2}$ for the quark masses we used), so the factors in the large
parentheses contribute at the same order of magnitude.  Note that the
contributions from the Thomas precession term are equal, but those from the
colour magnetic term differ by a factor of 2.  The model gives a large $\langle
H^{\rm SO+}_{q\bar{q}} \rangle$ of 47~MeV, but a small $\langle H^{\rm
SO-}_{q\bar{q}} \rangle$ of $-1$~MeV arising from a delicate cancellation of
the colour magnetic and Thomas precession terms.  It is thus conceivable that a
small change in the model could lead to a substantially larger $\langle H^{\rm
SO-}_{q\bar{q}} \rangle$, so we are unable to draw any conclusions about
whether the mixing is largely due to another mechanism, or the quark model
Hamiltonian just needs some retuning.

\chapter{Effects of Final State Interactions}
\markright{Chapter 4.  Effects of Final State Interactions}
\label{4:finalstate}

\section{Introduction to Final State Interactions}

Final state interactions (FSI's) refer to additional interactions between
particles in the final state that are not included in the basic
interaction.\footnote{ The techniques introduced here could also be used to
describe additional interactions between particles in the initial state.
However, since most physically interesting problems involve FSI's, the term
``final state interactions'' has become the name of choice.}  For example, in
Chapter \ref{5:ggpp}, we consider the basic interaction $\gamma \gamma \to
\pi^+ \pi^-$, which proceeds through a scalar QED vertex.  However, there will
also be a QCD interaction between the final state pions.  In Chapter
\ref{5:ggpp}, we attempt to calculate the effects of the FSI for
this process.  In this chapter, we introduce the techniques that we
will apply to the problem of FSI's
\cite{gillespie64:final}.

The techniques involve nonrelativistic quantum mechanics (including the
Schr\"{o}\-ding\-er equation), corrected to include relativistic phase space
where possible.  The basic interaction may still be calculated
relativistically: e.g.\ $\gamma \gamma \to \pi \pi$.  This mixture of
nonrelativistic and relativistic elements is typical of the quark model.

\section{Solving the Schr\"{o}dinger Equation with the FSI Potential}

Our first step in calculating the effects of an FSI is to find the wavefunction
describing the state of the final state particles due to the presence of the
FSI only.  We assume that the FSI can be described by a potential between the
final state particles, and that the wavefunction can be found by solving the
Schr\"{o}dinger equation.

The relative wavefunction for two bodies in a potential that only
depends on their relative position $\vec{r} \equiv \vec{r}_1-\vec{r}_2$ can be 
found from the Schr\"{o}dinger 
equation for a central-force potential
\begin{equation}
\left[\nabla^2_{\vec{r}}-2\mu V(r)+k^2\right]\psi(\vec{r})=0 \label{swe}
\end{equation}
where $\mu$ is the reduced mass of the particles, $V(r)$ is the potential 
between the two particles and $k$ is the relative momentum between the particles in the absence of a 
potential, given by $k^2\equiv 2\mu E$, where $E$ is the kinetic energy of the 
system.  Recall that we are using the 
particle physics convention that $c \equiv h\!\!\!^- \equiv 1$.

If we now select a solution with a particular orbital angular momentum
and set $\psi_{lm}(\vec{r}) \equiv R_l(r) \,Y_{lm}(\Omega_r)$, then 
$R_{l}(r)$ satisfies the radial equation
\begin{equation}
\left[ \frac{1}{r} \frac{{\rm d}^2}{{\rm d}r^2} r -\frac{l(l+1)}{r^2}-2\mu
V(r)+k^2\right]R_{l}(r) = 0. \label{radial1}
\end{equation}
If we further set $R_l(r) \equiv y_l(r)/r$, then $y_l(r)$ satisfies the 
equivalent radial equation 
\begin{equation}
\left[\frac{{\rm d}^2}{{\rm d}r^2} -\frac{l(l+1)}{r^2}-2\mu
V(r)+k^2\right]y_l(r) = 0. \label{radial2}
\end{equation}

The solution of Eq.~\ref{swe} for $V(r)=0$ is just a plane wave, which can be
broken up into its partial waves,
\begin{equation}
\phi_{\vec{k}}(\vec{r}) \equiv \dspexp{i\vec{k}\cdot\vec{r}} = 4 \pi 
\sum_{l=0}^\infty
\sum_{m_l=-l}^l i^l j_l(kr) \,Y_{lm_l}^*(\Omega_k) \,Y_{lm_l}(\Omega_r), 
\label{planewave}
\end{equation}
where $j_l(kr)$ is a spherical Bessel function and $\vec{k}$ gives the 
direction of the incident particle.  The equivalent solution of 
Eq.~\ref{radial1} for $V(r)=0$ (picking the only particular solution that is 
regular at the origin) is
\begin{equation}
R_l(r) = j_l(kr) \stackrel{kr\rightarrow \infty}{\longrightarrow} \frac{1}{kr}
\sin{(kr-{\scriptstyle\frac{1}{2}}l\pi)}.
\end{equation}

For the case with a potential, the stationary scattering waves
$\psi_{\vec{k}}^\pm(\vec{r})$ are solutions of Eq.~\ref{swe}.  Their asymptotic
forms consist of the incoming plane wave, and either an outgoing ($+$) or
incoming ($-$) spherical wave.  We will be interested in
$\psi_{\vec{k}}^-(\vec{r})$ because it will be describing the final state, and
hence must be propagated backwards in time as an incoming spherical wave to be
projected onto the initial state plane waves at an equal time.  In
nonrelativistic collision theory
\cite{messiah:quantum} it is defined to have the following asymptotic form
\begin{equation}
\psi_{\vec{k}}^-(\vec{r}) \stackrel{r\rightarrow \infty}{\longrightarrow}
\dspexp{i\vec{k}\cdot\vec{r}} + f_{\vec{k}}^-(\Omega_r) \:
\frac{\dspexp{- ikr}}{r}
\end{equation}
where $f_{\vec{k}}^-(\Omega_r)$ is called the scattering amplitude and 
describes the effect of the scattering.  

By symmetry, $\psi_{\vec{k}}^-(\vec{r})$ and $f_{\vec{k}}^-(\Omega_r)$ will be 
symmetric about the incident axis.  It is then 
reasonable to choose to set the incident axis along $\hat{z}$ (i.e.\ $\vec{k} =
k \hat{z}$), so that neither function is dependent on $\phi_r$.  We can then
make the following expansions in terms of Legendre polynomials
\begin{eqnarray}
\psi_{\vec{k}}^-(r,\theta_r) &=& \sum_{l=0}^\infty 
\frac{y_l(r)}{r} P_l(cos{\theta_r}), \nonumber \\
f_{\vec{k}}^-(\theta_r) &=& \sum_{l=0}^\infty f_l^- P_l(cos{\theta_r}).
\end{eqnarray}
where $y_l(r)$ is a regular solution of the equivalent radial equation 
(Eq.~\ref{radial2}), and has the following (un-normalized) asymptotic 
form\footnote{We make
the usual assumptions about the potential:  that it is analytic in the vicinity
of the origin, that ${\displaystyle \lim_{r\rightarrow 0}} \:r^2 V(r)=0$ and 
that 
${\displaystyle \lim_{r\rightarrow \infty}} \:r V(r)=0$.}
\begin{equation}
y_l(r) \stackrel{r\rightarrow \infty}{\longrightarrow} a_l 
\sin{(kr-{\scriptstyle\frac{1}{2}}l\pi+\delta_l)}.
\end{equation}

We can then match the two asymptotic forms of $\psi_{\vec{k}}^-(r,\theta_r)$
to obtain $a_l$ and $f_l^-$.  Then applying the spherical harmonic addition
theorem to the Legendre polynomial gives for the stationary scattering wave 
(with incoming spherical wave) solution of Eq.~\ref{swe}
\begin{equation}
\psi_{\vec{k}}^-(\vec{r}) = 4 \pi \sum_{l=0}^\infty
\sum_{m_l=-l}^l i^l \dspexp{-i\delta_l} \,u_l(k,r) \,Y_{lm_l}^*(\Omega_k) 
\,Y_{lm_l}(\Omega_r) \label{fsisol}
\end{equation}
where $u_l(k,r)$ is the regular solution of Eq.~\ref{radial1} with asymptotic 
form
\begin{equation}
u_l(k,r) \stackrel{r\rightarrow \infty}{\longrightarrow} \frac{1}{kr}
\sin{(kr-{\scriptstyle\frac{1}{2}}l\pi+\delta_l)}.
\end{equation}

\section{The Fermi Approximation}
\label{4:fermiapprox}

The simplest way to perform a correction due to FSI's is to use the Fermi
approximation \cite{fermi51:elementary}.  We assume that the FSI interaction is
very short range, and hence that only the value at the origin of the relative
wavefunction of the final state particles is significant.  Since the
wavefunction enters into the amplitude, the S-wave cross-section would be
enhanced by a factor of $|u_0(k,0)/j_0(0)|^2 = |u_0(k,0)|^2$, and the
cross-sections of other partial waves would be unaffected.

\section[Applying FSI's to the Flux-tube Breaking Model of Meson Decay]
{Applying FSI's to the Flux-tube Breaking\protect\\ Model of Meson Decay}

Geiger and Swanson \cite{geiger94:distinguishing} have applied FSI's to the
flux-tube breaking model of meson decay.  Although we do not apply their
methodology in this work, we give the details here for completeness and
because: it connects the two topics of this thesis (meson decay and FSI's), we
are aware of no {\em detailed} published discussion of the method, and our
derivation includes factors neglected by Geiger and Swanson.

Recognizing that the last exponential function of Eq.~\ref{fluxintegral} is the
wavefunction between the final state particles in the absence of an FSI
potential (Eq.~\ref{planewave}), Geiger and Swanson replace it with the
wavefunction in the presence of such a potential (Eq.~\ref{fsisol}).  By doing
so they are using a final state in which the FSI effects are included, instead
of simple plane waves in which the effects are ignored.  This should give a
final result that includes the FSI effects.

Let us first examine the partial wave amplitudes from the flux-tube breaking 
model in the absence of FSI's.  In Section~\ref{2:fluxcalc}, we used the 
Jacob-Wick formula to convert to partial-wave amplitudes (see 
Appendix~\ref{a:partialwaves}) -- here we use the recoupling 
calculation because we have a need for the 
$\int\!{\rm d}\Omega\; Y^*_{L M_L}\!(\Omega)$ integral.  
This gives us, from 
Eqs.~\ref{amplitude} and \ref{fluxintegral}, using Eqs.~\ref{lsymmetry} and 
\ref{recoupling}
\begin{eqnarray}
\lefteqn{M^{S L}(P) = \gamma_0\:\sqrt{8 E_A E_B E_C} \!\!
\sum_{\renewcommand{\arraystretch}{.5}\begin{array}[t]{l}
\scriptstyle M_{L_A},M_{S_A},M_{L_B},M_{S_B},M_{J_B},\\
\scriptstyle M_{L_C},M_{S_C},M_{J_C},M_L,M_S,m
\end{array}}\renewcommand{\arraystretch}{1} \!\!\!
\langle L M_L S M_S|J_A M_{J_A} \rangle } \nonumber \\
&& \times \langle J_B M_{J_B} J_C M_{J_C} | S M_S \rangle \:
\langle L_A M_{L_A} S_A M_{S_A} | J_A  (M_{J_B}\!+\!M_{J_C})\rangle \nonumber\\
&& \times \langle L_B M_{L_B} S_B M_{S_B} | J_B M_{J_B} \rangle\:
\langle L_C M_{L_C} S_C M_{S_C} | J_C M_{J_C} \rangle \:
\langle 1m\,1\!-\!\!m|00 \rangle \nonumber \\
&& \times \langle \chi^{1 4}_{S_B M_{S_B}} \chi^{3 2}_{S_C M_{S_C}} |
\chi^{1 2}_{S_A M_{S_A}} \chi^{3 4}_{1 -\!m} \rangle
\left[ \langle \phi^{1 4}_B \phi^{3 2}_C | \phi^{1 2}_A 
\phi^{3 4}_0 \rangle\:
I^{\rm FSI}(P,m_1,m_2,m_3) \right. \nonumber \\
&& +(-1)^{L_A+L_B+L_C+S_A+S_B+S_C} \:
\langle \left. \phi^{3 2}_B \phi^{1 4}_C | \phi^{1 2}_A \phi^{3 4}_0 \rangle\:
I^{\rm FSI}(P,m_2,m_1,m_3) \right]
\end{eqnarray}
where
\begin{eqnarray}
\lefteqn{I^{\rm FSI}(P,m_1,m_2,m_3)=-\frac{8}{(2\pi)^{\frac{3}{2}}}
\int\!{\rm d}^3\vec{r}\int\!{\rm d}^3\vec{w}\;\: 
\psi^*_{n_B L_B M_{L_B}}\!
(-\vec{w}-\vec{r})\;
\psi^*_{n_C L_C M_{L_C}}\!
(\vec{w}-\vec{r}) } \nonumber\\
&&\times \dspexp{-{\scriptstyle\frac{1}{2}}b w_{\rm min}^2}
\int\!{\rm d}\Omega_P\; Y^*_{L M_L}\!(\Omega_P)\;
{\cal Y}^m_1\!\!\left(\left[\left(\vec{P}+i\vec{\nabla}_{\vec{r}_A}\right)
\psi_{n_A L_A M_{L_A}}\!(\vec{r}_A)\right]_{\vec{r}_A=-2\vec{r}}\right)
\;\;\;\;\;\;\;\; \nonumber \\
&&\times \dspexp{i\vec{P}\cdot\left(m_+\vec{r} + m_-\vec{w}\right)}. 
\label{fsiint}
\end{eqnarray}
Note that in evaluating $M^{S L}(P)$ we can pick any value of $M_{J_A}$; 
alternatively, we could
sum over $M_{J_A}$ and divide by $(2 J_A +1)$, on the right side.

Next, set 
\begin{equation}
\vec{P} \dspexp{i\vec{P}\cdot\left(m_+\vec{r} + m_-\vec{w}\right)} =
\left[-\frac{i}{m_+} \vec{\nabla}_{\vec{r}_{BC}} 
\dspexp{i\vec{P}\cdot\left(m_+\vec{r}_{BC} + m_-\vec{w}\right)} \right]
_{\vec{r}_{BC}=\vec{r}},
\end{equation}
then use an alternate form of Eq.~\ref{planewave} to set
\begin{equation}
\dspexp{i\vec{P}\cdot\left(m_+\vec{r}_{BC} + m_-\vec{w}\right)} = 
4 \pi \sum_{l=0}^\infty
\sum_{m_l=-l}^l i^l j_l(Ps) \,Y_{lm_l}(\Omega_P) \,Y_{lm_l}^*(\Omega_s), 
\end{equation}
where $\vec{s}=m_+\vec{r}_{BC} + m_-\vec{w}$, and finally, apply the 
orthogonality of spherical harmonics to obtain 
\begin{eqnarray}
\lefteqn{I^{\rm FSI}(P,m_1,m_2,m_3)=-\frac{8}{(2\pi)^{\frac{3}{2}}}
\int\!{\rm d}^3\vec{r}\int\!{\rm d}^3\vec{w}\;\: 
\psi^*_{n_B L_B M_{L_B}}\!
(-\vec{w}-\vec{r})\;
\psi^*_{n_C L_C M_{L_C}}\!
(\vec{w}-\vec{r}) } \nonumber\\
&&\times 
{\cal Y}^m_1\!\!\left(\left[\left(-\frac{i}{m_+} \vec{\nabla}_{\vec{r}_{BC}}
+i\vec{\nabla}_{\vec{r}_A}\right)
\psi_{n_A L_A M_{L_A}}\!(\vec{r}_A)\;
\psi_{(BC)LM_L}^*(\vec{r}_{BC},\vec{w})
\right]_{\vec{r}_A=-2\vec{r},\vec{r}_{BC}=\vec{r}}\right) \nonumber\\
&&\times \dspexp{-{\scriptstyle\frac{1}{2}}b w_{\rm min}^2}
\end{eqnarray}
where, in the absence of FSI's, 
\begin{equation}
\psi_{(BC)LM_L}(\vec{r}_{BC},\vec{w}) = 4 \pi \,i^{-L}\, j_L(Ps) \:
Y_{LM_L}(\Omega_s). \label{psibcnofsi}
\end{equation}

Eq.~\ref{rot1} can be used to break $Y_{LM}(\Omega_s)$ up into spherical 
harmonics with arguments $\Omega_{\vec{r}_{BC}}$ and $\Omega_{\vec{w}}$ and of
course, $s$ is given by 
\begin{equation}
s=\sqrt{m_+^2r_{BC}^2+m_-^2w^2+2m_+m_-r_{BC}w\cos{\theta}} 
\end{equation}
where $\theta$ is the angle between $\vec{r}_{BC}$ and $\vec{w}$.

The above is just another way of expressing the flux-tube breaking model of
meson decay of Section~\ref{2:fluxtube}.  However, we
can include the effects of FSI's by replacing the plane wave 
$e^{i\vec{P}\cdot\left(m_+\vec{r}_{BC} + m_-\vec{w}\right)}$ 
with $\psi_{\vec{k}}^-(\vec{r})$ 
(Eq.~\ref{fsisol}), instead of just expanding it with Eq.~\ref{planewave}.  We
then get, instead of Eq.~\ref{psibcnofsi},\footnote{Note that Geiger and 
Swanson \cite{geiger94:distinguishing} 
associate a different constant factor with their definition of 
$\psi_{(BC)LM_L}(\vec{r}_{BC},\vec{w})$, they neglect the phase $i^{-L}$, and 
they do not explicitly mention the factor of $\dspexp{i\delta_L}$.}
\begin{equation}
\psi_{(BC)LM_L}(\vec{r}_{BC},\vec{w}) = 4 \pi \,i^{-L}\, \dspexp{i\delta_L}
\:u_L(P,s) \:Y_{LM_L}(\Omega_s).
\end{equation}
In the absence of an FSI potential, $\delta_L\Rightarrow 0$, 
$u_L(P,s)\Rightarrow j_L(Ps)$, and we recover Eq.~\ref{psibcnofsi}.

\section{Applying FSI's to QED Amplitudes}
\label{4:fullfsi}

In order to develop the expression for applying FSI corrections to QED
amplitudes, we first find the expression for correcting amplitudes in
nonrelativistic collision theory \cite{messiah:quantum}, in the Born
approximation.  Once the expression for the correction is derived, we use the
cross-section to relate the collision theory amplitude to the QED amplitude.
Applying the resulting relation gives us an expression for correcting QED
amplitudes for FSI effects.  We specialize to an interaction with two particles
in the final state.

In collision theory, consider the Hamiltonian $H=H_0+V$ where $H_0$ is the
kinetic energy term, and $V$ is the potential.  $V=W+U$ is divided into the
potential representing the basic interaction, $W$, and the FSI potential $U$.
The plane waves $\phi_{\vec{k}}(\vec{r})$ are the solutions of the
Schr\"{o}dinger equation in the absence of a potential,
\begin{equation}
(H_0-E)\:\phi_{\vec{k}}(\vec{r}) =0,
\end{equation}
the stationary scattering waves $\chi_{\vec{k}}^\pm(\vec{r})$ are the 
solutions with the full potential,
\begin{equation}
(H_0+W+U-E)\:\chi_{\vec{k}}^\pm(\vec{r}) = 0,
\end{equation}
and the stationary scattering waves $\psi_{\vec{k}}^\pm(\vec{r})$ are the 
solutions with the FSI potential only,
\begin{equation}
(H_0+U-E)\:\psi_{\vec{k}}^\pm(\vec{r}) = 0.
\end{equation}
The $\chi_{\vec{k}}^\pm(\vec{r})$ are also solutions of the equivalent 
integral equation, 
\begin{equation}
\chi_{\vec{k}}^\pm = \left(1+\frac{1}{E-H_0-V\pm i\varepsilon}\;
V\right)\phi_{\vec{k}}.
\end{equation}

The regular transition amplitude is defined in terms of the transition 
operator between
plane waves, or in terms of the potential between a plane wave and a stationary
scattering wave:
\begin{equation}
\langle f|T|i\rangle \equiv \langle \phi_{\vec{k}_f}|T|\phi_{\vec{k}_i} \rangle
=\langle \phi_{\vec{k}_f}|V|\chi_{\vec{k}_i}^+ \rangle
=\langle \chi_{\vec{k}_f}^-|V|\phi_{\vec{k}_i} \rangle.
\end{equation}
In the Born approximation, we assume that V is sufficiently small that 
$\chi_{\vec{k}}^\pm(\vec{r}) \simeq \phi_{\vec{k}}(\vec{r})$ and we can
replace $\chi_{\vec{k}}^\pm(\vec{r})$ by $\phi_{\vec{k}}(\vec{r})$ in the 
expression for $\langle f|T|i\rangle$:
\begin{equation}
\langle f|T|i\rangle \simeq \langle f|T|i\rangle^{\rm BORN} \equiv 
\langle \phi_{\vec{k}_f}|V|\phi_{\vec{k}_i} \rangle.
\end{equation}
We will use the Born approximation from this point on, but will drop the 
$^{\rm BORN}$ notation for convenience.

Now consider the problem of FSI's.  We have a very good theory (QED) that will
calculate the transition amplitude due to just the basic interaction between
plane waves, $\langle f|T|i\rangle=\langle \phi_{\vec{k}_f}|W|\phi_{\vec{k}_i}
\rangle$, but we need to find a way to include the FSI effects (the potential
U).  To do this we use a final state that includes the FSI effects,
$\psi_{\vec{k}}^-(\vec{r})$, so only the potential $W$ is left between the
initial and final states:
\begin{equation}
\langle f|T|i\rangle ^{\rm FSI} = 
\langle \psi_{\vec{k}_f}^-|W|\phi_{\vec{k}_i} \rangle.
\end{equation}
Next, we make use of the plane wave orthogonality and closure conditions
\begin{equation}
\langle\phi_{\vec{k}} |\phi_{\vec{k'}} \rangle = (2\pi)^3 
\delta^3(\vec{k}-\vec{k'}),  \nonumber 
\end{equation}
\vspace*{-1cm}
\begin{equation}
\frac{1}{(2\pi)^3}\int\!{\rm d}^3\vec{k} \:| \phi_{\vec{k}} \rangle \:
\langle \phi_{\vec{k}} | = I 
\end{equation}
to obtain
\begin{eqnarray}
\langle f|T|i\rangle ^{\rm FSI} &=& 
\frac{1}{(2\pi)^3}\int\!{\rm d}^3\vec{k} \;\langle \psi_{\vec{k}_f}^-| 
\phi_{\vec{k}} \rangle \:
\langle \phi_{\vec{k}} |W|\phi_{\vec{k}_i} \rangle \nonumber \\
&=& \frac{1}{(2\pi)^{\frac{3}{2}}}\int\!{\rm d}^3\vec{k} \;\psi_{\vec{k}_f}^{-*}(\vec{k})\:
\langle f|T|i\rangle, \label{fsiderivstep}
\end{eqnarray}
where $\psi_{\vec{k}_f}^{-*}(\vec{k})$ is the complex conjugate of the
momentum-space solution of the Schr\"{o}\-dinger equation (Eq.~\ref{swe}) with
the FSI potential $U$.

To be able to apply this to QED amplitudes, we need to know how the 
nonrelativistic scattering amplitudes relate to QED amplitudes.  A 
comparison of the expressions for the total cross-section 
reveals that $|{\cal M}|^2 \propto s|\langle f|T|i\rangle|^2$ where $s$ is the 
Mandelstam variable.\footnote{In the CM frame, the total energy is given by
$\sqrt{s}$.}  Fourier-transforming Eq.~\ref{fsisol} to momentum space, 
we find that the QED 
amplitude corrected for FSI effects is
\begin{eqnarray}
{\cal M}^{\rm FSI}(\vec{k}_f)&=& \frac{2}{\pi} \sqrt{s(k_f)} \:
\sum_{l=0}^{\infty} \sum_{m_l=-l}^l 
\dspexp{i\delta_l} \:Y_{lm_l}(\Omega_{k_f}) 
\int\!{\rm d}^3\vec{k} \:\frac{{\cal M}(\vec{k})}{\sqrt{s(k)}} 
\:Y_{lm_l}^*(\Omega_k) \nonumber \\
&&\times\int_0^\infty \!{\rm d}r \:r^2 j_l(kr) \:u_l(k_f,r)
\end{eqnarray}
where $s$ and ${\cal M}$ have been written as functions of the momenta in order to 
show
whether they are calculated with the variable of integration, $\vec{k}$, 
or the relative 
momentum between the final state particles in the CM frame, $\vec{k_f}$.  
There is 
ambiguity as to how to relate $k$ and $s(k)$ -- nonrelativistically or 
relativistically.  We follow the usual practice of keeping as much as possible
relativistic, and so use for equal-mass final state particles: 
$s(k) = 4k^2+4m^2$.

It is worthwhile to expand each ${\cal M}$ into partial waves,
\begin{equation}
{\cal M}(\vec{k})\equiv \sum_{L=0}^{\infty} \,\sum_{M_L=-L}^L f_{LM_L}(s(k)) \:
Y_{LM_L}(\Omega_k)
\end{equation}
in order to get the FSI correction for a single partial wave
\begin{equation}
f_{LM_L}^{\rm FSI}(s(k_f))=\frac{2}{\pi} \sqrt{s(k_f)} \;\dspexp{i\delta_L} 
\int_0^\infty \!{\rm d}k \int_0^\infty \!{\rm d}r \:r^2\: k^2 \:
\frac{f_{LM_L}(s(k))}{\sqrt{s(k)}} \,j_L(kr) \:u_L(k_f,r). \label{fcorrection}
\end{equation}
In the absence of an FSI potential, $\delta_L\Rightarrow 0$, 
$u_L(k_f,r)\Rightarrow j_L(k_f\,r)$, the orthogonality condition of the 
spherical
Bessel functions gives us a delta function, and we obtain 
$f_{LM_L}^{\rm FSI}(s(k_f))=f_{LM_L}(s(k_f))$ as we should.

\chapter{Final State Interactions:  An Application}
\markright{Chapter 5.  Final State Interactions:  An Application}
\label{5:ggpp}

\section{The Interaction $\gamma \gamma \rightarrow \pi \pi$}

In 1986, a group at Orsay (using DM1 on DCI) measured the cross-section of
$\gamma \gamma \rightarrow \pi^+ \pi^-$ near the $\pi$ $\pi$ threshold, and
found a cross-section about twice that expected from scalar QED\footnote{Scalar
QED refers to a subset of QED that deals with the interactions of just photons
and scalar (spin 0) particles.  Because the pions really consist of two
spin-$\frac{1}{2}$ quarks, scalar QED is used here as an effective theory that
is valid at low energies because the long wavelength photons cannot make out
the individual quarks.}
\cite{courau86:lepton}.  Although statistically not very significant (a
$2\sigma$ effect), the result aroused interest because it differed from
theoretical expectations.  By Low's low energy theorem
\cite{low54:unknown} the interaction of low energy photons depends only on the
static properties of the target.  Because at low energies the long wavelength
photons cannot make out the individual constituents of the pions, the
pion-photon coupling depends only on the charge of the pion, and the pion can
be treated as a point particle in scalar QED.  The strong final state
interactions (FSI's) between the outgoing pions were not expected to be as
significant as the electroweak effects.

\begin{figure}[t]
\vspace{-2cm}
\begin{center}
\makebox{\epsfxsize=6.5in\epsffile{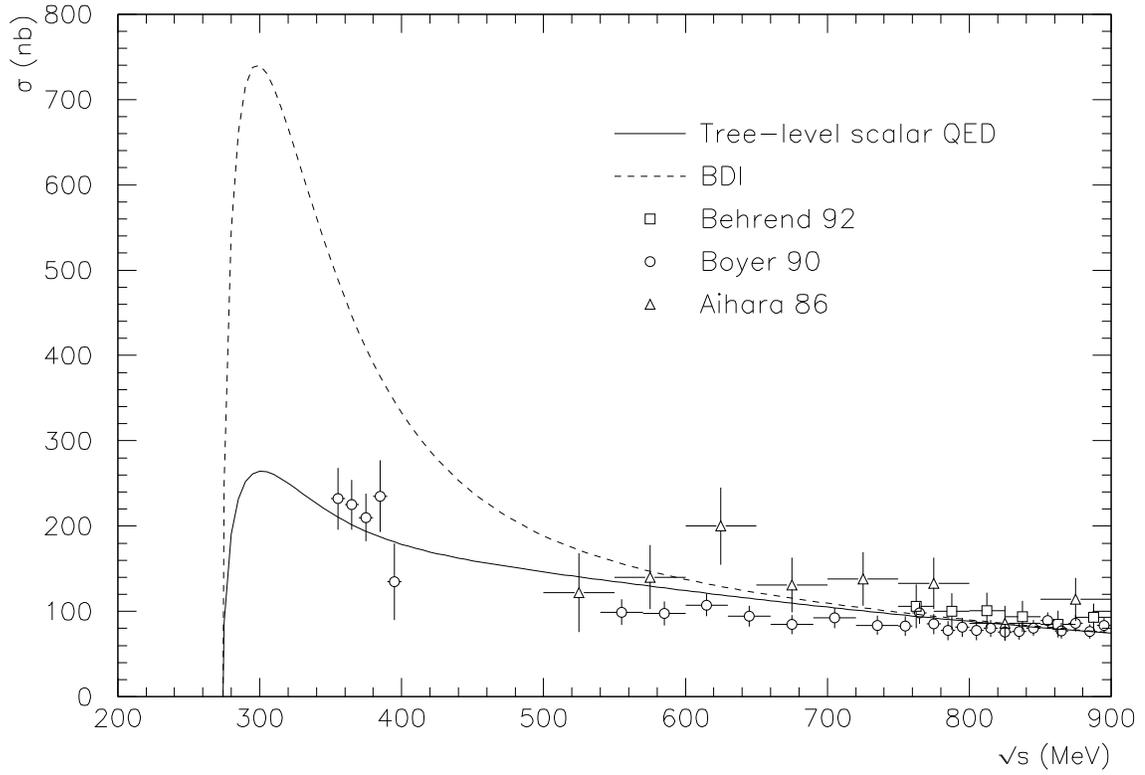}}
\end{center}
\vspace{-1.3cm}
\caption[Graph of the $\gamma \gamma \to \pi^+ \pi^-$ total cross-section vs.\ 
$\protect\sqrt{s}$, for the tree-level scalar QED and BDI predictions, and
recent data] {Graph of the $\gamma \gamma \to \pi^+ \pi^-$ total cross-section
vs.\ $\protect\sqrt{s}$, for the tree-level scalar QED and BDI predictions
(reproduced by us), and recent data.  The curves have been corrected to have a
limited polar acceptance to match the data: $|\cos{\theta}| \leq 0.6$.  The
data is from Behrend {\it et al.}\
\cite{behrend92:experimental}, Boyer {\it et al.}\ \cite{boyer90:two} and 
Aihara {\it et al.}\ \cite{aihara86:pion}.  Some of the data shown in this
chapter was obtained by us from References~\cite{durhamRALHEP} and
\cite{morgan94:compilation}.  The horizontal error bars on the data show the
bin sizes; for the vertical error bars all of the given errors were added
together in quadrature.}
\label{f:BDIcross}
\end{figure}
In 1987, Barnes, Dooley and Isgur (BDI) \cite{barnes87:final-state} estimated
the effects of these FSI's using the Fermi approximation (see Section 
\ref{4:fermiapprox}) with an effective potential extracted from a
quark model, and found a similar enhancement.  Unfortunately, the experimental
enhancement has since disappeared; the early experiments were apparently
plagued by problems with subtracting the lepton background due to low
statistics \cite{morgan91:is}, and the later experiments show no such
enhancement.  Figure~\ref{f:BDIcross} shows the two curves and the latest
experimental data.  Note that the data fits the tree-level scalar QED curve
well for the whole range shown (up to a CM energy of 900 MeV), even though we
would not necessarily expect it to be accurate at the higher energies.

Calculations of the $\gamma \gamma \to \pi \pi$ cross-sections have been done
using other methods.  One-loop chiral perturbation theory has produced mixed
results:  the $\gamma \gamma \to \pi^+ \pi^-$ cross-section curve
\cite{bijnens88:two} shows a small enhancement over the tree-level scalar QED
curve near threshold that is still compatible with the latest experimental
data, while the $\gamma \gamma \to \pi^0 \pi^0$ cross-section curve
\cite{bijnens88:two,donoghue88:reaction} has a shape that is rather different
from the data.  However, a two-loop chiral perturbation theory calculation of
the $\gamma \gamma \to \pi^0 \pi^0$ cross-section
\cite{bellucci94:low-energy} agrees with the data.  The difference is 
apparently mainly due to $\pi \pi$ rescattering and renormalization of the pion
decay constant.

Calculations of the $\gamma \gamma \to \pi \pi$ cross-sections have also been
done with dispersion relations, typically using phase shift data from
$\pi$--$\pi$ scattering, and constraints from unitarity, analyticity and
crossing.  The $\gamma \gamma \to \pi^+ \pi^-$ cross-section results are again
quite similar to both the tree-level scalar QED curve and the experimental data
near threshold
\cite{morgan87:low,donoghue93:photon}.  The $\gamma \gamma \to \pi^0 \pi^0$ 
cross-section results also do well \cite{morgan91:is,donoghue93:photon},
showing good agreement with the data.  It is believed that higher order
exchange effects, whose absence is problematic for the one-loop chiral
perturbation theory calculation, are included in the dispersive calculation
through the unitarity requirement \cite{donoghue93:photon}.

On the face of it, the failure of the quark model to do as well as these other
two approaches represents quite a blow to its reputation.  Being able to
predict the effects of FSI's is important if we are to understand QCD.  In
addition, if we understood the situation in $\gamma \gamma \rightarrow$ two
pseudoscalar mesons, we could then approach the situation in $\gamma \gamma
\rightarrow$ two vector mesons with more confidence; the structures seen in the
cross-sections of these processes are not well understood and are a
long-standing puzzle.  

However, there is room for improvement in the BDI calculation.  We will redo
the calculation using the full FSI apparatus of Section~\ref{4:fullfsi}, and
newer, better potentials.  In addition, we will calculate the cross-section for
the process $\gamma \gamma \rightarrow \pi^0 \pi^0$.  Since neutral pions do
not couple to long wavelength photons (there is no such process in scalar QED),
$\gamma \gamma \to \pi^0 \pi^0$ is expected to be suppressed relative to
$\gamma \gamma \to \pi^+ \pi^-$ making the effects of the FSI's relatively much
more significant.  In our calculation, it is the FSI potentials that mix the
$\pi^+ \pi^-$ final state into $\pi^0 \pi^0$.

In Section~\ref{5:scalarQED} we calculate the cross-section for $\gamma \gamma
\to \pi^+ \pi^-$ in tree-level scalar QED.  In Section~\ref{5:applyingFSI} we 
discuss the effective potentials and the application of the full FSI apparatus
(see Section~\ref{4:fullfsi}) to this problem.  In Section~\ref{5:results} we
give the results of our calculations.

\section{$\gamma \gamma \to \pi^+ \pi^-$ in Tree-level Scalar QED}
\label{5:scalarQED}

\begin{figure}[t]
\vspace{-0.7cm}
\begin{center}
\makebox{\epsfxsize=6.0in\epsffile{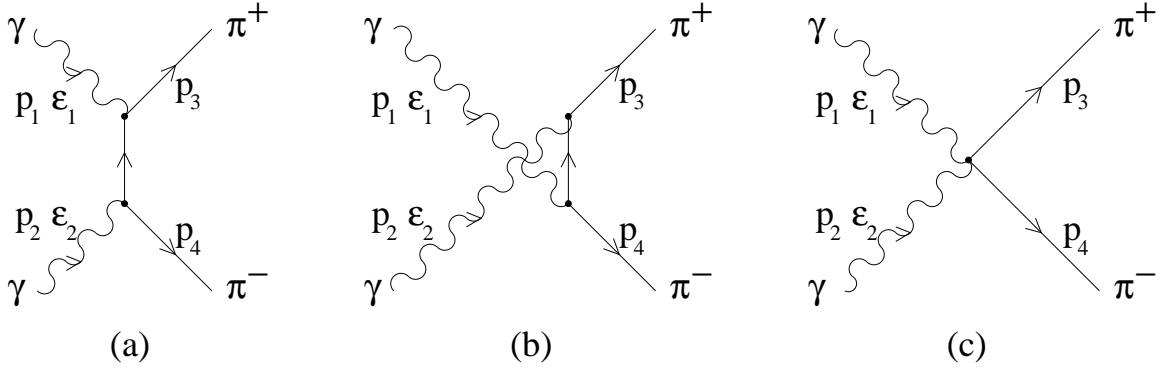}}
\end{center}
\vspace{-0.5cm}
\caption[The tree-level Feynman diagrams for the scalar QED 
interaction $\gamma \gamma \to \pi^+ \pi^-$] {The tree-level Feynman diagrams
for the scalar QED interaction $\gamma \gamma \to \pi^+ \pi^-$.  The labels of
the particle momenta ($p_i$) and photon polarization vectors ($\epsilon_i$)
are shown.}
\label{f:ggppdiag}
\end{figure}

The three tree-level scalar QED Feynman diagrams\footnote{See
Appendix~\ref{a:feynmandiag} for a brief explanation of Feynman diagrams.}
contributing to the interaction $\gamma \gamma \to \pi^+ \pi^-$ are shown in
Figure~\ref{f:ggppdiag}.

The corresponding Feynman
amplitude (using the momenta and photon polarization labels shown in 
Figure~\ref{f:ggppdiag}) is
\begin{equation}
{\cal M} = 4\pi\alpha \left[\frac{\epsilon_1\cdot(2p_3-p_1)\;
\epsilon_2\cdot(2p_4-p_2)}{(p_2-p_4)^2-m^2}+
\frac{\epsilon_1\cdot(2p_4-p_1)\;
\epsilon_2\cdot(2p_3-p_2)}{(p_1-p_4)^2-m^2}+2\epsilon_1\cdot\epsilon_2 \right]
\end{equation}
where $\alpha$ is the fine structure constant and $m$ is the pion mass.
Averaging the photon polarizations and integrating over phase space gives us
the total cross-section
\begin{equation}
\sigma = \frac{\pi\alpha^2}{4m^2}\left[2x(1+x)\sqrt{1-x} - x^2(2-x)
\ln\left(\frac{1+\sqrt{1-x}}{1-\sqrt{1-x}}\right)\right] \label{ggpptotcross}
\end{equation}
where $x\equiv \frac{4m^2}{s}$ and $s$ is the Mandelstam variable.

\begin{figure}[t]
\vspace{-0.5cm}
\begin{center}
\makebox{\epsfysize=1.5in\epsffile{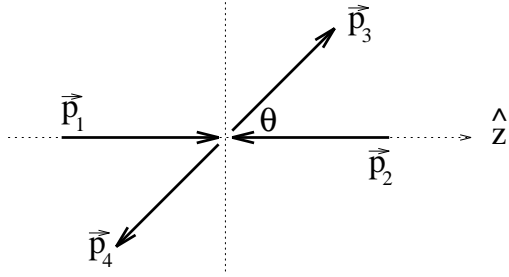}}
\end{center}
\vspace{-0.5cm}
\caption{The geometry of the $\gamma \gamma \to \pi^+ \pi^-$ interaction
in the CM frame.}
\label{f:ggppgeom}
\end{figure}

We would now like to work out the cross-sections corresponding to different 
partial waves in the final state.  In the CM frame, we take $\vec{p}_1$ to 
point in the positive $z$ direction, and the direction of $\vec{p}_3$ to be 
described by the spherical polar coordinates $\theta$ and $\phi$, as shown in  
Figure~\ref{f:ggppgeom}.
For the photons we take circular polarizations:
\begin{eqnarray}
\epsilon_1(+)=\epsilon_2(-)&=&\frac{1}{\sqrt{2}}(0,-1,-i,0), \nonumber \\
\epsilon_1(-)=\epsilon_2(+)&=&\frac{1}{\sqrt{2}}(0,1,-i,0). 
\end{eqnarray}

\begin{figure}[t]
\vspace{-2cm}
\begin{center}
\makebox{\epsfxsize=6.5in\epsffile{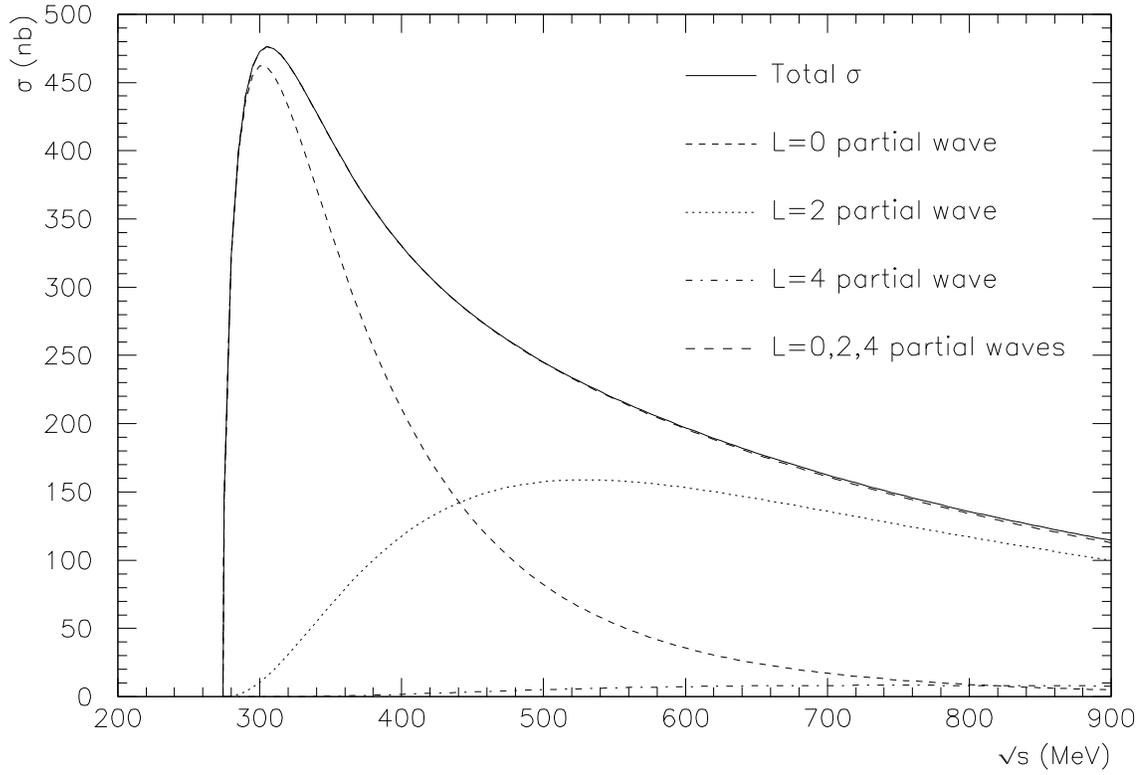}}
\end{center}
\vspace{-1.3cm}
\caption[Graph of the $\gamma \gamma \to \pi^+ \pi^-$ tree-level scalar QED 
cross-sections vs.\ $\protect\sqrt{s}$]{Graph of the $\gamma \gamma \to \pi^+
\pi^-$ tree-level scalar QED cross-sections vs.\ $\protect\sqrt{s}$.  The total
cross-section is shown, as well as the cross-sections of the $L=0,2,4$ partial
waves, and the cross-section corresponding to the sum of the $L=0,2,4$ partial
waves.  The curves are for full polar acceptance: $|\cos{\theta}| \leq 1$,
which is why they differ from those shown in Figure~\ref{f:BDIcross}.}
\label{f:uncorrected}
\end{figure}

We calculate the helicity amplitudes
\begin{eqnarray}
{\cal M}_{++}={\cal M}_{--} &=& \frac{8\pi \alpha \,x}{1-(1-x)\cos^2{\theta}},
\nonumber \\
{\cal M}_{\pm \mp} &=& \frac{8\pi \alpha \:(1-x) \sin^2{\theta} \;
\dspexp{\pm 2i\phi}}{1-(1-x)\cos^2{\theta}},
\end{eqnarray}
and then expand them in a spherical harmonic basis to obtain the partial waves
(e.g.\ for the $++$ case),
\begin{eqnarray}
{\cal M}_{++}&\equiv& \sum_{L=0}^{\infty} \,\sum_{M_L=-L}^L f_{LM_L}^{++} \:
Y_{LM_L}(\theta,\phi) \nonumber\\
\Rightarrow\;\;\;\; f_{LM_L}^{++} &=& \int \!{\rm d}\Omega \;Y_{LM_L}^*
(\theta,\phi)\:
{\cal M}_{++}. \label{partwaves}
\end{eqnarray}
We find that $f_{LM_L}^{++}=f_{LM_L}^{--}$ is only non-zero for $M_L=0$ and
that $f_{LM_L}^{+-}=f_{L\,-\!M_L}^{-+}$ is only non-zero for
$M_L=2$.\footnote{The orbital angular momentum $L$ between the pions is even
because of charge conjugation invariance.}  The cross-section is then written
in terms of the partial waves as
\begin{equation}
\sigma = \frac{\sqrt{1-x}}{128\pi^2 s}\left[ (f_{00}^{++})^2 + \sum_{L \geq2,
{\rm even}}^\infty \left[(f_{L0}^{++})^2 + (f_{L2}^{+-})^2 \right] \right].
\label{ggppcross}
\end{equation}
The needed partial waves (see below) are given by
\begin{eqnarray}
f_{00}^{++}&=& 8\pi^\frac{3}{2}\alpha \frac{x}{\sqrt{1-x}}
\ln\left(\frac{1+\sqrt{1-x}}{1-\sqrt{1-x}}\right), \nonumber \\
f_{20}^{++}&=& 4\sqrt{5} \pi^\frac{3}{2}\alpha\frac{x}{1-x} \left[
-6+\frac{2+x}{\sqrt{1-x}}\ln\left(\frac{1+\sqrt{1-x}}{1-\sqrt{1-x}}\right)
\right], \nonumber \\
f_{22}^{+-}&=& 4\sqrt{\frac{15}{2}} \pi^\frac{3}{2}\alpha
\left[\frac{10}{3}-\frac{2}{1-x} +\frac{x^2}{(1-x)^\frac{3}{2}}
\ln\left(\frac{1+\sqrt{1-x}}{1-\sqrt{1-x}}\right)
\right], \nonumber \\
f_{40}^{++}&=& 3\pi^\frac{3}{2}\alpha \frac{x}{1-x}\left[\frac{110}{3}-
\frac{70}{1-x}+\frac{3x^2+24x+8}{(1-x)^\frac{3}{2}}
\ln\left(\frac{1+\sqrt{1-x}}{1-\sqrt{1-x}}\right)\right], \nonumber \\
f_{42}^{+-}&=& 3\sqrt{10}\pi^\frac{3}{2}\alpha
\left[\frac{-54}{5}+\frac{76}{3(1-x)}-\frac{14}{(1-x)^2}+\frac{x^2(6+x)}
{(1-x)^\frac{5}{2}}
\ln\left(\frac{1+\sqrt{1-x}}{1-\sqrt{1-x}}\right)
\right], \nonumber \\
f_{60}^{++}&=& 8\sqrt{13}\pi^\frac{3}{2}\alpha \frac{x}{1-x}
\left[-\frac{231}{40}+
\frac{7(4-15x)}{8(1-x)}
-\frac{21(1+5x+5x^2)}{8(1-x)^2} \right.\nonumber \\
&&\left.+\frac{16+120x+90x^2+5x^3}
{16(1-x)^\frac{5}{2}}\ln\left(\frac{1+\sqrt{1-x}}{1-\sqrt{1-x}}\right)
\right].
\end{eqnarray}
The total cross-section and the cross-sections of the $L=0,2,4$ partial waves 
are shown in Figure~\ref{f:uncorrected}.
 
\section{Applying the FSI Correction}
\label{5:applyingFSI}

\subsection{The Effective Potentials We Use}
\label{5:fsipot}

The $\pi$--$\pi$ potentials that we use were calculated by Swanson {\it et
al.}\
\cite{swanson92:inter-meson,li94:Iequals0,swanson96:private} using techniques 
developed by Barnes and Swanson \cite{barnes91:diagrammatic}.  The potentials
come from equating the T-matrix giving the leading order interactions in a
perturbation theory for a nonrelativistic quark model, to the T-matrix for the
interactions of point-like mesons, written in terms of the effective potential
they were trying to find.

The potentials are different for the different isospin states of the two pions.
Because of charge conjugation invariance, the orbital angular momentum $L$
between the two pions must be even.  Because of Bose-Einstein symmetry (which
holds for members of the same isospin multiplet in the context of the strong
interaction \cite{nguyen-khac68:applications}), the sum $L+I$ must be even, and
hence the total isospin $I$ must be even as well.  It then follows that the
isospin states (written as $|II_z\rangle$) of the pions are
\begin{eqnarray}
|\pi^+\pi^-\rangle &=& \sqrt{\frac{2}{3}}\:|00\rangle+\frac{1}{\sqrt{3}}\:|20
\rangle, \nonumber \\
|\pi^0\pi^0\rangle &=& -\frac{1}{\sqrt{3}}\:|00\rangle+\sqrt{\frac{2}{3}}\:|20
\rangle. \label{pidefns}
\end{eqnarray}

\begin{figure}[t]
\vspace{-0.5cm}
\begin{center}
\makebox{\epsfxsize=6.0in\epsffile{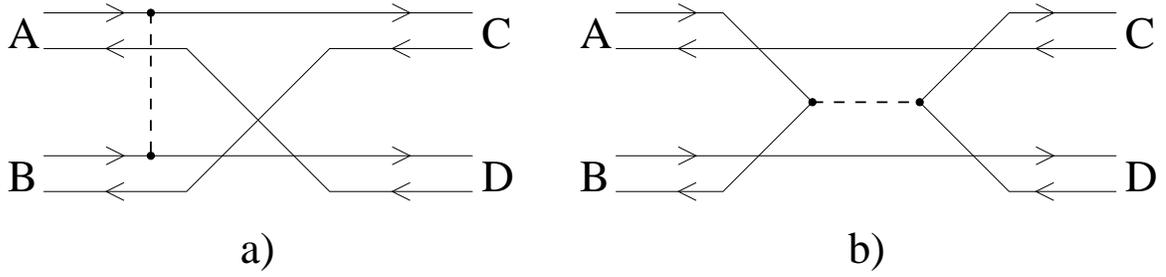}}
\end{center}
\vspace{-0.5cm}
\caption[Example diagrams of the two classes of interactions included in the 
quark model used to find the effective potentials]
{Example diagrams of the two classes of interactions included in the 
quark model used to find the effective potentials:  a) t-channel gluon 
exchange, b) s-channel gluon exchange.}
\label{f:potdiag}
\end{figure}
There are two classes of interaction included in the quark model used to find
the effective potentials.  The first, referred to as t-channel gluon exchange
by Swanson {\it et al.}, involves both one-gluon-exchange and a linear
confining potential, followed by quark rearrangement back into colour singlet
states.  This can occur for both $I=0$ and $I=2$ states.  The second, referred
to as s-channel gluon exchange, involves the annihilation of two of the quarks,
producing a meson hybrid in the intermediate state.  Because of the
intermediate state, this interaction can only occur for $I=0$ states.  Example
diagrams of the two classes of interaction are shown in Figure~\ref{f:potdiag}.

We only present results for FSI corrections to the $L=0$ partial wave of
$\gamma \gamma \to \pi \pi$ because the $L=2$ potentials available from the
same sources are smaller than those for $L=0$, and we found that their effect
on the cross-section was negligible.

Although\label{discussionetastart} relativistic phase space is included in the
quark model used to find the effective potentials, the potentials themselves
are essentially nonrelativistic.  It is our hope that they will contain enough
physics to correctly calculate the distortion of the final state wavefunction
leading to FSI effects.  However, it is known
\cite{weinstein90:kkbar,barnes91:diagrammatic,swanson92:inter-meson} that the 
t-channel gluon exchange potential is unable to accurately
predict the phase shifts resulting from $I=2$ $\pi$--$\pi$ scattering,
apparently because it lacks relativistic phase space.  It is expected that the 
other potentials suffer from the same problem.
\begin{figure}[t]
\vspace{-2cm}
\begin{center}
\makebox{\epsfxsize=6.5in\epsffile{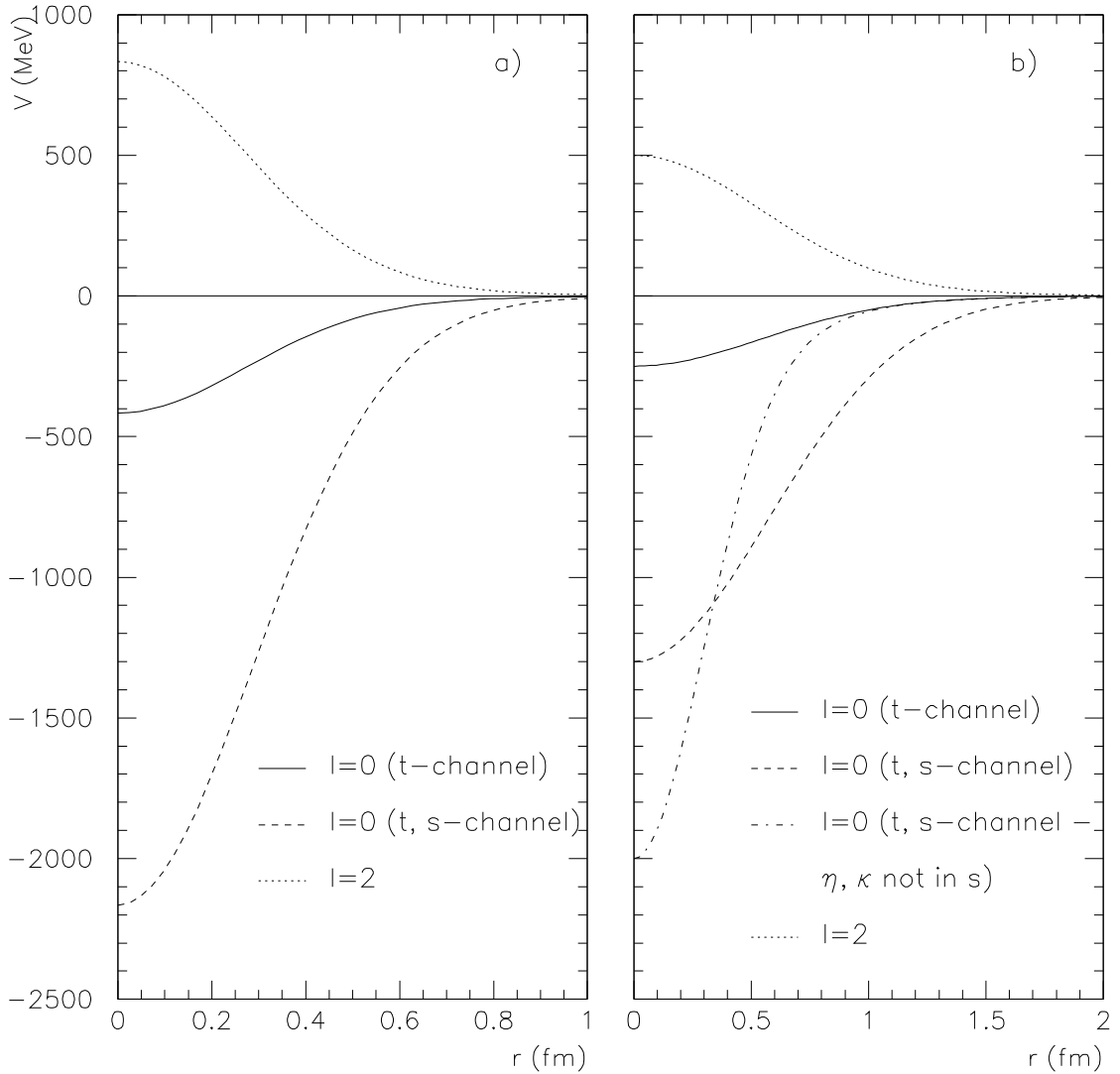}}
\end{center}
\vspace{-1.3cm}
\caption[Graphs of the $\pi$--$\pi$ potentials used in this work vs.\ $r$]
{Graphs of the $\pi$--$\pi$ potentials used in this work vs.\ $r$.  Note the
different scales on the two $r$ axes.  T-channel gluon exchange is present in
all potentials.  For $I=0$, the potentials are plotted both with and without
the contribution from s-channel gluon exchange.  In a), the potentials without
the factors $\eta$ and $\kappa$ are plotted.  In b), the potentials with
$\eta=0.6$ and $\kappa=2.0$ are plotted, except for one curve where the
contribution from t-channel gluon exchange contains $\eta$ and $\kappa$, but
that from s-channel gluon exchange does not.}
\label{f:potplot}
\end{figure}

Weinstein and Isgur \cite{weinstein90:kkbar} have argued that this is related
to the inability of the nonrelativistic quark model to accurately predict the
charge radii of the ground state mesons and baryons, and so apply a correction
factor $\kappa=2.0$ to the ranges of the $\pi$--$\pi$ potentials, and an
overall normalization factor $\eta=0.6$ which is fitted to experiment.  To be
specific, a potential originally parametrized by 
\begin{equation}
V(r)=V_0\:\dspexp{\scriptstyle -\frac{1}{2}\left(\frac{r}{r_0}\right)^2} 
\label{potparam1}
\end{equation}
is written as 
\begin{equation}
V(r)=\eta V_0\:\dspexp{\scriptstyle -\frac{1}{2}\left(\frac{r}{\kappa \,r_0}
\right)^2} \label{potparam2}
\end{equation}
with the corrections.  However, it is not clear whether a factor that corrects
the phase shift for the nonrelativistic phase space will necessarily give the
correct dynamics for the distortion of the wavefunction.  A better approach
might be to use the potentials without $\eta$ and $\kappa$ to obtain the
wavefunctions, but to use the phase shifts calculated directly in the quark
model\footnote{We refer to the specific quark model used by Swanson {\it et
al.}\ to extract the effective potentials.} to predict the FSI effects (in
Eq.~\ref{fcorrection} for example).  Swanson {\it et al.}\
\cite{barnes91:diagrammatic,swanson92:inter-meson,li94:Iequals0} found that the
quark model, with its relativistic phase space, does accurately reproduce the
$\pi$--$\pi$ scattering phase shifts away from threshold (although resonance
effects must be added for $I=0$ -- see below).

In addition, since $\eta$ and $\kappa$ have previously been applied only to
potentials representing t-channel gluon exchange, it is not entirely clear if
they should be applied to the potential representing s-channel gluon exchange.
We will examine these three possibilities below. \label{discussionetaend} The
various potentials, with and without the $\eta$ and $\kappa$, are shown in
Figure~\ref{f:potplot}.  The parameters $V_0$ and $r_0$ of the potentials
considered (including the $L=2$ potentials, for completeness) are given in
Table~\ref{t:potparam}.

\begin{table}[t]
\vspace{-0.3cm}
\begin{center}
\begin{tabular}{|c|c||c|c|c|c|c|c|} \hline
\multicolumn{2}{|c||}{} & \multicolumn{6}{c|}{$\pi$--$\pi$ Potentials 
($V_0$ in GeV, $r_0$ in ${\rm GeV}^{-1}$)} \\ \cline{3-8}
\multicolumn{2}{|c||}{Pion State} & 
\multicolumn{4}{c|}{t-channel gluon exchange} & 
\multicolumn{2}{c|}{s-channel} \\ \cline{1-6}
 & & \multicolumn{2}{c|}{hyperfine} & 
\multicolumn{2}{c|}{confinement} & 
\multicolumn{2}{c|}{ gluon exchange} \\ \cline{3-8}
\hspace{0.25cm}L\hspace{0.25cm} & I & $V_0$ & $r_0$ & $V_0$ & $r_0$ & $V_0$ & 
$r_0$ \\ \hline\hline
0 & 0 & $-0.392$ & 1.36 & $-0.024$ & 2.29 & $-1.75$ & 1.48 \\
0 & 2 & 0.786 & 1.36 & 0.047 & 2.29 &  & \\
2 & 0 & $-0.044$ & 1.40 & 0.0175 & 1.49 &  & \\
2 & 2 & 0.088 & 1.40 & $-0.035$ & 1.49 &  & \\\hline
\end{tabular}
\end{center}
\vspace{-0.4cm}
\caption[The parameters of the potentials used in this work]
{The parameters of the potentials used in this
work \cite{swanson92:inter-meson,li94:Iequals0,swanson96:private}.  The
t-channel gluon exchange potentials have two contributions, due to
colour-hyperfine and confinement terms, which must be summed.  The potentials
are parametrized as in Eqs.~\ref{potparam1} and \ref{potparam2}, depending on
whether $\eta$ and $\kappa$ are to be included or not.}
\label{t:potparam}
\end{table}

\subsection{The FSI Apparatus Applied to the Pions}

Calculating the effects of FSI's on the $\gamma \gamma \to \pi^+ \pi^-$
cross-sections is complicated by the fact that there are two isospin states
involved.  Because the $I=0$ and $I=2$ states feel different potentials,
they will be affected differently by the FSI's.  This will unbalance the
isospin combination that is $\pi^+ \pi^-$, leading to some $\pi^0 \pi^0$
production.  BDI used a single $\pi^+ \pi^-$ potential, and hence neglected
conversion to $\pi^0 \pi^0$.

We have two choices as to how to proceed.  We could transform the potentials
from the $I=0$ and $I=2$ basis to the $\pi^+ \pi^-$ and $\pi^0 \pi^0$ basis
(leading to off-diagonal potential terms that would mix the two states), and
then solve a coupled-channel Schr\"{o}dinger equation to find the final state
wavefunctions needed for the FSI calculations (see for example 
Reference~\cite{weinstein96:multichannel}).  Alternatively, we could take
advantage of the facts that the potential matrix is diagonal in the isospin
basis (since isospin is conserved in strong interactions), and that we are
assuming the pion states to be degenerate in mass (at 137 MeV), and just solve
uncoupled Schr\"{o}dinger equations in the isospin basis.  We opt for this
second choice.

The cross-section for $\gamma \gamma \to \pi \pi$, corrected for FSI effects,
is given by \footnote{Note that we have dropped the $^{++}$ and $^{+-}$ 
helicity notation on the $f$'s -- it can be identified from the value of $M_L$
anyway ($M_L=0 \Rightarrow ^{++}$, $M_L=2 \Rightarrow ^{+-}$).} ({\it cf.} 
Eq.~\ref{ggppcross})
\begin{equation}
\sigma^{\rm FSI}_{\pi\pi} = \frac{{\cal S}\sqrt{1-x}}{128\pi^2 s}\left[ 
\left|f_{00\,\pi\pi}^{\rm FSI}\right|^2 + \sum_{L \geq2, {\rm even}}
^\infty 
\left[\left|f_{L0\,\pi\pi}^{\rm FSI}\right|^2 + 
\left|f_{L2\,\pi\pi}^{\rm FSI}\right|^2 
\right] \right].  \label{ggppcrossfsi}
\end{equation}
Note that we have not yet specified the
charges of the pions -- ${\cal S}
\equiv 1/(1+\delta_{\pi\pi})$ is a statistical factor which is needed if the
pions are identical particles (i.e.\ $\pi^0 \pi^0$).

The calculation of $f_{LM_L\,\pi\pi}^{\rm FSI}$ proceeds similarly to that
in Section~\ref{4:fullfsi} -- the first divergence is in
Eq.~\ref{fsiderivstep}, which is replaced by
\begin{eqnarray}
\langle f|T|i\rangle ^{\rm FSI} &=& 
\frac{1}{(2\pi)^3}\int\!{\rm d}^3\vec{k} \;\langle \psi_{\vec{k}_f}^{-\pi\pi}|
\left[|\phi_{\vec{k}}^{\pi^+\pi^-} \rangle \:\langle \phi_{\vec{k}}
^{\pi^+\pi^-} |+
|\phi_{\vec{k}}^{\pi^0\pi^0} \rangle \:\langle \phi_{\vec{k}}^{\pi^0\pi^0}| 
\right] W|\phi_{\vec{k}_i} \rangle \nonumber \\
&=&  \frac{1}{(2\pi)^3}\int\!{\rm d}^3\vec{k} \;\langle \psi_{\vec{k}_f}
^{-\pi\pi}|\phi_{\vec{k}}^{\pi^+\pi^-} \rangle \:\langle \phi_{\vec{k}}
^{\pi^+\pi^-} |W|\phi_{\vec{k}_i} \rangle 
\end{eqnarray}
where we have inserted plane waves for both $\pi^+ \pi^-$ and $\pi^0 \pi^0$ --
the $\pi^0 \pi^0$ term is then dropped because $\pi^0 \pi^0$ isn't produced
from $\gamma \gamma$ in tree-level scalar QED, which is what the potential $W$
represents.

We can proceed as before, and relate the final bra-ket to the scalar QED
amplitude, and the left-hand side to the corrected scalar QED amplitude.  The
first bra-ket can be written in the isospin basis using Eq.~\ref{pidefns} as
one of
\begin{eqnarray}
\langle \psi_{\vec{k}_f}^{-\pi^+\pi^-}|\phi_{\vec{k}}^{\pi^+\pi^-} \rangle &=& 
\left[ \sqrt{\frac{2}{3}}\:\langle \psi_{\vec{k}_f}^{-0}|+\frac{1}{\sqrt{3}}
\:\langle \psi_{\vec{k}_f}^{-2}| \right]
\left[ \sqrt{\frac{2}{3}}\:|\phi_{\vec{k}}^0\rangle+\frac{1}{\sqrt{3}}\:
|\phi_{\vec{k}}^2 \rangle \right]
\nonumber \\
&=&(2\pi)^{\frac{3}{2}} \left[\frac{2}{3}\: \psi_{\vec{k}_f}^{-0*}(\vec{k}) +
\frac{1}{3}\: \psi_{\vec{k}_f}^{-2*}(\vec{k}) \right], \nonumber \\
\langle \psi_{\vec{k}_f}^{-\pi^0\pi^0}|\phi_{\vec{k}}^{\pi^+\pi^-} \rangle &=&
(2\pi)^{\frac{3}{2}} \left[-\frac{\sqrt{2}}{3}\:\psi_{\vec{k}_f}^{-0*}(\vec{k})
+\frac{\sqrt{2}}{3}\: \psi_{\vec{k}_f}^{-2*}(\vec{k}) \right],
\end{eqnarray}
depending on which pion state is produced in the end.  Here
$\psi_{\vec{k}_f}^{-I*}(\vec{k})$ is the complex conjugate of the
momentum-space solution of the Schr\"{o}dinger equation (Eq.~\ref{swe}) with
the potential for isospin $I$.  If $u_L^I(k_f,r)$ is the solution of the equivalent radial equation (Eq.~\ref{radial1}) with orbital angular momentum $L$, then Eq.~\ref{fcorrection} becomes for $\pi^+ \pi^-$ production,
\begin{eqnarray}
f_{LM_L\,\pi^+\pi^-}^{\rm FSI}(s(k_f))&=&\frac{2}{\pi} \sqrt{s(k_f)}  
\int_0^\infty \!{\rm d}k \int_0^\infty \!{\rm d}r \:\:r^2\: k^2 \:
\frac{f_{LM_L}(s(k))}{\sqrt{s(k)}} \,j_L(kr) \nonumber \\
&& \times \left[\frac{2}{3}\: 
\dspexp{i\delta_L^0} \:u_L^0(k_f,r)+\frac{1}{3}\:\dspexp{i\delta_L^2}\:
u_L^2(k_f,r) \right],
\end{eqnarray}
where $\delta_L^I$ is the phase shift of $u_L^I(k_f,r)$.

In order to keep things real, we define
\begin{equation}
g_{LM_L}^I(s(k_f))\equiv \frac{2}{\pi} \sqrt{s(k_f)}  
\int_0^\infty \!{\rm d}k \int_0^\infty \!{\rm d}r \;r^2\: k^2 \:
\frac{f_{LM_L}(s(k))}{\sqrt{s(k)}} \,j_L(kr)\:u_L^I(k_f,r),
\end{equation}
which gives
\begin{equation}
\left|f_{LM_L\,\pi^+\pi^-}^{\rm FSI}\right|^2 = \frac{4}{9} \left(g_{LM_L}^0
\right)^2+\frac{1}{9}
\left(g_{LM_L}^2\right)^2+\frac{4}{9} \:g_{LM_L}^0\: g_{LM_L}^2 \,
\cos{(\delta_L^0-\delta_L^2)} .
\end{equation}

Similarly, for $\pi^0 \pi^0$ production, we get 
\begin{equation}
\left|f_{LM_L\,\pi^0\pi^0}^{\rm FSI}\right|^2 = \frac{2}{9} \left(g_{LM_L}^0
\right)^2+\frac{2}{9}
\left(g_{LM_L}^2\right)^2-\frac{4}{9} \:g_{LM_L}^0\: g_{LM_L}^2 \,
\cos{(\delta_L^0-\delta_L^2)} .
\end{equation}

Note that if there is no FSI potential, $\delta_L^I\Rightarrow 0$,
$u_L^I(k_f,r)\Rightarrow j_L(k_f\,r)$, and the orthogonality condition of the
spherical Bessel functions gives us a delta function, so $g_{LM_L}^I
\Rightarrow f_{LM_L}$ and we obtain $f_{LM_L\,\pi^+\pi^-}^{\rm FSI}= f_{LM_L}$
and $f_{LM_L\,\pi^0\pi^0}^{\rm FSI}= 0$, as expected.

In this work, we only correct the $L=0$ partial wave, because higher partial
waves have negligible FSI effects.  Thus in Eq.~\ref{ggppcrossfsi}
only the $f_{00\,\pi\pi}^{\rm FSI}$ term survives for $\pi^0 \pi^0$, and for
$\pi^+ \pi^-$ it is the only term that differs from the uncorrected value.

\subsection{The Problem of Limited Polar Acceptance}

Expanding the scalar QED amplitudes in a spherical harmonic basis
(Eq.~\ref{partwaves}) to obtain cross-sections for particular partial waves is
essential to our calculation, but it requires that we be able to integrate over
the entire $4\pi$ angular range when calculating the total cross-section.
Unfortunately, experiments don't measure data over the whole polar angle
($|\cos{\theta}| \leq 0.6 - 0.8$ is typical), so we cannot easily compare our
results with experiment.

Integrating to get the total uncorrected cross-section for a limited polar
acceptance ($-\cos{\theta_{\rm acc}} \leq \cos{\theta} \leq \cos{\theta_{\rm
acc}}$) is not a problem: Eq.~\ref{ggpptotcross} is replaced by
\begin{eqnarray}
\sigma_{\rm acc} &=& \frac{\pi\alpha^2}{4m^2}\left[2\cos{\theta_{\rm acc}}\;
x\sqrt{1-x}\left(\frac{x^2}{1-(1-x)\cos^2{\theta_{\rm acc}}}+1 \right) 
\right. \nonumber \\
&& \left. - x^2(2-x) \ln\left(\frac{1+\sqrt{1-x}\,\cos{\theta_{\rm acc}}}
{1-\sqrt{1-x}\,\cos{\theta_{\rm acc}}}\right)\right].
\end{eqnarray}

However, the expression in terms of partial waves is affected more drastically.
Eq.~\ref{ggppcross} becomes (where once again we drop the explicit helicity
labels)
\begin{eqnarray}
\sigma_{\rm acc} &=& \frac{\sqrt{1-x}}{256\pi^2 s}
\int_0^{2\pi} \!\!{\rm d}\phi \int_{-\cos{\theta_{\rm acc}}}
^{\cos{\theta_{\rm acc}}} \!\!{\rm d}(\cos{\theta}) \,
\left[ 2\left| \sum_{L \geq 0,{\rm even}}^\infty f_{L0} \,Y_{L0}(\theta,\phi)
\right|^2  \right. \nonumber \\
&& + \left. \left| \sum_{L \geq 2,{\rm even}}^\infty f_{L2}
\,Y_{L2}(\theta,\phi)
\right|^2+ \left| \sum_{L \geq 2,{\rm even}}^\infty f_{L2} \,
Y_{L\,-\!2}(\theta,\phi)\right|^2 \right]. \label{fsiinftyterms}
\end{eqnarray}
For the limited polar acceptance, the spherical harmonics are not orthogonal to
each other, so each magnitude contains an infinite number of terms.  As well,
the contributions of the partial waves to the total cross-section can no longer
be separated.

Because the $L=0$ partial wave is expected to be the only one affected
significantly by FSI's, we assume that all of the observed $\gamma \gamma \to
\pi^0 \pi^0$ events are in an $L=0$ state.  Since that distribution is 
spherically symmetric, we can correct the data to full polar acceptance by
simply dividing it by the value of $\cos{\theta_{\rm acc}}$ appropriate for the
particular experiment.

For $\gamma \gamma \to \pi^+ \pi^-$ the total cross-section observed is a
mixture of all partial waves; again, we assume that only the $L=0$ wave is 
affected by FSI's.  Define
\begin{equation}
h_{l_1l_2}^m \equiv \int_0^{2\pi} \!\!{\rm d}\phi 
\int_{-\cos{\theta_{\rm acc}}}
^{\cos{\theta_{\rm acc}}} \!\!{\rm d}(\cos{\theta}) \:Y_{l_1m}^*(\theta,\phi)
\,Y_{l_2m}(\theta,\phi)
\end{equation}
and note that $h_{l_1l_2}^m = h_{l_2l_1}^m$ and $h_{l_1l_2}^{-\!2}=
h_{l_1l_2}^{2}$.  We can then write the total cross-section with only the $L=0$
partial wave corrected for FSI's as the total uncorrected cross-section, plus
an infinite number of correction terms, each of which is the difference between
the corrected and uncorrected values of a term in Eq.~\ref{fsiinftyterms}.
Only those terms involving $f_{00}$, which goes to $f_{00\,\pi^+\pi^-}^{\rm
FSI}$ in the correction, need correcting.  Eq.~\ref{ggppcrossfsi} is replaced
by
\begin{eqnarray}
\lefteqn{\sigma^{\rm FSI}_{{\rm acc},\pi^+\pi^-} = 
\sigma_{\rm acc} +
\frac{\sqrt{1-x}}{128\pi^2 s}\left[ \rule[-.5cm]{0cm}{1.0cm}
h_{00}^0\left\{\left|f_{00\,\pi^+\pi^-}^{\rm FSI}\right|^2 -(f_{00})^2 \right\}
\right.} \nonumber \\
&&\left. +\sum_{L \geq 2,{\rm even}}^\infty h_{L0}^0 \left\{
f_{00\,\pi^+\pi^-}^{\rm FSI} \,f_{L0} + 
\left(f_{00\,\pi^+\pi^-}^{\rm FSI}\right)^* f_{L0} 
-2 f_{00} f_{L0} \right\}\right] \nonumber \\
&=&  \sigma_{\rm acc} +
\frac{\sqrt{1-x}}{128\pi^2 s}\left[ \rule[-.5cm]{0cm}{1.0cm}
h_{00}^0\left\{\frac{4}{9} \left(g_{00}^0
\right)^2+\frac{1}{9}
\left(g_{00}^2\right)^2+\frac{4}{9} \:g_{00}^0\: g_{00}^2 \,
\cos{(\delta_0^0-\delta_0^2)}-(f_{00})^2 \right\}
\right. \nonumber \\
&&\left. +\sum_{L \geq 2,{\rm even}}^\infty 2\,h_{L0}^0 \:f_{L0} \left\{
\frac{2}{3} \,g_{00}^0\:\cos{\delta_0^0} +
\frac{1}{3} \,g_{00}^2\:\cos{\delta_0^2} 
- f_{00} \right\}\right]. 
\end{eqnarray}
We find that we get excellent convergence over the energy range we are
interested in when we truncate the series after $L=6$.  Fortunately, the three
experiments whose data we use for comparisons all have the same limited polar
acceptance ($|\cos{\theta}| \leq 0.6$), so we only have to correct our results
for one value of $\cos{\theta_{\rm acc}}$.

\subsection{Numerical Details}

We solved the radial Schr\"{o}dinger equation (Eq.~\ref{radial2}) with the
appropriate potentials using the Bulirsch-Stoer method for differential
equations (with Stoermer's rule for $2^{\rm nd}$ order conservative equations)
\cite{press92:numerical}.  The solutions were started off from the origin with
the forms of $r$ times the spherical Bessel functions of the same $L$, since
the effects of the potentials are insignificant at the origin.  The amplitudes
and phase shifts of the solutions were extracted by fitting the last
oscillatory cycle to a general sinusoidal function.  We carried out the
integrals over $r$ and $k$ using an extended Simpson's rule.

\section{Results}
\label{5:results}

We have calculated the effects of the FSI's on $\gamma \gamma \rightarrow \pi
\pi$ for three primary cases of the FSI potential (see Section~\ref{5:fsipot} 
for a full discussion).  
The three cases are:
\begin{description}
\item{Case A.} In this case we do not use $\eta$ and $\kappa$ in the 
potentials.  We know that the potentials will then fail to accurately predict
the phase shifts.  In order to avoid problems from this, we use the phase
shifts directly calculated in the quark model in our expressions for FSI
corrections.  For more information, see the detailed discussion on $\eta$ and
$\kappa$ on pages~\pageref{discussionetastart} through
\pageref{discussionetaend} .
\item{Case B.} In this case we use $\eta=0.6$ and $\kappa=2.0$ in all of the 
potentials, and we use the resulting phase shifts in our expressions for FSI
corrections.
\item{Case B1.} This case is the same as case B, except that $\eta$ and 
$\kappa$ are used in the t-channel gluon exchange contribution to the
potentials, but not in the contribution from s-channel gluon exchange.
\end{description}
In addition, we have three secondary cases, A$'$, B$'$ and B1$'$.  Each of
these is similar to the corresponding unprimed case, but does not use the same
phase shifts in our expressions for FSI corrections.  Instead, they use a
simple (straight-line) expression\footnote{The straight-line expressions we
use to describe the phase shifts are:\\ 
I=0:  $\delta_0^0 = 0.0027 \;{\rm rad/MeV} \times (\sqrt{s}- 250 
\;{\rm MeV})$ \\
I=2:  $\delta_0^2 = -0.00062 \;{\rm rad/MeV} \times (\sqrt{s}- 250 
\;{\rm MeV})$.} 
for the phase shifts that attempts
to describe the experimental data.  We do this in order to get an idea of how
much our results are affected by inaccurate phase shifts.  If our results are
poor because we have ignored a particular contribution to the potential (such
as resonance effects), then using accurate phase shifts in the expressions for
FSI corrections would partly correct for this.  Of course, we would still need
to determine the effect of the neglected contribution on the distortion of the
wavefunction, and hence on the FSI effects.

\begin{figure}[t]
\vspace{-2cm}
\begin{center}
\makebox{\epsfxsize=6.5in\epsffile{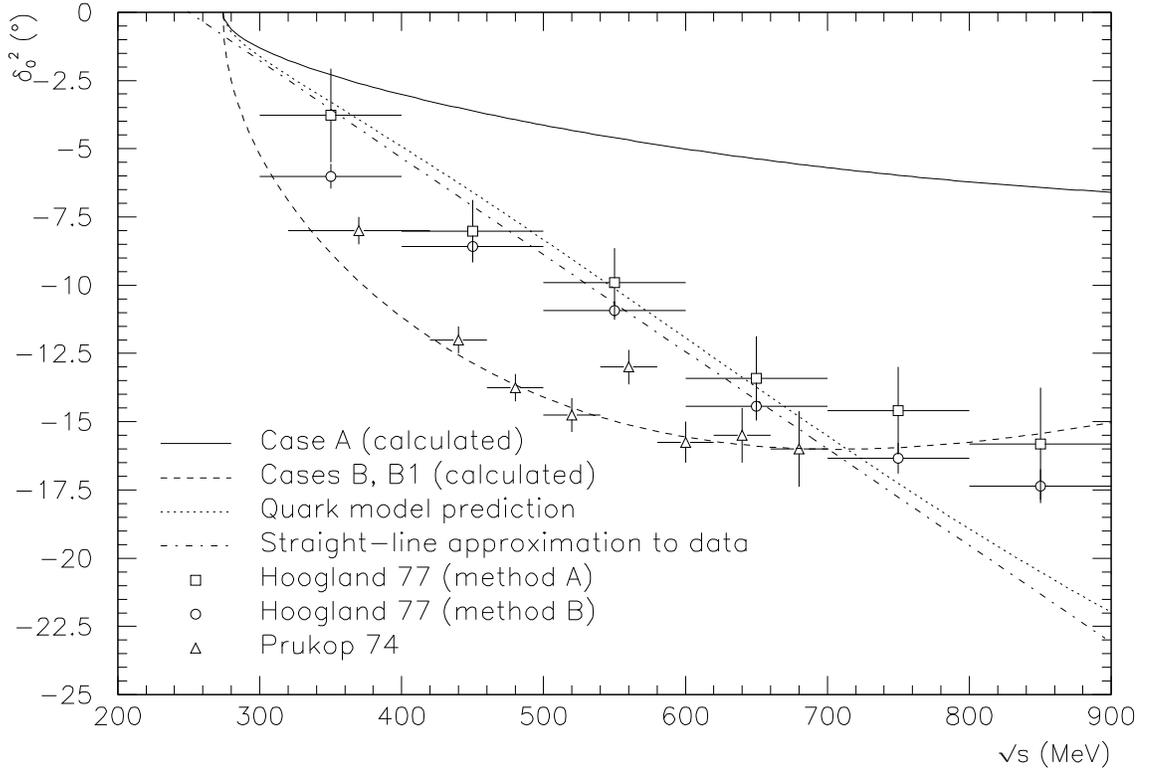}}
\end{center}
\vspace{-1.3cm}
\caption[Graphs of the $I=2$ $\pi$--$\pi$ scattering phase shift vs.\ 
$\protect\sqrt{s}$, for our predictions, the quark model prediction, and 
experimental data] {Graphs of the
$I=2$ $\pi$--$\pi$ scattering phase shift vs.\ $\protect\sqrt{s}$, for our
predictions, the quark model prediction, and experimental data.  The data is 
from Hoogland {\it et al.}\
\cite{hoogland77:measurement} -- we show the results of both of their methods; 
and Prukop {\it et al.}\ \cite{prukop74:unknown} -- we show the results of 
their first fit.
For additional comments see Figure~\ref{f:BDIcross}.}
\label{f:phases2}
\end{figure}
\begin{figure}[t]
\vspace{-2cm}
\begin{center}
\makebox{\epsfxsize=6.5in\epsffile{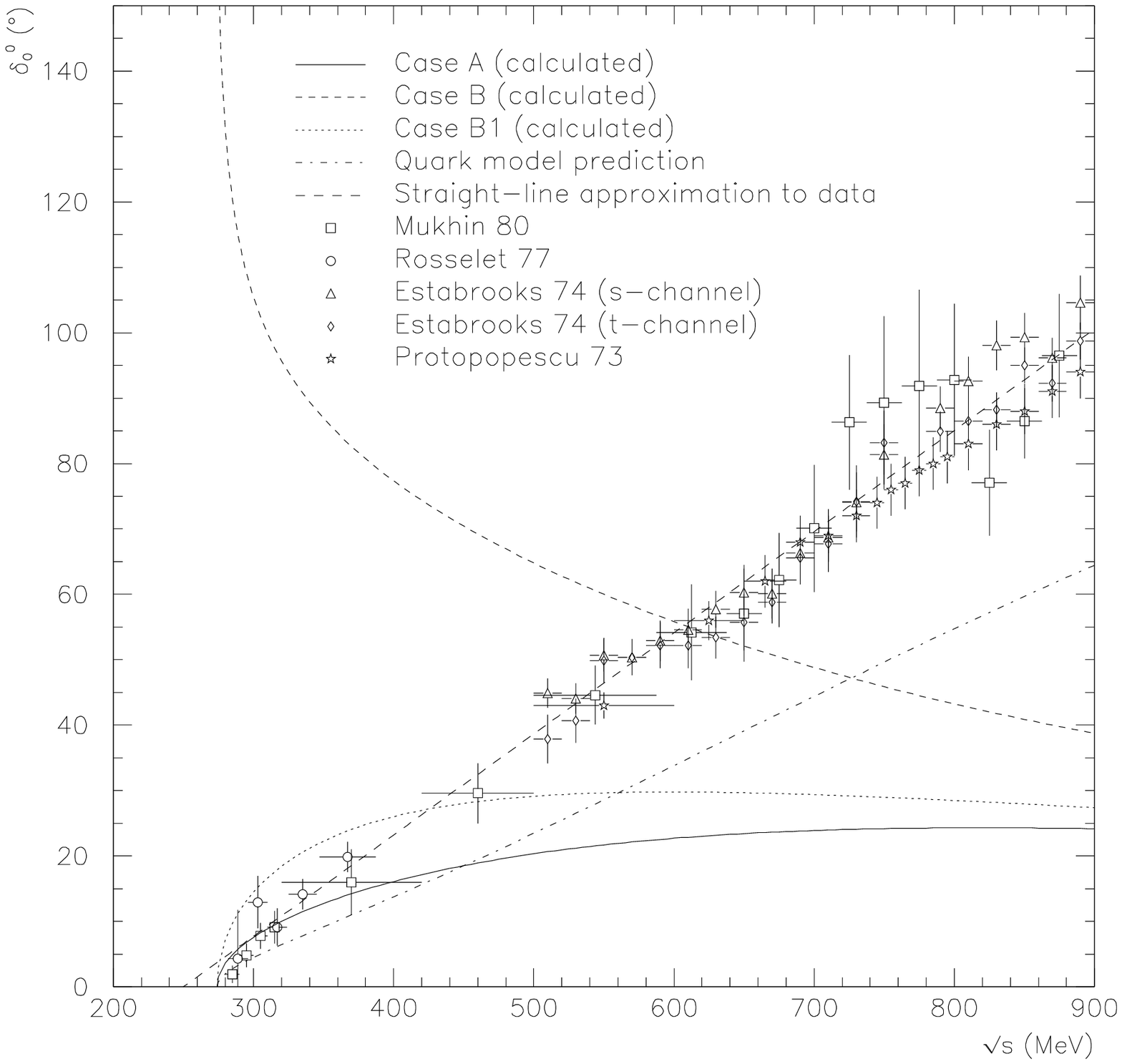}}
\end{center}
\vspace{-1.3cm}
\caption[Graphs of the $I=0$ $\pi$--$\pi$ scattering phase shift vs.\ 
$\protect\sqrt{s}$, for our predictions, the quark model prediction, and 
experimental data] {Graphs of the
$I=0$ $\pi$--$\pi$ scattering phase shift vs.\ $\protect\sqrt{s}$, for our
predictions, the quark model prediction, and experimental data.  The data is 
from Mukhin {\it et al.}\
\cite{mukhin80:values}; Rosselet {\it et al.}\ \cite{rosselet77:experimental} --
the horizontal bars only approximate their bins, and their data is actually 
for $\delta_0^0-\delta_1^1$ -- we have used the $\delta_0^0$ data
extracted from it by Li {\it et al.}\ \cite{li94:Iequals0}; Estabrooks and 
Martin \cite{estabrooks74:pipi} -- we show the results of both their s- and 
t-channel fits; and Protopopescu {\it et al.}\ \cite{protopopescu73:pipi} -- we 
show the results of their case 1.
For additional comments see Figure~\ref{f:BDIcross}.}
\label{f:phases0}
\end{figure}
Let us first examine the phase shift results.  The $I=2$ phase shifts are shown
in Figure~\ref{f:phases2}.  For the three cases A, B and B1, the curves shown
are the phase shifts that come from solving the Schr\"{o}dinger equation with
the potentials.  The phase shift for case A is poor, as expected; in the
expressions for FSI corrections we will replace it with the curve calculated
directly in the quark model.  The phase shifts for cases B and B1 are identical
because they only differ in whether or not $\eta$ and $\kappa$ are included in
the contribution to the potential from s-channel gluon exchange, which is not
present for $I=2$.  For the range of data shown, it is not clear which is
better: the B and B1 phase shifts from the Schr\"{o}dinger equation, or that
from the quark model.  However, data at somewhat higher energies favours the
quark model curve.  The remaining curve shown is just a straight-line which we
used to describe the data for the primed cases.

The $I=0$ phase shifts are shown in Figure~\ref{f:phases0}.  Note the rather
unfortunate behaviour of the phase shift for case B, which suggests that
$\eta$ and $\kappa$ should not be used in the s-channel gluon exchange
contribution to the potential.  Since there is no {\it a priori} good reason
for not using them in this part of the potential, we must question their use at
all.  However, we will continue to show the results arising from cases B and
B1.  Again, the phase shift for case A is poor, and for FSI corrections we will
replace it with the curve arising directly from the quark model.  Note that
this time the quark model curve is low as well; the reason is as follows.  When
the quark model predictions for the $I=0$ phase shifts were calculated
\cite{li94:Iequals0} three contributions were considered: t-channel gluon 
exchange, s-channel gluon exchange, and contributions from interactions with
the $f_0(980)$ and $f_0(1300)$ resonances.  These contributions were fit to the
data, but because we are not including resonance effects in this work, we only
include the contributions from the s- and t-channel gluon exchange in our quark
model curve.  When resonance effects are included, the quark model curve fits
the data rather well.  Again, the remaining curve shown is just a straight-line
which we have used to describe the data for the primed cases.

\begin{figure}[t]
\vspace{-2cm}
\begin{center}
\makebox{\epsfxsize=6.5in\epsffile{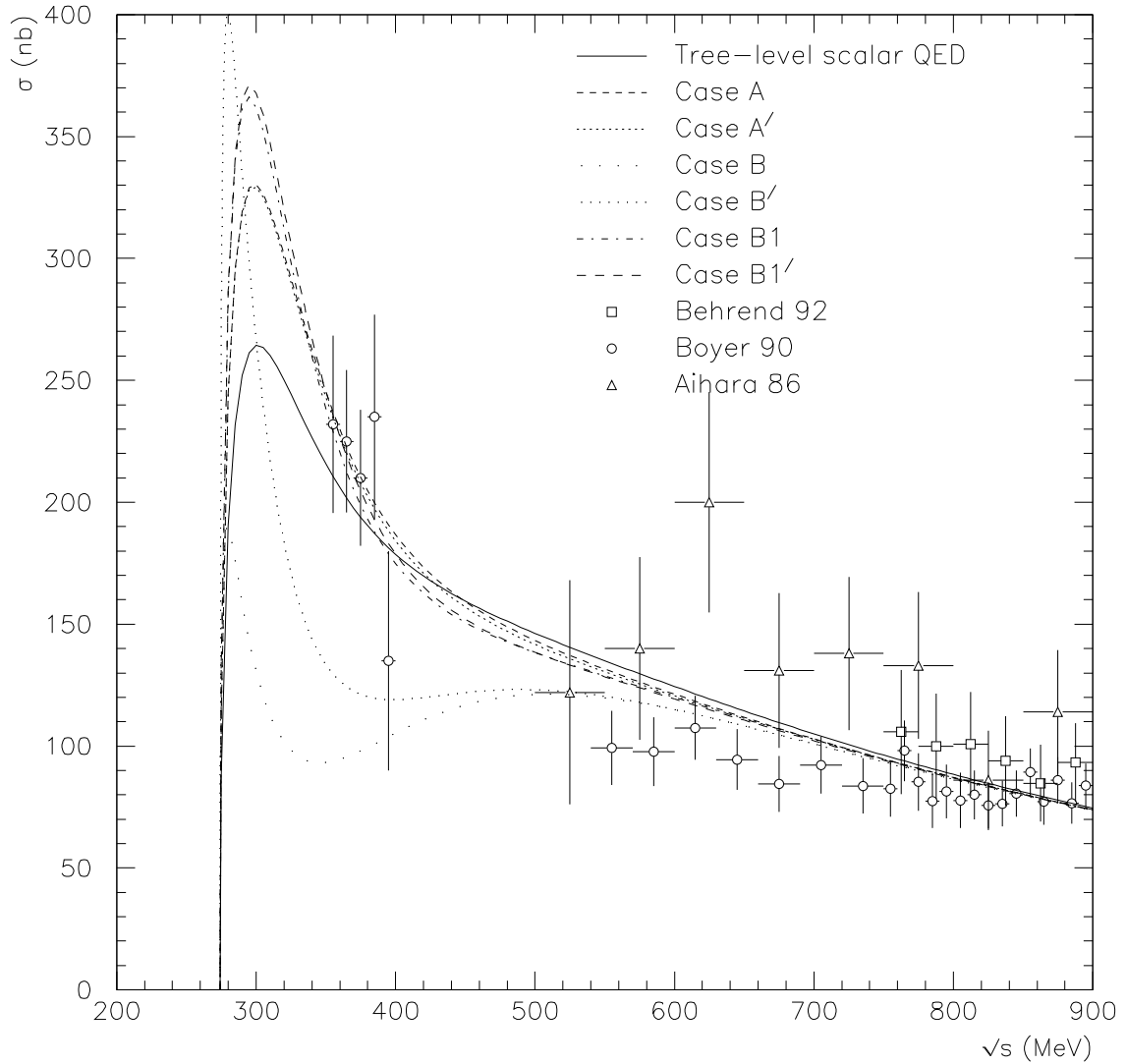}}
\end{center}
\vspace{-1.3cm}
\caption[Graph of the $\gamma \gamma \to \pi^+ \pi^-$ total cross-section vs.\ 
$\protect\sqrt{s}$, for our predictions, the tree-level scalar QED prediction,
and experimental data] {Graph of the $\gamma \gamma \to \pi^+ \pi^-$ total
cross-section vs.\ $\protect\sqrt{s}$, for our predictions, the tree-level
scalar QED prediction, and experimental data.  The curves have been corrected
to have a limited polar acceptance to match the data: $|\cos{\theta}| \leq
0.6$.  For references to the data and additional comments see
Figure~\ref{f:BDIcross}.}
\label{f:ggppcrosspm1}
\end{figure}
Now let us look at the $\gamma \gamma \rightarrow \pi^+ \pi^-$ cross-section
results, shown in Figure~\ref{f:ggppcrosspm1}.  Cases A, A$'$, B1 and B1$'$ all
fit the data well (as does the tree-level scalar QED prediction).  The effects
of using the quark model phase shifts are not significant for these cases.
Better data would be needed to differentiate between case A, case B1 and the
tree-level scalar QED prediction.  Cases B and B$'$ are disfavoured by the
data, as might be expected from the poor performance of case B in predicting
the phase shifts.  Because that prediction differed drastically from the data,
cases B and B$'$ do differ appreciably in their predictions of the
cross-section.

\begin{figure}[t]
\vspace{-2cm}
\begin{center}
\makebox{\epsfxsize=6.5in\epsffile{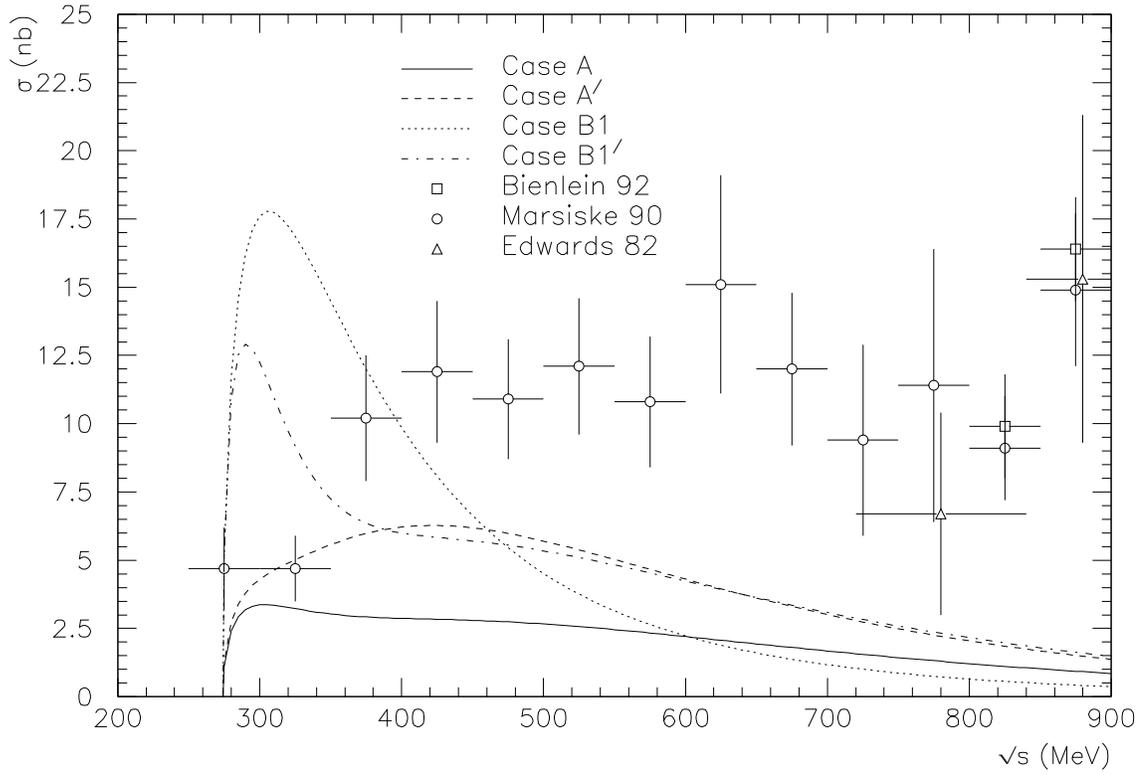}}
\end{center}
\vspace{-1.3cm}
\caption[Graphs of the $\gamma \gamma \to \pi^0 \pi^0$ cross-section vs.\ 
$\protect\sqrt{s}$, for our predictions (cases A, A$'$, B1 and B1$'$) and
experimental data] {Graphs of the $\gamma
\gamma \to \pi^0 \pi^0$ cross-section vs.\  $\protect\sqrt{s}$, for our
predictions (cases A, A$'$, B1 and B1$'$) and experimental data.  The data is
from Bienlein ({\it et al.}\ )
\cite{bienlein92:new}, Marsiske {\it et al.}\ \cite{marsiske90:measurement} and
Edwards {\it et al.}\ \cite{edwards82:production}.  The data has been corrected
to full polar acceptance: $|\cos{\theta}| \leq 1.0$.  For additional comments
see Figure~\ref{f:BDIcross}.}
\label{f:ggppcross001}
\end{figure}
\begin{figure}[t]
\vspace{-2cm}
\begin{center}
\makebox{\epsfxsize=6.5in\epsffile{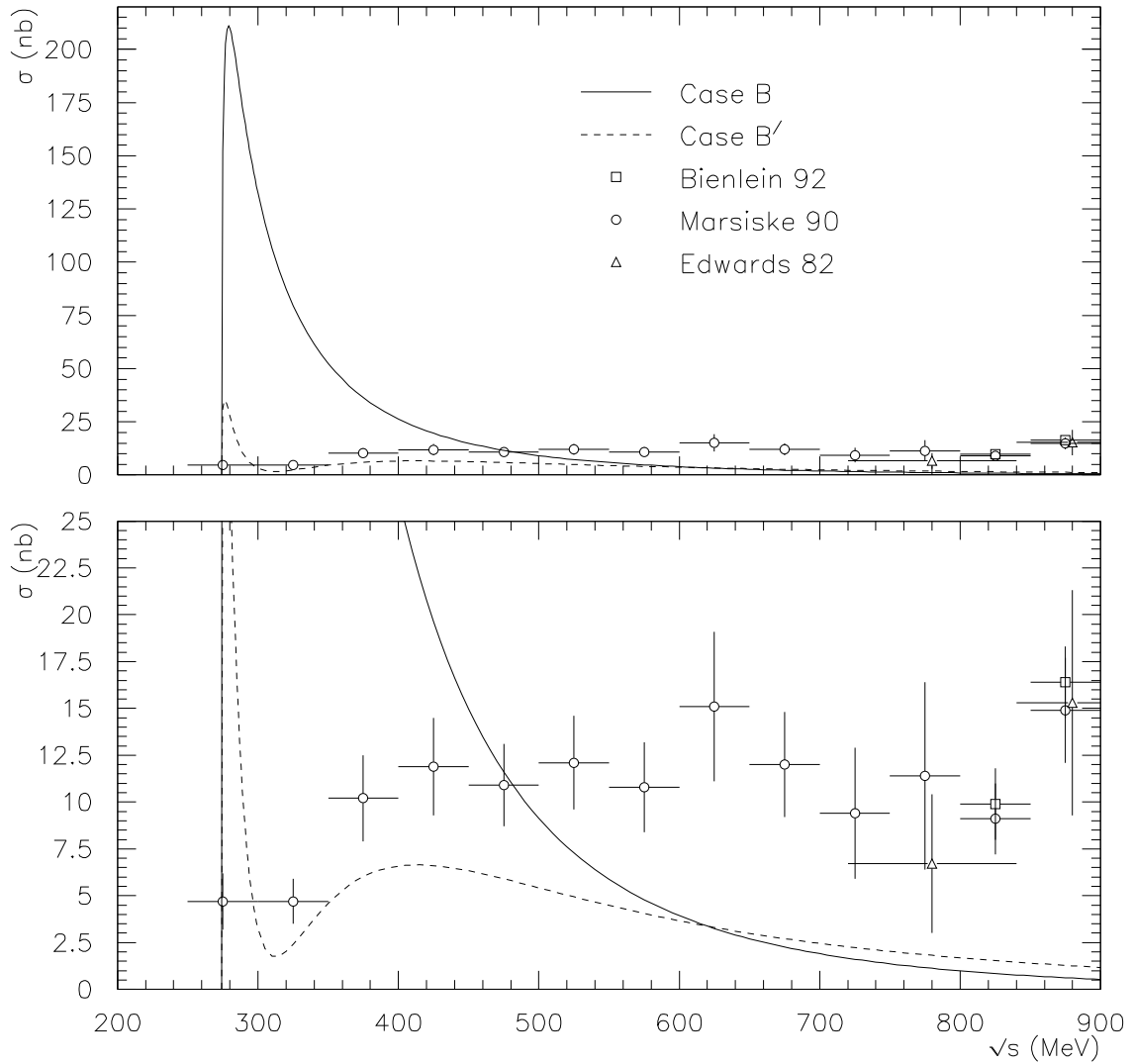}}
\end{center}
\vspace{-1.3cm}
\caption[Graphs of the $\gamma \gamma \to \pi^0 \pi^0$ cross-section vs.\ 
$\protect\sqrt{s}$, for our predictions (cases B and B$'$) and experimental
data] {Graphs of the $\gamma
\gamma \to \pi^0 \pi^0$ cross-section vs.\  $\protect\sqrt{s}$, for our
predictions (cases B and B$'$) and
experimental data.  The upper and lower plots differ only in their vertical
scales.  The data has been corrected to full polar acceptance: $|\cos{\theta}|
\leq 1.0$.  For references to the data see Figure~\ref{f:ggppcross001}.  For
additional comments see Figure~\ref{f:BDIcross}.}
\label{f:ggppcross002}
\end{figure}
The $\gamma \gamma \rightarrow \pi^0 \pi^0$ cross-section results are shown in
Figures~\ref{f:ggppcross001} and \ref{f:ggppcross002}.  Here our results do
not do as well.  This is partly because our $\gamma \gamma \rightarrow \pi^0
\pi^0$ cross-section is wholly due to FSI's while in the 
$\gamma \gamma \rightarrow \pi^+ \pi^-$ case the FSI effects are corrections to
a more significant curve.  This also means that the primed and unprimed cases
differ substantially.  None of our cases are consistent with the data.  Cases B
and B$'$ do particularly badly as expected from our previous results.  Of the
others, case A$'$ comes closest, being roughly a factor of 2 too low.  It also
shows a local maximum in the cross-section at higher energies, in keeping with
the data.  It is interesting to note that using the experimental phase shifts
instead of the quark model predictions in our expressions for the FSI
corrections greatly improves curve A.  Since the difference in these phase
shifts is due to resonance effects, we tend to believe that including the
effects of resonances in the wavefunction distortion as well would improve the
curve still further.

We have demonstrated that the quark model can predict the $\gamma \gamma
\rightarrow \pi^+ \pi^-$ cross-section in agreement with experimental data when
a sufficiently detailed calculation is performed.  However, with the present
potentials it is unable to accurately predict the $\gamma \gamma \rightarrow
\pi^0 \pi^0$ cross-section.  We find that using Weinstein and Isgur's $\eta$
and $\kappa$ factors in the potentials is not helpful when calculating FSI
effects; it is better to use the expressions for the phase shifts calculated
directly in the quark model in the expressions for the FSI corrections.  We
believe that the next logical step in this program of work is to include
potentials allowing the $I=0$ $\pi
\pi$ state to convert to the $f_0(980)$ and $f_0(1300)$ resonances (and back).
This would mean solving a coupled-channel Schr\"{o}dinger equation in the $I=0$
channel.

\chapter{Conclusions}
\markright{Chapter 6.  Conclusions}
\label{6:conclusions}

In this thesis we have examined three problems, split between two general areas
of the quark model.  

Both of the first two problems involved two models of meson decay: the $3P_0$
model and the flux-tube breaking model.  We have developed general routines for
these models that can be used for any OZI-allowed strong decay of a meson into
two other mesons, as long as the radial portion of the meson wavefunctions can
be expanded in terms of SHO radial wavefunctions, and sufficient computer
resources are available.  The general nature of these routines make them useful
tools in the study of meson phenomenology.

While fitting the free parameter of the models and investigating a number of
choices that must be made in their application, we evaluated the models by
comparing their predictions for the decay widths of 28 of the best known meson
decays to the experimental values.  We found that these models are not very
accurate -- the best they can hope for is to predict a decay width to within a
factor of 2, and even larger deviations are common.  Since they are coarse
models of a complicated theory, this is not surprising.  We also found that
both the use of SHO wavefunctions with the effective $\beta$'s of Kokoski and
Isgur \cite{kokoski87:meson}, and the use of nonrelativistic phase
space/normalization give less accurate results than can be obtained with other
choices.

The first problem that we investigated was the identity of the $f_4(2220)$.
Although tentatively identified as the $^3F_4$ $s\bar{s}$ meson by the Particle
Data Group, its identity has been uncertain since its discovery in 1983.  We
carried out detailed calculations of the decays of the $^3F_2$ and $^3F_4$
$s\bar{s}$ mesons in order to see if a meson identification for the $f_4(2220)$
is tenable.  Despite the uncertainties of the decay models, we found that the
$f_4(2220)$, with a measured decay width of approximately 30~MeV, cannot be the
$^3F_2$ $s\bar{s}$, whose width we expect to be \approxge~400~MeV.  We expect
the width of the $^3F_4$ $s\bar{s}$ meson to be \approxge~140~MeV and
\approxle~600~MeV, but the uncertainty of the models precludes us from ruling
out the identification of the $f_4(2220)$ as the $^3F_4$ $s\bar{s}$ with any
surety.  We do feel it is unlikely however, and propose the following
explanation: that the broad state seen in hadron beam experiments is the
$^3F_4$ $s\bar{s}$ meson, and the narrow state seen in $J/\psi$ radiative decay
is a glueball.  Further experimental data is needed to finally identify the
$f_4(2220)$.

The second problem that we investigated was the determination of the mixing
angle between the $K_1(1270)$ and $K_1(1400)$ mesons.  This was done by
comparing predictions of the meson decay models for five partial decay widths
and two ratios of D to S amplitudes with experimental data.  We found the
mixing angle $\theta_K$ to be approximately $44^\circ$.  If the mixing is due
only to the spin-orbit interactions already included in the quark model of
Godfrey and Isgur, then this implies that the responsible term in the
Hamiltonian, $\langle H^{\rm SO-}_{q\bar{q}} \rangle$, has a value far larger
than calculated.  However, because of the delicate cancellation that leads to
the small calculated value, we are unable to draw any conclusions about whether
the mixing is largely due to another mechanism, or the quark model Hamiltonian
just needs retuning somewhat.

The third problem that we investigated was the effect of the strong final state
interactions (FSI's) in the reaction $\gamma \gamma \to \pi \pi$ near
threshold.  Using effective potentials extracted from a quark model, we found
that our preferred set of potentials (without the $\eta$ and $\kappa$
correction factors) successfully described the experimental data for the
$\gamma \gamma \to \pi^+ \pi^-$ cross-section.  However, the prediction in the
absence of FSI's also described the data, so better data is need to distinguish
between the two.  This still represents a vindication for the use of effective
potentials extracted from a quark model, which has been in disrepute because
the latest data does not agree with the results of a previous,
less-comprehensive calculation.  We were also able to conclude that the use of
Weinstein and Isgur's $\eta$ and $\kappa$ correction factors for the potentials
is inappropriate for the calculation of FSI effects.

Unfortunately, we
were unable to reproduce the data for the $\gamma \gamma \to \pi^0 \pi^0$
cross-section.  This is partly because the $\gamma \gamma \rightarrow \pi^0
\pi^0$ cross-section is wholly due to FSI's while in the 
$\gamma \gamma \rightarrow \pi^+ \pi^-$ case the FSI effects are corrections to
a more significant quantity.  However, we have indications that including the
effects of resonance production in the $I=0$ channel may bring our results in
line with the data.   The next step in our program would be to solve a
coupled-channel Schr\"{o}dinger equation to include these effects.  In light of
our poor results for the $\gamma \gamma \to \pi^0 \pi^0$ cross-section, it
would be premature to investigate the structures seen in the cross-sections of
$\gamma \gamma \rightarrow$ two vector mesons at this time.

\appendix

\chapter{Some Tools of Particle Physics}
\markright{Appendix A.  Some Tools of Particle Physics}
\label{a:tools}

In this appendix we introduce some of the terms and tools of particle physics,
to help specialists in other fields in understanding the work of this thesis.
For an excellent introduction to the field, see Griffiths
\cite{griffiths87:introduction}.

\section{Cross-section and Width}
\label{a:crosssecwid}

In most particle physics experiments two particles are collided and the
outgoing particles are collected and identified.  Due to the interactions
between the particles, they may be scattered in different directions, and/or
they may be transformed into other particles.  The probability of a particular
reaction occurring is related to the cross-section $\sigma$ which is a function
of the total energy of the particles and is measured in units of barns, or b 
(1~b = $10^{-28}\;{\rm m}^2$).

A particle will decay eventually if there is an interaction that permits it to
be transformed into other particles with less total mass.  The average lifetime
of the particle is denoted by $\tau$; by Heisenberg's uncertainty principle,
this finite lifetime leads to an uncertainty in the mass of the particle.  In a
collision of two particles, the production of a short-lived particle in the
intermediate state (which then decays to give the final state particles) shows
up in the cross-section plot as a resonance described by a Breit-Wigner curve
with width $\Gamma=1/\tau$ (see for example Figure~\ref{f:xidata}).

For a particle that can decay into two or more different final states we can
define its branching ratios, which are the probabilities that the particle will
decay into the various final states.  From these we can define the partial
width for a particular decay mode as the product of the branching ratio for
that mode, and the total width.  The sum of all the partial widths gives the
total width, which is what is used to get the particle's average lifetime.  We
may sometimes refer to a partial width as the width for a particular decay
mode.

Similarly for the cross-section for a particular initial state, there are the
cross-sections going to specific final states, which sum to give the total
cross-section.  Note that in a plot of the cross-section to a specific final
state, a resonance in the intermediate state still produces a Breit-Wigner
curve with a width equal to the total width for that resonance.  (Even though
we are only looking at a particular decay mode, the lifetime of the resonance
is still that determined by all of the possible decay modes.)

\section{Feynman Diagrams}
\label{a:feynmandiag}

In a perturbation theory such as that normally used to calculate quantities in
QED, each order of perturbation is represented by Feynman diagrams, such as
those shown in Figure~\ref{f:ggppdiag} for $\gamma
\gamma \to \pi^+ \pi^-$ in scalar QED.  In
a Feynman diagram the particles are represented by lines, and basic
interactions by vertices where a number of lines meet.  Since for QED each
vertex corresponds to another factor of $\sqrt{\alpha}$ in the Feynman
amplitude\footnote{The Feynman amplitude is similar to the amplitude of quantum
mechanics.  Its magnitude squared is combined with the phase space of the final
state particles to obtain the cross-section or width.} the order of the
perturbation is given by the number of vertices in the diagram.

A tree-level diagram is one in which there is only one path of lines connecting
any two vertices (i.e.\ there are no loops).  For QED at least, if a tree-level
diagram exists for a particular process then it is the lowest order diagram.

\section{Clebsch-Gordan Coefficients and the Wigner $nj$ Symbols}
\label{a:cgcoeffsnj}

An angular momentum state, be it spin, orbital angular momentum, or a vector
sum of some combination of these, is represented (in an SU(2) algebra) by the
usual quantum numbers $|jm\rangle$ where the magnitude of the total angular
momentum is $\sqrt{j(j+1)}$ and the $z$-component is $m$.  To combine two
angular momentum states we use the Clebsch-Gordan coefficients, defined by
\begin{equation}
|j_1m_1\rangle \:|j_2m_2\rangle \equiv \sum_{j=|j_1-j_2|}^{j_1+j_2} 
\langle j_1m_1j_2m_2|JM\rangle \:|JM\rangle
\end{equation}
where $\langle j_1m_1j_2m_2|JM\rangle$ is the Clebsch-Gordan coefficient and $M
=m_1+m_2$.

The Wigner $3j$ symbol is related to the Clebsch-Gordan coefficient by 
\begin{equation}
\left(\begin{array}{ccc}j_1 & j_2 & J \\ m_1 & m_2 & -M \end{array} \right)
\equiv \frac{(-1)^{j_1-j_2+M}}{\sqrt{2J+1}} \:\langle j_1m_1j_2m_2|JM\rangle.
\end{equation}
The Wigner $6j$ symbol
\[
\left\{ \begin{array}{ccc} j_1 & j_2 & j_3 \\ J_1 & J_2 & J_3 \end{array} 
\right\}
\]
is used to combine three angular momentum states, and the Wigner $9j$ symbol
\[
\left\{\begin{array}{ccc} j_1 & j_2 & J_{12} \\
j_3 & j_4 & J_{34} \\ J_{13} & J_{24} & J
\end{array} \right\} 
\]
is used to combine four angular momentum states.  We will not define these
here, though the $9j$ symbol is defined in Appendix~\ref{a:spinoverlap}.  For
more information, see for example Reference~\cite{messiah:quantum}.

\section{Conservation Laws and Invariance Principles}
\label{a:conslaws}

Because of symmetries in the theories of the interactions, the interactions
themselves obey a number of conservation laws and invariance principles (see
for example Reference \cite{nguyen-khac68:applications}).  We will only discuss
the ones of interest here.

There is the usual conservation of energy, momentum, angular momentum and
charge.

If one thinks of an antiparticle as being $-1$ particles, then the numbers of
leptons and quarks are separately conserved as well -- this is enforced by
the allowed interaction vertices (e.g.\ if an $e^+e^-$ pair is created, the
change in lepton number is $(-1)+(+1)=0$).

Because the strong interaction is independent of the type of quark involved,
and the $u$ and $d$ quarks have very similar masses, we can think of them as
different projections of the same state which we represent by the angular
momentum-like quantity called isospin (where $u=|\frac{1}{2}\frac{1}{2}\rangle$
and $d=|\frac{1}{2}-\!\!\frac{1}{2}\rangle$).  Isospin is conserved in the
strong interaction to the extent that the masses of the $u$ and $d$ quarks are
degenerate.

The spin-statistics theorem governs how systems of identical particles behave
when two of the particles are exchanged (boson wavefunctions are unaffected,
fermion wavefunctions pick up a factor of $-1$).  It is important that the 
wavefunctions we construct obey these symmetries.  In the context of the
strong interactions, members of the same isospin multiplet also obey the
spin-statistics theorem, even though they are not strictly identical.

The parity of a system indicates how the wavefunction behaves under the space
inversion $\vec{r} \to -\vec{r}$; the wavefunction is either unchanged (${\cal
P}=+1$) or picks up a factor of $-1$ (${\cal P}=-1$).  The total parity of a
system is conserved in the electromagnetic and strong interactions, but not the
weak interaction.

Under the charge conjugation operator, particles are changed to
their antiparticles (and vice versa).  Systems which are eigenstates of ${\cal
C}$ (such as $\pi^0$, or the $\pi^+ \pi^-$ system) may have ${\cal C}=+1$ or
$-1$.  Charge conjugation is also conserved in the electromagnetic and
strong interactions, but not the weak interaction.  Charge conjugation can be
extended by combining it with isospin to get G-parity, which is conserved in
the strong interaction only.

\section{Relative Coordinates}
\label{a:relcoord}

The wavefunction of two particles in position-space depends on the coordinates
of the two particles, $\vec{r}_1$ and $\vec{r}_2$.  When solving the
Schr\"{o}dinger equation it is convenient to separate it into equations
involving the centre of mass (CM) coordinate of the particles, $R$, and their
relative coordinate $r$:
\renewcommand{\arraystretch}{1.5}
\begin{equation}
\begin{array}{rcc}
  \vec{R} &=& {\displaystyle \frac{m_1 \vec{r}_1 +m_2 \vec{r}_2}{m_1+m_2}} \\
  \vec{r} &=& \vec{r}_1 - \vec{r}_2
\end{array} 
\Longleftrightarrow
\begin{array}{rcl}
  \vec{r}_1 &=& \vec{R} + {\displaystyle \frac{m_2}{m_1+m_2}} \vec{r}\\
  \vec{r}_2 &=& \vec{R} - {\displaystyle \frac{m_1}{m_1+m_2}} \vec{r}
\end{array} .
\end{equation}  
The conjugate momenta $\vec{p}_1$ and $\vec{p}_2$, and $\vec{P}$ and $\vec{p}$,
are related by:
\begin{equation}
\begin{array}{rcc}
  \vec{P} &=& \vec{p}_1 + \vec{p}_2\\
  \vec{p} &=& {\displaystyle \frac{m_2 \vec{p}_1 -m_1 \vec{p}_2}{m_1+m_2}} 
\end{array} 
\Longleftrightarrow
\begin{array}{rcl}
  \vec{p}_1 &=& {\displaystyle \frac{m_1}{m_1+m_2}} \vec{P} + \vec{p}\\
  \vec{p}_2 &=& {\displaystyle \frac{m_2}{m_1+m_2}} \vec{P} - \vec{p}
\end{array} .
\end{equation}  
\renewcommand{\arraystretch}{1}
This is a nonrelativistic construction: $p$ remains unchanged when both
particles are given the same velocity boost only for nonrelativistic
velocities.

The CM frame refers to the inertial reference frame in which $\vec{P}=0$;
i.e.\ in which the centre of mass of the system is motionless.

\section{The OZI Rule}
\label{a:ozirule}

Okubo, Zweig and Iizuka (OZI) noticed that QCD interactions in which all of the
initial quarks are annihilated, and are not present in the final state, are
suppressed.  This is thought to be because for interactions of this type,
asymptotic freedom means that the strong coupling constant $\alpha_s$ (which
enters the cross-section as at least $\alpha_s^4$) becomes small since all of
the energy of the initial state particles must be included in the momentum
transferred through the gluons (see for example Reference
\cite{leyaouanc88:hadron}).

\section{Meson Mixings}
\label{a:mixings}

Particles with the same ``good'' quantum numbers (i.e.\ a complete set of
commuting observables) ``mix'' if there exists an interaction through which one
particle can be transformed into the other, and vice versa.  For example, the
strange mesons ($-u\bar{s}, -d\bar{s}, -s\bar{d}, s\bar{u}$) with states
$^1P_1$ and $^3P_1$ only differ in the sum of the quark spins, and can mix via
the spin-orbit interaction (and possibly others) to produce the physical states
$K_1(1270)$ and $K_1(1400)$.  Another example is found in the flavour
wavefunctions: the $u\bar{u}$, $d\bar{d}$ and $s\bar{s}$ are all flavourless
(since the flavour of the $u$ cancels that of the $\bar{u}$) and so there is
nothing to prevent the flavour states in a $^1S_0$ spin/space state (for
example) from mixing by an annihilation interaction to produce the physical
states $\pi^0$, $\eta$ and $\eta'$.

By mix, we mean that the basis of quantum number states (e.g.\ $K(^1P_1)$ and
K($^3P_1)$) differs from the basis of the physical states (e.g.\ $K_1(1270)$ and
$K_1(1400)$).  When we are only talking about two states, we can represent the
mixing by a rotation:
\begin{equation}
\left( \begin{array}{c} K_1(1270)^+ \\ K_1(1400)^+ \end{array} \right) =
\left( \begin{array}{cc} \cos{\theta} & \sin{\theta} \\
                         -\sin{\theta} & \cos{\theta} \end{array} \right)
\left( \begin{array}{c} K(^1P_1)^+ \\ K(^3P_1)^+ \end{array} \right).
\label{mixingeg}
\end{equation}
For this particular mixing, we take $\theta=45^\circ$ \cite{montanet94:review}
for the calculations of Section~\ref{3:xi}.  In Section~\ref{3:k1}, we attempt
to determine this mixing angle by comparing predictions of meson decay models
to experimental data.

In terms of the amplitudes and widths, Eq.~\ref{partwavwidth} becomes with 
mixing, using a decay involving the $K_1(1270)^+$ for example,
\begin{equation}
\Gamma^{S L}_{K_1(1270)^+} = \frac{\pi}{4} \frac{P {\cal S}}{M_A^2} 
|M^{S L}_{K_1(1270)^+}|^2 =  \frac{\pi}{4} \frac{P {\cal S}}{M_A^2}
|\cos{\theta}\:M^{S L}_{K(^1P_1)^+}+\sin{\theta} \:M^{S L}_{K(^3P_1)^+}|^2.
\end{equation}

In Eq.~\ref{mixingeg}, note that we specified the positively charged mesons,
because the flavour makes a difference to the mixing angle for these particular
mesons.  This comes about in the quark model Hamiltonian -- the term that
causes the mixing, $\langle H^{SO-}_{q\bar{q}} \rangle$, changes sign when the
antiquark instead of the quark is the heavier strange, and this leads to a
mixing angle of opposite sign when the $^1P_1$--$^3P_1$ Hamiltonian is
diagonalized (see Eqs.~\ref{massmatrix} and \ref{diagcond}).  Thus if we have a
mixing angle of $45^\circ$ for the $K_1(1270)^+$ and $K_1(1400)^+$ (and
$K_1(1270)^0$ and $K_1(1400)^0$), we must use $-45^\circ$ for the $K_1(1270)^-$
and $K_1(1400)^-$ (and $\bar{K}_1(1270)^0$ and $\bar{K}_1(1400)^0$).

The other meson mixing that we use in this work\footnote{We did
initially consider similar mixings of other mesons, but the widths of
the decays involving these mesons were too small to be included in our
results.} is due to the tensor part of the colour-hyperfine
interaction \cite{godfrey85:mesons}:
\begin{equation}
\left( \begin{array}{c} K^*(1410) \\ K^*(1680) \end{array} \right) =
\left( \begin{array}{cc} 1.00 & 0.04 \\
                         -0.04 & 1.00 \end{array} \right)
\left( \begin{array}{c} K(2^3S_1) \\ K(1^3D_1) \end{array} \right).
\end{equation}

\chapter{Meson Wavefunctions Used in this Work}
\markright{Appendix B.  Meson Wavefunctions Used in this Work}

As can be seen from Eq.~\ref{mockmeson}, meson wavefunctions contain four
different component wavefunctions: space, spin, flavour and colour.  The space
and flavour wavefunctions are given below.  The spin wavefunctions are given in
Appendix~\ref{a:spinoverlap}.  The colour wavefunction is trivially a colour
singlet (see Appendix~\ref{a:colouroverlap}).

\section{Space Wavefunctions}
\label{a:space}

In this work we investigate three different sets of meson space wavefunctions,
though only two are used for calculations.  All three are expressed in terms of
simple harmonic oscillator (SHO) wavefunctions.  In momentum-space the SHO
wavefunctions are
\begin{equation}
\psi^{\rm SHO}_{n L M_L}\!(\vec{p})=R_{nL}^{\rm SHO}(p) \:
Y_{L M_L}(\Omega_p)
\end{equation}
where the radial wavefunctions are given by 
\begin{equation}
R_{nL}^{\rm SHO}(p) = \frac{(-1)^n (-i)^L}{\beta^\frac{3}{2}}
\sqrt{\frac{2 \:n!}{\Gamma(n+L+{\scriptstyle \frac{3}{2}})}} 
\left(\frac{p}{\beta}\right)^L
\dspexp{-p^2/(2\beta^2)} \:L_n^{L+\frac{1}{2}}(p^2/\beta^2),
\end{equation}
the oscillator parameter $\beta$ contains all the parameters of the
Hamiltonian, and $L_n^{L+\frac{1}{2}}(p^2/\beta^2)$ is an associated Laguerre
polynomial.  The quantum numbers $n$ and $L$ are unrelated, and have the
ranges: $n$: 0,1,2,..., and $L$: 0,1,2,...

In position-space we have the Fourier transform of the above,
\begin{equation}
\psi^{\rm SHO}_{n L M_L}\!(\vec{r})=R_{nL}^{\rm SHO}(r) \:
Y_{L M_L}(\Omega_r),
\end{equation}
where the radial wavefunctions are given by 
\begin{equation}
R_{nL}^{\rm SHO}(r) = \beta^\frac{3}{2}
\sqrt{\frac{2 \:n!}{\Gamma(n+L+{\scriptstyle \frac{3}{2}})}} 
\left(\beta r\right)^L
\dspexp{-\beta^2 r^2/2} \:L_n^{L+\frac{1}{2}}(\beta^2 r^2).
\end{equation}

In some cases it may be convenient to pull the $p^L$ or $r^L$ out of the radial
wavefunction, and associate it with the spherical harmonic to create a solid 
harmonic.

The first set of meson space wavefunctions that we use for this work are just
the SHO wavefunction with the same $L$ and $M_L$ as the meson, and the same
degree of radial excitation (so a meson in its radial ground state, normally
denoted with $n=1$ in the $n^{2S+1}L_J$ spectroscopic notation, uses $\psi^{\rm
SHO}_{n=0, L M_L}$).  The oscillator parameter $\beta$ is taken to be the same
for all mesons, and we examine the choice of it in Section~\ref{3:parameters}.
SHO wavefunctions are expected to be coarse approximations to the true
wavefunctions, since they neglect all but the linear confining part of the
meson potential, and that they only approximate.  However, they are
qualitatively similar, and are useful for generating analytical results.

The second set of meson space wavefunctions that we use are those of Godfrey
and Isgur \cite{godfrey85:mesons}, calculated in the highly successful
relativized quark model described in Section~\ref{1:quarkmodel}.  We
will label these wavefunctions by RQM.  The radial parts of these wavefunctions
are expressed as linear combinations of the first $N+1$ radial SHO
wavefunctions (as in Eq.~\ref{shoexpansion}), where $N$ was chosen to give a
good description of the particular wavefunction.  The oscillator parameter was
fit individually for each meson, and the quark masses were also fitted: $m_u =
220$~MeV, $m_d = 220$~MeV, and $m_s = 419$~MeV.

The third set of meson space wavefunctions that we use are single SHO
wavefunctions like the first set, but with the effective oscillator parameters
(different for each meson) of Kokoski and Isgur \cite{kokoski87:meson}.  The
effective parameters were obtained by fitting the rms momenta of the SHO
wavefunctions to be equal to those obtained from the second set of
wavefunctions.  We do not use this third set of wavefunctions for actual
calculations because of their poor performance in Section~\ref{3:parameters}.

\section{Flavour Wavefunctions}
\label{a:flavours}

We use the following meson flavour wavefunctions in this work:

\noindent For the isovectors:
\begin{eqnarray*}
\pi^+ &=& -u\bar{d} \\
\pi^0 &=& \frac{1}{\sqrt{2}}(u\bar{u}-d\bar{d}) \\
\pi^- &=& d\bar{u}.
\end{eqnarray*}
For the strange mesons:
\begin{eqnarray*}
K^+ &=& -u\bar{s} \\
K^0 &=& -d\bar{s} \\
\bar{K}^0 &=& -s\bar{d} \\
K^- &=& s\bar{u}.
\end{eqnarray*}
For the isoscalars we assumed ``ideal mixing'',
\begin{eqnarray*}
\phi_{\rm non strange} &=& \frac{1}{\sqrt{2}}(u \bar{u}+d \bar{d}) \\
\phi_{\rm strange} &=& s \bar{s},
\end{eqnarray*}
except for the pseudoscalars in a radial ground state, where we assumed
``perfect mixing'',
\begin{eqnarray*}
\eta &=& \frac{1}{\sqrt{2}}(\phi_{\rm non strange}-\phi_{\rm strange})
\;\;=\;\; \frac{1}{2} (u \bar{u}+d \bar{d}) - \frac{1}{\sqrt{2}} 
s \bar{s} \\
\eta' &=& \frac{1}{\sqrt{2}}(\phi_{\rm non strange}+\phi_{\rm strange}) 
\;\;=\;\; 
\frac{1}{2} (u \bar{u}+d \bar{d}) + \frac{1}{\sqrt{2}} s \bar{s}. 
\end{eqnarray*}
In terms of the SU(3) flavour representation \cite{montanet94:review}, 
\begin{eqnarray*}
\eta &=& \eta_8 \cos{\theta_P}-\eta_1 \sin{\theta_P}\\
\eta' &=& \eta_8 \sin{\theta_P}+\eta_1 \cos{\theta_P}
\end{eqnarray*}
where
\begin{eqnarray*}
\eta_1 &=& \sqrt{\frac{2}{3}}\phi_{\rm non strange}+\frac{1}{\sqrt{3}}
\phi_{\rm strange}\\
\eta_8 &=& \frac{1}{\sqrt{3}}\phi_{\rm non strange}-\sqrt{\frac{2}{3}}
\phi_{\rm strange}, 
\end{eqnarray*}
ideal mixing corresponds to $\theta_P=-54.7^\circ$, while perfect mixing
corresponds to $\theta_P=-9.7^\circ$.  The latter number is consistent with the
angle obtained from the $\eta$--$\eta'$ mass matrix, while data from decay to
two photons suggests $\theta_P\simeq-20^\circ$ \cite{montanet94:review}.  The
use of this second value is considered in Chapter~\ref{3:mesonapp}.

\chapter{Field Theory Conventions Used in this Work}
\markright{Appendix C.  Field Theory Conventions Used in this Work}
\label{a:fieldtheorycon}

We use the following field theory conventions in this work:

For the spinor normalizations,
\begin{eqnarray*}
u^{(\alpha)\dagger}(p) \, u^{(\beta)}(p) = v^{(\alpha)\dagger}(p) \, 
v^{(\beta)}(p) &=& \frac{E}{m} \, \delta_{\alpha \beta} \\
u^{(\alpha)\dagger}(p) \, v^{(\beta)}(\bar{p}) = v^{(\alpha)\dagger}(p) 
\, u^{(\beta)}(\bar{p}) &=& 0,
\end{eqnarray*}
where if $p=(E,\vec{p})$, then $\bar{p}\equiv(E,-\vec{p})$.

We define the spinor field
\[
\Psi(x) = \sum_{\alpha} \int\!\frac{{\rm d}^3\vec{p}}{(2\pi)^{\frac{3}{2}}}\;
\sqrt{\frac{m}{E}} \;[b_\alpha(p) \:u^{(\alpha)}(p) \:\dspexp{-i p\cdot x}+
d_\alpha^\dagger(p) \:v^{(\alpha)}(p) \:\dspexp{i p\cdot x}],
\]
and use the anticommutators
\begin{eqnarray*}
\{\Psi_i(t,\vec{x}),\Psi_j^\dagger(t,\vec{x}')\} &=& 
\delta^3(\vec{x}-\vec{x}') \,\delta_{ij} \\
\{\Psi_i(t,\vec{x}),\Psi_j(t,\vec{x}')\} = 
\{\Psi_i^\dagger(t,\vec{x}),\Psi_j^\dagger(t,\vec{x}')\} &=& 0 \\
\{b_\alpha(p),b_{\alpha'}^\dagger(p')\} = 
\{d_\alpha(p),d_{\alpha'}^\dagger(p')\} &=&  
\delta^3(\vec{p}-\vec{p}') \,\delta_{\alpha\alpha'} \\
\{b_\alpha(p),b_{\alpha'}(p')\} = 
\{d_\alpha(p),d_{\alpha'}(p')\} &=& 0 \\
\{b_\alpha^\dagger(p),b_{\alpha'}^\dagger(p')\} = 
\{d_\alpha^\dagger(p),d_{\alpha'}^\dagger(p')\} &=& 0, 
\end{eqnarray*}
which gives for the normalization of one particle states
\[
\langle b_\alpha(p)|b_{\alpha'}(p')\rangle = 
\langle d_\alpha(p)|d_{\alpha'}(p')\rangle = 
\delta^3(\vec{p}-\vec{p}') \,\delta_{\alpha\alpha'}.
\]

\chapter[Evaluating the Colour, Flavour and Spin Overlaps for Models \protect\\
of Meson Decay]{Evaluating the Colour, Flavour and Spin Overlaps for
Models of Meson Decay}
\markright{Appendix D.  Evaluating the Colour, Flavour and Spin Overlaps...}
\label{a:overlaps}

In the $^3P_0$ and flux-tube breaking models of meson decay, the overlaps of 
the colour, flavour, spin and space wavefunctions of the mesons and the 
created pair must be calculated.  The overlap of the space wavefunctions is 
accomplished by the integrals of Eqs.~\ref{3P0integral} and \ref{fluxintegral}.
The other overlaps are discussed below.

\section{Colour Overlap}
\label{a:colouroverlap}

The calculation of the colour overlap is particularly simple, because 
the mesons and the created pair are all colour 
singlets, with the wavefunction 
$\omega = \frac{1}{\sqrt{3}} (R \bar{R} + G \bar{G} + B \bar{B})$ (using
the arbitrary three colours Red, Green, Blue). 

This may be represented in matrix form (this may seem excessive now, but will
be useful for the flavour overlaps as well) by 
\begin{equation}
\omega = \frac{1}{\sqrt{3}} \left(\begin{array}{ccc}
1 & 0 & 0 \\
0 & 1 & 0 \\
0 & 0 & 1 
\end{array}\right)
\end{equation}
where the rows indicate a quark of colour $R$, $G$, $B$, and the columns
indicate antiquarks of color $\bar{R}$, $\bar{G}$, $\bar{B}$.

Consider a simple case with only one meson on each side, $\langle \omega_A^{12}
| \omega_B^{12} \rangle$.  The superscripts indicate that the quark of $A$
(labelled 1) is also the quark of $B$, and similarly for the antiquark
(labelled 2).  A particular colour combination (e.g.\ $R \bar{R}$) will only
contribute to the overlap if it is found on both sides of the bra-ket, and the
resulting term is just the product of the coefficients in the matrices for that
colour combination.  Then the total result is just the sum of these terms over
the possible colours,
\begin{equation}
\langle \omega_A^{12}| \omega_B^{12} \rangle = \sum_{c_1,c_2} (\omega_A)_
{c_1 c_2} (\omega_B)_{c_1 c_2} = \sum_{c_1,c_2} (\omega_A)_
{c_1 c_2} (\omega_B^T)_{c_2 c_1} = {\rm Tr}[\omega_A \omega_B^T],
\end{equation}
where $c_1$ ($c_2$) runs over the quark (antiquark) colours.

This is easily expanded to the case where there are two mesons on each side 
(e.g.\ for the first diagram of Figure~\ref{f:twodiag}):
\begin{eqnarray}
\langle \omega_B^{14} \omega_C^{32} | \omega_A^{12} \omega_0^{34}\rangle &=& 
\sum_{c_1,c_2,c_3,c_4} (\omega_B)_{c_1 c_4} (\omega_C)_{c_3 c_2} 
(\omega_A)_{c_1 c_2} (\omega_0)_{c_3 c_4} \label{colouroverlap} \\
&=& \sum_{c_1,c_2,c_3,c_4} (\omega_A^T)_{c_2 c_1} (\omega_B)_{c_1 c_4} 
(\omega_0^T)_{c_4 c_3} (\omega_C)_{c_3 c_2}  
= {\rm Tr}[\omega_A^T \omega_B \omega_0^T \omega_C].\nonumber 
\end{eqnarray}
For the case where everything is a colour singlet, the results are
\begin{equation}
\langle \omega_B^{14} \omega_C^{32} | \omega_A^{12} \omega_0^{34}\rangle =
\langle \omega_B^{32} \omega_C^{14} | \omega_A^{12} \omega_0^{34}\rangle =
\frac{1}{3}. 
\end{equation}

\section{Flavour Overlap}
\label{a:flavouroverlap}

The overlap of the flavour wavefunctions of the mesons and the created pair can
be calculated using the matrix notation introduced in
Appendix~\ref{a:colouroverlap}.  The flavour of a meson may be represented by a
matrix where the rows indicate a quark of flavour $u$, $d$, $s$,... and the
columns indicate antiquarks of color $\bar{u}$, $\bar{d}$, $\bar{s}$,...  For
example, consider the $\pi^0$,
\begin{equation}
\ \phi_{\pi^0}= \frac{1}{\sqrt{2}}(u\bar{u}-d\bar{d}) = \frac{1}{\sqrt{2}} 
\left(\begin{array}{ccc}
1 & 0 & 0 \\
0 & -1 & 0 \\
0 & 0 & 0 
\end{array}\right),
\end{equation}
and the $K^+$,
\begin{equation}
\ \phi_{K^+}= -u\bar{s} = 
\left(\begin{array}{ccc}
0 & 0 & -1 \\
0 & 0 & 0 \\
0 & 0 & 0 
\end{array}\right).
\end{equation}

The $^3P_0$ and flux-tube breaking models assume that the pair is created in 
an SU(N) flavour singlet;
we take N = 3, since we are not concerned with any mesons involving c, b or t 
quarks:
\begin{equation}
\ \phi_0= \frac{1}{\sqrt{3}}(u\bar{u}+d\bar{d}+s\bar{s}) = \frac{1}{\sqrt{3}} 
\left(\begin{array}{ccc}
1 & 0 & 0 \\
0 & 1 & 0 \\
0 & 0 & 1  
\end{array}\right). \label{flavoursinglet}
\end{equation}
Taking N to be some other number would change the value of $\gamma$
($\gamma_0$) needed to fit the data (because of the normalization in
Eq.~\ref{flavoursinglet}), but would not change the calculated decay widths.

The overlap is given by expressions similar to Eq.~\ref{colouroverlap}: 
\begin{eqnarray}
\langle \phi_B^{14} \phi_C^{32} | \phi_A^{12} \phi_0^{34}\rangle &=&
{\rm Tr}[\phi_A^T \phi_B \phi_0^T \phi_C] \;\;=\;\; \frac{1}{\sqrt{3}} \:
{\rm Tr}[\phi_A^T \phi_B \phi_C] \\
\langle \phi_B^{32} \phi_C^{14} | \phi_A^{12} \phi_0^{34}\rangle &=&
{\rm Tr}[\phi_A^T \phi_C \phi_0^T \phi_B] \;\;=\;\; \frac{1}{\sqrt{3}}\:
{\rm Tr}[\phi_A^T \phi_C \phi_B].
\end{eqnarray}
In many cases only one of these overlaps will be non-zero.

As an example, consider the flavour overlaps in the decay $K^*(892)^+\to K^+
\pi^0$; $\phi_{K^*(892)^+} =\phi_{K^+}$, giving us
\begin{eqnarray}
\langle \phi_B^{14} \phi_C^{32} | \phi_A^{12} \phi_0^{34}\rangle &=&
0 \nonumber \\
\langle \phi_B^{32} \phi_C^{14} | \phi_A^{12} \phi_0^{34}\rangle &=&
\frac{1}{\sqrt{6}}.
\end{eqnarray}

\section{Spin Overlap}
\label{a:spinoverlap}

Since quarks have spin $\frac{1}{2}$, their spins may only combine to give 0 or
1 in a meson.  They combine in the usual singlet and triplet states (given as
e.g.\ $|S_A M_{S_A}\rangle$):
\begin{eqnarray}
|00\rangle &=& \frac{1}{\sqrt{2}} \left[
|{\scriptstyle\frac{1}{2}\frac{1}{2}}\rangle
|{\scriptstyle\frac{1}{2}-\!\frac{1}{2}}\rangle - 
|{\scriptstyle\frac{1}{2}-\!\frac{1}{2}}\rangle
|{\scriptstyle\frac{1}{2}\frac{1}{2}}\rangle \right] \nonumber \\
\left\{\begin{array}{c}
|11\rangle \\ |10\rangle \\ |1\!-\!\!1\rangle \end{array} \right. \!\!\!\!\!
&\begin{array}{c} = \\ = \\ = \end{array} 
&\begin{array}{l} 
|{\scriptstyle\frac{1}{2}\frac{1}{2}}\rangle 
|{\scriptstyle\frac{1}{2}\frac{1}{2}}\rangle \\
{\displaystyle \frac{1}{\sqrt{2}}} \left[
|{\scriptstyle\frac{1}{2}\frac{1}{2}}\rangle
|{\scriptstyle\frac{1}{2}-\!\frac{1}{2}}\rangle + 
|{\scriptstyle\frac{1}{2}-\!\frac{1}{2}}\rangle
|{\scriptstyle\frac{1}{2}\frac{1}{2}}\rangle \right] \\
|{\scriptstyle\frac{1}{2}-\!\frac{1}{2}}\rangle 
|{\scriptstyle\frac{1}{2}-\!\frac{1}{2}}\rangle
\end{array}.
\end{eqnarray} 

The overlap of the spin wavefunctions of the mesons and the created pair can be
calculated by angular momentum algebra.  It can be expanded in a sum over
possible total spins (e.g.\ for the first diagram of Figure~\ref{f:twodiag})
\begin{eqnarray}
\lefteqn{\langle \chi^{1 4}_{S_B M_{S_B}} \chi^{3 2}_{S_C M_{S_C}} |
\chi^{1 2}_{S_A M_{S_A}} \chi^{3 4}_{1 -\!m} \rangle =} \nonumber \\
&& \sum_{S,M_S}
\langle(j_1 j_4) S_B,(j_3 j_2) S_C;S M_S |(j_1 j_2) S_A,(j_3 j_4) 1;S M_S
\rangle \nonumber \\
&& \times \langle S_B M_{S_B} S_C M_{S_C} | S M_S \rangle \:
\langle S_A M_{S_A} 1 \!-\!\!m |S M_S \rangle 
\end{eqnarray}
where the $j_i$'s are all $\frac{1}{2}$, but are necessarily written this way
to show the connection between the left and right sides of the first factor.
The first factor is a transformation between two different bases for combining
four angular momenta, and is independent of $M_S$.  On the left side, $j_1$ and
$j_4$ are added together to obtain $S_B$, which is added to $S_C$ to obtain $S$
-- a similar procedure is carried out on the right side.  The $9j$ symbol is
defined in terms of this transformation,
\begin{eqnarray}
\lefteqn{\langle(j_1 j_4) J_{14},(j_3 j_2) J_{32};J M |(j_1 j_2) J_{12},
(j_3 j_4) J_{34}; J M \rangle =} \\ 
&&(-1)^{2J_{12}+2J_{14}+j_2-j_4+J_{32}-J_{34}} 
\sqrt{(2J_{12}+1)(2J_{34}+1)(2J_{14}+1)(2J_{32}+1)} \nonumber \\
&&\times \left\{
\begin{array}{ccc}
j_1 & j_2 & J_{12} \\
j_4 & j_3 & J_{34} \\
J_{14} & J_{32} & J
\end{array} \right\}, \nonumber 
\end{eqnarray}
which allows us to write the spin overlap in terms of a $9j$ symbol:
\begin{eqnarray}
\lefteqn{\langle \chi^{1 4}_{S_B M_{S_B}} \chi^{3 2}_{S_C M_{S_C}} |
\chi^{1 2}_{S_A M_{S_A}} \chi^{3 4}_{1 -\!m} \rangle =} \nonumber \\
&& (-1)^{S_C+1} \sqrt{3(2S_A+1)(2S_B+1)(2S_C+1)} \sum_{S,M_S} 
\langle S_B M_{S_B} S_C M_{S_C} | S M_S \rangle \nonumber \\
&& \times \langle S_A M_{S_A} 1 \!-\!\!m |S M_S \rangle 
\left\{
\begin{array}{ccc}
\frac{1}{2} & \frac{1}{2} & S_A \\
\frac{1}{2} & \frac{1}{2} & 1 \\
S_B & S_C & S
\end{array} \right\}.
\end{eqnarray}

Using an alternative definition of the $9j$ symbol that couples the quarks 
differently, we can get a similar expression for the spin overlap for the 
second diagram of Figure~\ref{f:twodiag} which satisfies
\begin{equation}
\langle \chi^{3 2}_{S_B M_{S_B}} \chi^{1 4}_{S_C M_{S_C}} |
\chi^{1 2}_{S_A M_{S_A}} \chi^{3 4}_{1 -\!m} \rangle = 
(-1)^{1+S_A+S_B+S_C} \: \langle \chi^{1 4}_{S_B M_{S_B}} 
\chi^{3 2}_{S_C M_{S_C}} |
\chi^{1 2}_{S_A M_{S_A}} \chi^{3 4}_{1 -\!m} \rangle. \label{spintrans}
\end{equation}
Eq.~\ref{spintrans} was used to simplify Eq.~\ref{amplitude}.

\chapter[Converting to Partial Wave Amplitudes for Models of Meson\protect\\ 
Decay]{Converting to Partial Wave Amplitudes for Models of Meson Decay}
\markright{Appendix E.  Converting to Partial Wave Amplitudes...}
\label{a:partialwaves}

The decay amplitudes of the $^3P_0$ and flux-tube breaking models derived in
Sections \ref{2:3P0} and \ref{2:fluxtube}, $M^{M_{J_A} M_{J_B} M_{J_C}}$, are
given for a particular basis of the final state:
$|\theta,\phi,M_{J_B},M_{J_C}\rangle
\equiv |\Omega,M_{J_B},M_{J_C}\rangle$.  Here $\theta$ and $\phi$ are the
spherical polar angles of the outgoing momentum of meson B ($\vec{P}\equiv
\vec{P_B}$) 
in the CM frame.

We would prefer to calculate amplitudes for particular outgoing partial waves,
$|J,M,S,L\rangle$, since they are what are measured experimentally.  Here
$|J,M\rangle$ are the quantum numbers of the total angular momentum of the
final state, $|S,M_S\rangle$ are the quantum numbers for the sum of the total
angular momenta of B and C, and $|L,M_L\rangle$ are the quantum numbers for the
orbital angular momentum between B and C.

The formula for the decay width in terms of partial wave amplitudes is 
different from Eq.~\ref{width}:
\begin{equation}
\Gamma = \sum_{S,L} \Gamma^{S L}
\end{equation}
where
\begin{equation}
\Gamma^{S L} = \frac{\pi}{4} \frac{P {\cal S}}{M_A^2} |M^{S L}|^2. 
\label{partwavwidth}
\end{equation}
$M^{S L}$ is a partial wave amplitude, and $\Gamma^{S L}$ is the partial width 
of that partial wave.

We use two methods to convert our calculated amplitudes to the partial wave
basis \cite{chung71:spin}:  by use of a recoupling calculation, and by use of 
the Jacob-Wick Formula.

\section{Converting by a Recoupling Calculation}

The recoupling transformation directly relates the bases
$|\Omega,M_{J_B},M_{J_C}\rangle$
and $|J,M,S,L\rangle$:
\begin{eqnarray}
| J,M,S,L\rangle &=& \!\!\! \sum_{M_{J_B},M_{J_C},M_S,M_L} \!
\langle L M_L S M_S|J M \rangle \:
\langle J_B M_{J_B} J_C M_{J_C} | S M_S \rangle \nonumber\\
&&\times \int\!{\rm d}\Omega\;
Y_{L M_L}\!(\Omega) \;| \Omega,M_{J_B},M_{J_C}\rangle. 
\end{eqnarray}
In terms of the amplitudes, the net result we want is the transformation
\begin{eqnarray}
M^{S L}(P) &=& \sum_{M_{J_B},M_{J_C},M_S,M_L} \!\!\!
\langle L M_L S M_S|J_A M_{J_A} \rangle \:
\langle J_B M_{J_B} J_C M_{J_C} | S M_S \rangle \nonumber\\
&&\times \int\!{\rm d}\Omega\; 
Y^*_{L M_L}\!(\Omega)\; M^{M_{J_A} M_{J_B} M_{J_C}}(\vec{P}). 
\label{recoupling}
\end{eqnarray}
Note that this result holds for any value of $M_{J_A}$, since $M^{S L}(P)$ is
independent of $M_{J_A}$.  If we wanted to avoid favouring one particular
value, we could sum over $M_{J_A}$, and convert the sum on the left side to a
factor of $1/(2J_A+1)$ on the right side.

\section{Converting with the Jacob-Wick Formula}

The Jacob-Wick formula
\begin{equation}
| J,M,S,L\rangle = \! \sum_{\lambda_B,\lambda_C} \! 
\sqrt{\frac{2L+1}{2J+1}}\:
\langle L 0 S \lambda|J \lambda \rangle \:
\langle J_B \lambda_B J_C -\!\!\lambda_C  | S \lambda \rangle \:
|J,M,\lambda_B,\lambda_C\rangle
\end{equation}
relates the partial wave basis $|J,M,S,L\rangle$ to the 
helicity basis $|J,M,\lambda_B,\lambda_C\rangle$, where $\lambda_B$ and 
$\lambda_C$ are the helicities of B and C, respectively.  To use it we must 
first relate the helicity basis
to the basis $|\Omega,\lambda_B,\lambda_C\rangle$,
\begin{equation}
|J,M,\lambda_B,\lambda_C\rangle = \sqrt{\frac{2J+1}{4\pi}} \:
\int\!{\rm d}\Omega\; D^{J*}_{M\lambda}(\phi,\theta,0) \;
|\Omega,\lambda_B,\lambda_C\rangle
\end{equation}
where $D^{J*}_{M\lambda}(\phi,\theta,0)$ is a rotation matrix, 
and then choose $\vec{P}$ to lie 
along the positive z
axis (in the CM frame still), so that $\lambda_B = M_{J_B}$, $\lambda_C = -
M_{J_C}$, and $|\Omega,M_{J_B},M_{J_C}\rangle = 
|\Omega,\lambda_B,-\lambda_C\rangle$.  

The final transformation between the amplitudes is
\begin{eqnarray}
M^{S L}(P) &=& \frac{\sqrt{4 \pi (2 L+1)}}{2 J_A +1} \!\!
\sum_{M_{J_B},M_{J_C}} 
\langle L 0 S (M_{J_B}\!+\!M_{J_C})|J_A (M_{J_B}\!+\!M_{J_C})\rangle \\
&&\times \langle J_B M_{J_B} J_C M_{J_C} | S (M_{J_B}\!+\!M_{J_C}) \rangle \:
M^{(M_{J_A}=M_{J_B}+M_{J_C}) M_{J_B} M_{J_C}}(P \hat{z}). \nonumber 
\end{eqnarray}
Note that $M_{J_A}$ in the calculated amplitude is replaced by 
$M_{J_B}+M_{J_C}$.

\markright{Bibliography}
\bibliographystyle{unsrt}
\addcontentsline{toc}{chapter}{Bibliography}
\bibliography{mesons}
\end{document}